\documentclass[reprint,amsmath,amssymb,aps,prb,showkeys,nofootinbib]{revtex4-2}
\usepackage[utf8]{inputenc}
\usepackage{hyperref}
\usepackage{listings}
\usepackage{graphicx}
\usepackage{float}
\usepackage{subfigure}
\usepackage{color}
\usepackage{comment}

\newcommand{\blue}[1]{{\color{blue} {#1}}}

\newcommand{\metric}{\mathfrak{d}}

\newcommand{\ZZ}{\mathbb{Z}}
\newcommand\RR{\mathbb{R}}
\newcommand\DeltaPhi{\boldsymbol{\Delta}\boldsymbol{\varphi}}
\newcommand\DeltaPhiHat{\widehat{\boldsymbol{\Delta \varphi}}}
\newcommand\DeltaPsi{\boldsymbol{\Delta \psi}}
\newcommand\NablaBF{\boldsymbol{\nabla}}
\newcommand\NablaHat{\hat{\boldsymbol{\nabla}}}
\newcommand\NablaTilde{\widetilde{\boldsymbol{\nabla}}}
\newcommand\NablaTildePhi{\widetilde{\boldsymbol{\nabla}}\phi}

\newcommand\Etilde{\hat{\boldsymbol{\nabla}} \times \Qvec}
\newcommand\Ebar{\bar{\mathbf{E}}}
\newcommand{\Evec}{\mathbf{E}}

\newcommand{\Fvec}{\mathbf{F}}

\newcommand{\hvec}{\mathbf{h}}
\newcommand{\kvec}{\mathbf{k}}
\newcommand{\lvec}{\mathbf{l}}

\newcommand{\mvec}{\mathbf{m}}
\newcommand{\nvec}{\mathbf{n}}
\newcommand{\Pvec}{\mathbf{P}}
\newcommand{\Qvec}{\mathbf{Q}}
\newcommand{\qvec}{\mathbf{q}}
\newcommand{\rvec}{\mathbf{r}}
\newcommand{\tvec}{\mathbf{t}}
\newcommand{\vvec}{\mathbf{v}}
\newcommand{\wvec}{\mathbf{w}}
\newcommand{\xvec}{\mathbf{x}}

\newcommand{\UnitX}{\mathbf{e}_x}
\newcommand{\UnitY}{\mathbf{e}_y}
\newcommand{\UnitZ}{\mathbf{e}_z}

\newcommand{\wrm}{\textrm{w}}

\newcommand{\glb}{\left(}  % ' group left b' 
\newcommand{\grb}{\right)}  % ' group right b' 
\newcommand{\glc}{\left[}  % ' group left c'
\newcommand{\grc}{\right]}  % ' group right c' 
\newcommand{\gld}{\left\{}  % ' group left d' 
\newcommand{\grd}{\right\}}  % ' group right d'

\begin{document}

%\title{Emergent electrostatics in planar XY spin models: the bridge between topological order/nonergodicity and $U(1)$ symmetry breaking}
%\title{Emergent electrostatics in planar XY spin models: the bridge connecting topological order/nonergodicity with broken $U(1)$ symmetry}
\title{Emergent electrostatics in planar XY spin models: the bridge connecting topological order with broken $U(1)$ symmetry}
\author{Michael F. Faulkner}
\affiliation{Warwick Centre for Predictive Modelling and School of Engineering, University of Warwick, UK}
\affiliation{HH Wills Physics Laboratory, University of Bristol, UK}
\date{\today}

\begin{abstract}
Topological phases have been a central focus of condensed-matter physics for over 50 years.  Along with many experimental applications, they have provided much intellectual interest due to their characterization via some form of topological ordering, as opposed to the symmetry-breaking ordering of conventional continuous phase transitions.  This distinction is most subtle in the case of the Berezinskii--Kosterlitz--Thouless (BKT) transition as its experimental realizations appear to break $U(1)$ symmetry at low temperature.  It also presents two further paradoxes: i) its prototypical \emph{short-range} interacting planar XY spin model behaves as an emergent \emph{long-range} interacting electrolyte; ii) its topological ordering is not accompanied by a topological nonergodicity within the BKT picture.  This review paper addresses these three interconnected questions.  We review a series of papers that demonstrate that $U(1)$ symmetry is indeed broken, but within a broader framework than that traditionally used to characterize broken symmetry.  We discuss recovery of this symmetry by breaking velocity-symmetry in a deterministic Markov process.  We then expand on a modern field theory of the emergent electrolyte that maps directly from the spin field to an emergent lattice electric field governed by an augmented electrostatic Boltzmann distribution.  This local model of electrolyte physics resolves both the short-range--long-range paradox and the question of topological nonergodicity -- as in contrast with the BKT picture, it describes global topological defects and their nonergodic freezing by the topological ordering.  It also connects the broken $U(1)$ symmetry with the topological ordering, providing a comprehensive framework for broken symmetry at the transition.  We introduce long-time topological stability as a measure of topological nonergodicity -- within a general framework for weakly broken ergodicity.
\end{abstract}

\keywords{emergent electrostatics, topological nonergodicity, symmetry breaking, topological phases, Berezinskii--Kosterlitz--Thouless transition, 2DXY model, 2D electrolyte, superconducting films, superfluid films, cold-atom films, monolayer magnets, Josephson junctions, piecewise deterministic Markov processes, event-chain Monte Carlo}



\maketitle

%\tableofcontents

\section{Introduction}
\label{sec:Intro}

Emergent electrostatics has played a key role in the resolution of many complex phenomena in condensed-matter physics.  The recent discovery of magnetic monopoles~\cite{Castelnovo2008MagneticMonopoles,Jaubert2009SignatureOfMagneticMonopole} and magnetricity~\cite{Bramwell2009Measurement} in the frustrated magnet spin ice~\cite{harris1997geometrical} explained many of its macroscopic properties -- including its non-linear response to an external magnetic field in terms of the second Wien effect of electrolyte physics~\cite{Onsager1934Deviations,Kaiser2013OnsagersWienEffect,Kaiser2015ACWienEffectInSpinIce}.  The monopoles are quasiparticle divergences in the magnetic spin field.  They interact via an emergent analogue of the Coulomb law of electrostatics, and respond to an external magnetic field in analogy with the response of electrical charges to an electric field.  This picture of an emergent electrolyte -- or \emph{magnetolyte} -- of magnetic monopoles provided the theoretical framework for the non-linear response and many other Coulombic properties of spin-ice materials.  

This success was due to the magnetic monopoles being local topological defects in the magnetic spin field.  Emergent electrostatics has been similarly successful in resolving other complex phenomena governed by topological defects, capturing departures from the mean-field description of Landau theory.  In two spatial dimensions, it explains the Berezinskii--Kosterlitz--Thouless (BKT) phase transition~\cite{Berezinskii1973DestructionLongRangeOrder,Kosterlitz1973OrderingMetastability} in the two-dimensional (2D) XY spin/condensate-phase model.  The transition governs physics as varied as the melting of films of hard disks and weakly pinned superconducting vortices~\cite{Kosterlitz1973OrderingMetastability,Halperin1978TheoryTwoDimensionalMelting,Young1979Melting,Bernard2011TwoStepMelting,Thorneywork2017TwoDimensionalMelting,Maccari2023TransportSignatures}; quantum gases on a one-dimensional ring lattice~\cite{Roscilde2016FromQuantum}; 2D arrays of superconducting qubits~\cite{King2018ObservationOfTopological}, Josephson junctions~\cite{Wolf1981TwoDimensionalPhaseTransition,Resnick1981KosterlitzThouless} and Bose--Einstein condensates~\cite{Trombettoni}; and planar superfluids~\cite{Bishop1978StudySuperfluid,Bramwell2015PhaseOrder},  superconductors~\cite{Baity2016EffectiveTwoDimensionalThickness,Shi2016EvidenceCorrelatedDynamics}, magnets~\cite{Bramwell1993Magnetization,Huang1994MagnetismFewMonolayersLimit,Elmers1996CriticalPhenomenaTwoDimensionalMagnet,BedoyaPinto2021Intrinsic2DXYFerromagnetism} and cold-atom systems~\cite{Hadzibabic2006KosterlitzThouless,Fletcher2015ConnectingBKTAndBEC,Christodoulou2021Observation}.  Elsewhere, evidence was found for a BKT-like phase transition with respect to quarternions in four spatial dimensions~\cite{Gorham2018SU2OrientationalOrdering}, and further extensions to octonions may have consequences for phase transitions in the standard model of particle physics.  The spin/condensate phases describe either the relative orientations of the hard disks, the phases of the magnetic spins or the local phase of the condensate wavefunction, but for simplicity we will tend to refer to spins and spin phases.

The low-temperature thermodynamic phase is characterized by algebraic spin-spin correlations~\cite{Berezinskii1973DestructionLongRangeOrder,Kosterlitz1973OrderingMetastability}, while the transition to the high-temperature disordered phase is induced by the thermal dissociation of bound pairs of vortices -- or local topological defects -- in the phase-difference field~\cite{Kosterlitz1973OrderingMetastability} (n.b., the phase-difference field is defined in Section~\ref{sec:HXYModel} and measures spatial variations in the spin-phase field).  KT~\cite{Kosterlitz1973OrderingMetastability} observed that the vortex pairs map to charge-neutral pairs of particles in the 2D electrolyte, in which the confinement-deconfinement transition was first discovered by Salzberg \& Prager~\cite{Salzberg1963EquationOfStateTwoDimensional}.  Villain~\cite{Villain1975TheoryOneAndTwoDimensionalMagnets} then constructed an approximation of the 2DXY model with a purely quadratic potential, which Jos\'{e} {\it et al}.~\cite{Jose1977Renormalization} used to decouple the %phase vortices 
emergent electrostatic charges from the continuous spin-wave fluctuations -- the second of the two local principal excitations.  This combined with the work of Nelson \& Kosterlitz~\cite{Nelson1977UniversalJump} to characterize the transition in terms of a universal jump in the spin stiffness -- used by Bishop \& Reppy~\cite{Bishop1978StudySuperfluid} to model the superfluid stiffness in superfluid films.

%This emergent-electrostatic picture describes the topological nature of the transition, with Jos\'{e} {\it et al}. characterizing the topological ordering of the low-temperature phase in terms of a nonzero \emph{spin stiffness}.  This presented, however, a paradox: order typically induces a loss of ergodicity under certain dynamics~\cite{Palmer1982}, but topological order under the above framework does not identify a topological nonergodicity.
The above emergent-electrostatic picture was hugely successful in %describing the topological nature of the transition 
characterizing both the phase transition and the notion of topological order in terms of the spin stiffness, but it immediately presented two paradoxes.  i) The \emph{short-range} interacting 2DXY model maps to an electrolyte of \emph{long-range} interacting electrostatic charges.  ii) Order typically induces a nonergodicity under certain dynamics~\cite{Palmer1982}, but the above framework could not discern whether the onset of topological order does indeed break some form of topological ergodicity.
%Moreover, while the above pioneering work was fundamental to modelling the superfluid experiments of Bishop \& Reppy~\cite{Bishop1978StudySuperfluid}, connection with other experimental systems remained elusive... 
In addition, the Mermin--Wagner--Hohenberg theorem~\cite{Mermin1966AbsenceFerromagnetism,Hohenberg1967ExistenceOfLong-RangeOrder} states that the algebraic correlations preclude spontaneous symmetry breaking at nonzero temperature -- reflected in the expected norm of the $U(1)$ symmetry-breaking order parameter going to zero in the thermodynamic limit.  Phenomena consistent with low-temperature broken symmetry have been measured, however, across a broad and diverse array of BKT experimental systems~\cite{Baity2016EffectiveTwoDimensionalThickness,Shi2016EvidenceCorrelatedDynamics,Bishop1978StudySuperfluid,Bramwell2015PhaseOrder,Wolf1981TwoDimensionalPhaseTransition,Resnick1981KosterlitzThouless,Bramwell1993Magnetization,Huang1994MagnetismFewMonolayersLimit,Elmers1996CriticalPhenomenaTwoDimensionalMagnet,BedoyaPinto2021Intrinsic2DXYFerromagnetism,Hadzibabic2006KosterlitzThouless,Fletcher2015ConnectingBKTAndBEC,Christodoulou2021Observation}, presenting a paradox between theory and experiment.  This was partially resolved in finite systems by Archambault, Bramwell \& Holdsworth~\cite{Archambault1997MagneticFluctuations} (built on later with Pinton~\cite{Bramwell1998Universality}) who showed that the expected low-temperature norm goes to zero very slowly and at the same rate as its fluctuations.  This led to great success in describing the expected norm in magnetic-film experiments~\cite{Bramwell1993Magnetization,Huang1994MagnetismFewMonolayersLimit,Elmers1996CriticalPhenomenaTwoDimensionalMagnet,Chung1999EssentialFiniteSizeEffect,BedoyaPinto2021Intrinsic2DXYFerromagnetism,Venus2022RenormalizationGroup} but did not address the thermodynamic limit itself, nor the dynamics of the directional phase of the order parameter in either finite or thermodynamic systems.  The latter is foundational to the characterization of broken symmetry, and crucially, may explain the strongly autocorrelated electrical resistance recently measured at the BKT transition in superconducting films~\cite{Shi2016EvidenceCorrelatedDynamics}.  

The above paradoxes were resolved by the development of an emergent electrostatic-field theory of the model system.  Vallat \& Beck~\cite{Vallat1994CoulombGas} began by reformulating the topological framework of BKT, Villain, Jos\'{e} {\it et al}. and  Nelson \& Kosterlitz on the torus.  % Faulkner, Bramwell \& Holdsworth~\cite{Faulkner2015TSFandErgodicityBreaking,Faulkner2017AnElectricFieldRepresentation} then took this electrostatic picture further with a field theory in which the vortices appear as divergences in an emergent electrostatic field, with the spin waves mapping to a purely rotational auxiliary gauge field.
This electrostatic picture was then taken further with a field theory~\cite{Faulkner2015TSFandErgodicityBreaking,Faulkner2017AnElectricFieldRepresentation} in which the vortices appear as divergences in an emergent electrostatic field, with the spin waves mapping to a purely rotational auxiliary gauge field.  
This resolved the short-range--long-range paradox, as the electrostatic picture had been reformulated in terms of a model of \emph{local} electrostatic fields.  In addition, the field theory augmented the BKT phase-vortex representation to include the topological sector of the emergent field.  These global topological defects are additional degrees of freedom that correspond to internal global twists in the spin field.  % Faulkner, Bramwell \& Holdsworth~\cite{Faulkner2015TSFandErgodicityBreaking,Faulkner2017AnElectricFieldRepresentation} showed that they are nonergodically frozen in the low-temperature phase for systems restricted to local Brownian dynamics.  
They are nonergodically frozen in the low-temperature phase for systems restricted to local Brownian dynamics~\cite{Faulkner2015TSFandErgodicityBreaking,Faulkner2017AnElectricFieldRepresentation}.  
This reframed topological order in terms of a topological nonergodicity.  The Mermin--Wagner--Hohenberg paradox was then fully resolved by introducing the concept of general symmetry breaking~\cite{Faulkner2024SymmetryBreakingBKT}.  This broadened the elegant yet restrictive framework of spontaneous symmetry breaking by allowing the expected norm of the $U(1)$ order parameter to go to zero in the thermodynamic limit, provided the fluctuations in its directional phase are asymptotically smaller.  This asymptotically slow directional mixing of the $U(1)$ order parameter was demonstrated in the low-temperature BKT phase, corresponding to the low-temperature $U(1)$ order parameter arbitrarily choosing some well-defined direction in the thermodynamic limit -- as in spontaneous symmetry breaking.  The field theory then revealed the intimate connection between the topological ergodicity and $U(1)$ symmetry: %, as the latter can be ensured at all nonzero temperatures by non-physical global-twist dynamics that also ensure topological ergodicity (on timescales that do not diverge with system size, in both cases). 
both can be ensured (on timescales that do not diverge with system size) at all nonzero temperatures by non-physical global-twist dynamics that tunnel through the $U(1)$ sombrero potential via high-energy global topological defects.  This implied that topological order (defined by the topological nonergodicity) induces the general symmetry breaking observed in experiment.  

%Both effects had eluded theorists prior to the development of the modern field theory, due in part to the Mermin--Wagner--Hohenberg theorem~\cite{Mermin1966AbsenceFerromagnetism,Hohenberg1967ExistenceOfLong-RangeOrder} precluding spontaneous symmetry breaking at nonzero temperature.  This resolved the paradox of measurements consistent with symmetry-breaking phenomena across a diverse array of experimental BKT systems~\cite{Baity2016EffectiveTwoDimensionalThickness,Shi2016EvidenceCorrelatedDynamics,Bishop1978StudySuperfluid,Bramwell2015PhaseOrder,Wolf1981TwoDimensionalPhaseTransition,Resnick1981KosterlitzThouless,Bramwell1993Magnetization,Huang1994MagnetismFewMonolayersLimit,Elmers1996CriticalPhenomenaTwoDimensionalMagnet,BedoyaPinto2021Intrinsic2DXYFerromagnetism,Hadzibabic2006KosterlitzThouless,Fletcher2015ConnectingBKTAndBEC,Christodoulou2021Observation} in spite of the Mermin--Wagner--Hohenberg theorem.  Indeed, general symmetry breaking is a more general form of symmetry breaking than that described by the mathematically elegant yet physically restrictive definition of spontaneous symmetry breaking.

Here we present a comprehensive review of general symmetry breaking, topological order reframed as a topological nonergodicity, and the connection between the two via the modern emergent electrostatic-field theory.  We begin with some key definitions in Section~\ref{sec:Definitions}.  In Section~\ref{sec:IsingModelAndSSB}, we review spontaneous symmetry breaking via the prototypical 2D Ising model -- with a particular focus on its correspondence to asymptotically slow directional mixing of the symmetry-breaking order parameter under single-spin-flip dynamics.  In Section~\ref{sec:XYModelAndGSB}, we present the asymptotically slow directional mixing of the symmetry-breaking order parameter of the 2DXY model under Brownian spin dynamics.  This general symmetry breaking follows reviews of both the Mermin--Wagner--Hohenberg theorem and the %Bramwell--Holdsworth theory of the norm of the order parameter.  
finite-system work of Bramwell \& Holdsworth.  We then present the concept of non-physical global-twist dynamics -- and demonstrate that they ensure $U(1)$ symmetry at all nonzero temperatures.  We start Section~\ref{sec:EmergentElectrolyte} by decomposing the model into its three principal excitations, before transforming from the spin field to the emergent electrostatic-field theory.  We use this field theory to reframe topological order %and demonstrate its associated low-temperature topological nonergodicity 
in terms of the topological nonergodicity, which we demonstrate at low temperature under local Brownian spin/emergent-field dynamics.  %This then elucidates the close connection between topological order/nonergodicity and general symmetry breaking  
We then elucidate the close connection between topological order and general symmetry breaking -- by demonstrating that topological ergodicity can also be ensured at all nonzero temperatures by the global-twist dynamics.  Finally, we discuss the implications for experiment, critical slowing down and computational statistics in Section \ref{sec:Discussion}.  We note that this paper should be viewed as a complement to the illuminating review article of Minnhagen~\cite{Minnhagen1987Review2DCG} which covered many key concepts of emergent electrostatics in planar superfluids and superconductors.

\section{Key definitions}
\label{sec:Definitions}

\subsection{Some foundational mathematical symbols}

\begin{enumerate}
    \item $\mapsto$ denotes a mapping between quantities.
    \item When placed between sets, $\to$ denotes a mapping between the sets.  Elsewhere, it denotes a limit.
    \item $\uparrow\downarrow$ denotes limits taken from below/above.
    \item $\| \mathbf{f} \|$ denotes the norm of some vector $\mathbf{f} \in \mathbb{R}^d$.
\end{enumerate}

\subsection{Statistical quantities}

\begin{enumerate}
    \item $\pi(\xvec) \propto e^{-\beta U(\xvec)}$ is the \emph{Boltzmann (probability) distribution}, with $U(\xvec) \in \RR$ the \emph{interaction potential} of some microstate $\xvec$ and $\beta > 0$ the \emph{inverse temperature}.
    \item The expectation $\mathbb{E}\mathbf{f} := \int \mathbf{f}(\xvec) \pi(\xvec) d\xvec$ of some observable $\mathbf{f} \in \mathbb{R}^d$ is the expected value of $\mathbf{f}$ as predicted by the Boltzmann distribution $\pi(\xvec) \propto e^{-\beta U(\xvec)}$.  If $\mathbf{f}$ is also a function of some vector of hyperparameters $\boldsymbol{\lambda}$, we may denote its expectation as $\mathbb{E}\mathbf{f}(\boldsymbol{\lambda})$, but the expectation is not taken over $\boldsymbol{\lambda}$.
    \item ${\rm Var}[\mathbf{f}] := \mathbb{E} \| \mathbf{f} - \mathbb{E}\mathbf{f} \|^2$ is the expected variance of $\mathbf{f}$.
    \item $s_{\mathbf{f}}^2(\beta, \tau, N, \boldsymbol{\Xi}) := [n / (n - 1)] \overline{\| \mathbf{f} - \overline{\mathbf{f}} \|^2}$ is the (unbiased) simulation variance (of $\mathbf{f}$) of some $N$-spin simulation of \emph{simulation timescale} $\tau = n \Delta t$, inverse temperature $\beta$ and vector $\boldsymbol{\Xi}$ of additional model and simulation parameters.  Here, the bar denotes a simulation mean, $n$ is the sample size and $\Delta t$ is the simulation time step (which is proportional to the number of spins $N$).
    \item $\langle \cdot \rangle$ denotes a mean over an infinitely large number of independent simulations at fixed $\beta$, $\tau$, $N$, $\boldsymbol{\Xi}$.
    \item The indicator function $\mathbb{I}(A)$ is one/zero if $A$ does/does not hold.
    \item The Heaviside step function $\Theta : \RR \to \{0, 1\}$ is defined as
    \begin{align}
        \Theta(x) := \begin{cases}
            1 \,\,\, \forall \, x \ge 0, \\
            0 \,\,\, \forall \, x < 0 .
        \end{cases}
    \end{align}
    The Heaviside function can always be represented by an indicator function [$\Theta(x - a) = \mathbb{I}(x \ge a)$ for any fixed $a \in \RR$] but we use the Heaviside function wherever it is standard in physics.
    \item In a slight abuse of notation, $\delta(x) dx$ is the Dirac measure $\delta(dx)$ for $x \in \RR$ and $\delta(\Fvec) \equiv \delta(F_x) \delta(F_y)$ for $\Fvec \in \RR^2$.
\end{enumerate}

\subsection{Geometric quantities and discrete vector calculus}

\begin{enumerate}
    \item $\metric(\rvec, \rvec') > 0$ is the shortest distance between points $\rvec$ and $\rvec'$ on any topologically toroidal surface or lattice.
    \item $\oplus$ and $\ominus$ define (respectively) addition and subtraction on any topologically toroidal surface or lattice, i.e., addition and subtraction while accounting for periodic boundaries.
    \item $a > 0$ and $L > 0$ are (respectively) the \emph{lattice spacing} and \emph{linear size} of some regular square lattice.
    \item The \emph{charge lattice} $D$ is defined as the set $\{ (a / 2, a / 2), (3a / 2, a / 2), \dots , (L - a / 2, L - a / 2) \}$ with toroidal topology.
    \item The \emph{spin lattice} $D'$ is defined as the set $\{ (0, 0), (a, 0), \dots , (L - a, L - a) \}$ with toroidal topology.
    \item As in the framework for discrete vector calculus presented by Chew~\cite{Chew1994ElectromagneticTheoryLattice}, all vector functions $\Fvec : \Omega \to \RR^2$ on some regular, square and topologically toroidal lattice $\Omega$ are defined as discrete counterparts of smooth vector fields, i.e., 
    \begin{align}
        \Fvec(\rvec) := \sum_{\mu \in \{ x, y \}} F_\mu \glb \rvec + \frac{a}{2} \mathbf{e}_\mu \grb \mathbf{e}_\mu , 
    \end{align}
    with each Cartesian component $F_\mu \in \RR$ defined equidistant between neighboring lattice sites $\rvec$ and $\rvec \oplus a \mathbf{e}_\mu$ (the 3D case is defined analogously). 
    \item $\widetilde{\boldsymbol{\nabla}}$ and $\hat{\boldsymbol{\nabla}}$ are (respectively) the \emph{forwards} and \emph{backwards finite-difference operators}~\cite{Chew1994ElectromagneticTheoryLattice}, i.e., for any scalar function $f : \Omega \to \RR$, 
    \begin{align}
        \widetilde{\boldsymbol{\nabla}}f(\rvec) := \sum_{\mu \in \{ x, y \}} \frac{f(\rvec \oplus a \mathbf{e}_\mu) - f(\rvec)}{a} \mathbf{e}_\mu 
    \end{align}
    and 
    \begin{align}
        \hat{\boldsymbol{\nabla}}f(\rvec) := \sum_{\mu \in \{ x, y \}} \frac{f(\rvec) - f(\rvec \ominus a \mathbf{e}_\mu)}{a} \mathbf{e}_\mu .
    \end{align} 
    \item $\boldsymbol{\nabla}^2 := \hat{\boldsymbol{\nabla}} \cdot \widetilde{\boldsymbol{\nabla}}$ is the \emph{lattice Laplacian operator}~\cite{Chew1994ElectromagneticTheoryLattice}.
    \item For any vector function $\Fvec : \Omega \to \RR^2$, 
    \begin{align}
        \widetilde{\boldsymbol{\nabla}} & \times \Fvec(\rvec) := \nonumber\\
        & \left[ \widetilde{\boldsymbol{\nabla}}_x F_y \left( \rvec + \frac{a}{2} \mathbf{e}_y \right) - \widetilde{\boldsymbol{\nabla}}_y F_x \left( \rvec + \frac{a}{2} \mathbf{e}_x \right) \right] \mathbf{e}_z 
    \end{align}
    is its curl (the 3D case is defined analogously)~\cite{Chew1994ElectromagneticTheoryLattice}.
    \item $\int \mathcal{D}\Fvec := \prod_{\rvec \in \Omega} \int d\Fvec(\rvec)$ denotes a functional integral with respect to $\Fvec$.
\end{enumerate}

\subsection{Glossary of key subject-specific terms}

\begin{enumerate}
    \item The \emph{symmetry-breaking order parameter} $\mvec$ is defined as the intensive quantity whose expectation $\mathbb{E}\mvec$ is proportional to the gradient of the free energy with respect to the symmetry-breaking field.
    \item \emph{Mixing} (of the entire system or some quantity) is defined as ergodic exploration of the state space (of the system or with respect to the quantity). % -- corresponding to simulation means reaching their expected values within some small $\varepsilon > 0$.  
    In this paper, the corresponding \emph{mixing timescale} is the simulation timescale on which this is guaranteed from an arbitrary region of high probability mass.  We assume such initial configurations unless otherwise stated.  %We may also refer to mixing between states -- which describes mixing between states/regions around states on a discrete/continuous state space.
    %\item Some quantity is defined to be mixing if it is ergodically exploring its state space.
    \item \emph{Directional mixing} refers to the mixing of the directional component of $\mvec$.  For the 2DXY model, this corresponds to ergodic exploration of all angles $\phi_\mvec \in [-\pi, \pi)$ of the 2D vector $\mvec$.
    \item \emph{Asymptotically slow mixing} is defined as a divergence (with the number of spins $N$) of the corresponding mixing timescale. %ergodically explores its state space.  
    This defines (weakly) \emph{broken ergodicity} in classical statistical physics.
    \item The dynamical phenomenon of \emph{broken symmetry} is a special case of broken ergodicity.  In the present context, it coincides with asymptotically slow directional mixing.  More generally, it corresponds to an asymptotically slow mixing between equilibrium states/regions of equal (measure and) probability mass on a discrete/continuous state space.
    \item \emph{Spontaneous symmetry breaking} is a framework for characterizing broken symmetry.  It is expressed mathematically by calculating $\mathbb{E}\mvec$ under the influence of a fixed-direction symmetry-breaking field before taking the thermodynamic and then zero-field limits -- resulting in a nonzero vector that aligns with the direction of the field.
    \item \emph{Long-time directional stability} is defined by the directional fluctuations of $\mvec$ going to zero in the thermodynamic %limit (at any fixed timescale).  
    and then long-time limits.  This corresponds to asymptotically slow directional mixing.
    \item \emph{General symmetry breaking} is an alternative framework for characterizing broken symmetry.  It is defined as the directional fluctuations of $\mvec$ going to zero in the thermodynamic %limit (i.e., long-time directional stability) 
    limit provided they are asymptotically smaller than $\mathbb{E}\|\mvec\|$.  %This broadens the framework of spontaneous symmetry breaking but is not distinct from it.
    Spontaneous symmetry breaking is a special case of general symmetry breaking.
    \item A \emph{topological defect} in some vector field is an object that cannot be removed by continuously deforming the field lines.  \emph{Local/global topological defects} are topological defects that can be defined by loops on/around the torus.  For example, electrical charges are local topological defects in the electric field, as reflected in the Gauss law.  
    \item In the present context, an \emph{electrolyte} is a (charge-neutral, unless otherwise stated) system composed of electrical charges.  We restrict to charges existing on a single-occupancy lattice.
    \item An \emph{emergent electrolyte} is some spin model whose local topological defects interact with each other in analogy with the electrical charges of an electrolyte.  The following electrostatic quantities all have emergent XY analogues in this paper.   %An analogous emergent electric field can also be defined.
    \item In the present context, the \emph{harmonic mode} of an electrolyte is the spatial mean of its electric field.
    \item For any given charge distribution, the \emph{topological sector} describes any discrepancy between the harmonic mode and its low-energy solution.  Each nonzero Cartesian component then corresponds to a global topological defect -- and the topological sector changes when a charge traces a closed path around the torus.
    \item \emph{Long-time topological stability} is defined by zero topological-sector fluctuations in the thermodynamic %limit.  
    and then long-time limits.  This implies that charges are confined in neutral pairs.  As the expected variance of the topological sector is nonzero at all nonzero temperatures, \emph{topological nonergodicity} is defined by long-time topological stability.  This equivalently defines \emph{topological order} within the present framework.
    \item A \emph{global spin twist} rotates the XY spin field by an integer multiple of $2\pi$ along all closed paths around some Cartesian dimension of the lattice.
    \item \emph{Global-twist dynamics} are Monte Carlo moves that attempt one (externally applied) global spin twist along each Cartesian dimension after each Monte Carlo time step.
    \item An \emph{internal global spin twist} is the spin field removed from an annealed XY system after performing global spin twists until the system reaches its low-energy state [Section~\ref{sec:HarmonicMode} discusses the subtle use of equation~\eqref{eq:HXYHamiltonian} (in this recipe) for both XY models].  This quantity maps precisely to the topological sector.  Moreover, topological nonergodicity coincides with nonergodicity analogously defined with respect to this quantity.
    %\item If some concept holds in an \emph{arbitrarily large system}, then no matter how large the system in which the concept holds, there exists some larger system in which the concept still holds.  This is distinct from the notion of infinitely large.
    \item If some concept holds in an \emph{arbitrarily large} system, then if it holds at some finite size, there exists some larger finite size at which it still holds.  This is distinct from the notion of infinitely large.
\end{enumerate}

\section{Two-dimensional Ising model and spontaneous symmetry breaking}
\label{sec:IsingModelAndSSB}

Spontaneous symmetry breaking is a characteristic property of conventional continuous phase transitions.  As it is fundamental to this review, we begin by exploring the concept via its prototypical case of the 2D Ising model of magnetism.  This is a set of %integer-valued (taking values on $\{ -1, +1 \}$) 
$Z_2$ magnetic spins fixed at the $N$ sites of a topologically toroidal, square lattice with interaction potential 
\begin{align}
U_{\rm Ising} := - J \sum_{\langle \rvec, \rvec' \rangle} \tilde{s}_\rvec \tilde{s}_{\rvec'} - h \sum_\rvec \tilde{s}_\rvec .
\label{eq:IsingModelDefinition}
\end{align}
Here, $J > 0$ is the exchange constant, $h \in \mathbb{R}$ is the symmetry-breaking field, $\tilde{s}_\rvec \in \{ -1, +1 \}$ is the spin at site $\rvec \in D'$ and the sum $\sum_{\langle \rvec, \rvec' \rangle}$ is over all nearest-neighbor pairs of the spin lattice $D'$.  %We set $h = 0$ below unless otherwise stated.  
The \emph{Ising magnetization} 
\begin{align}
    m := \frac{1}{N} \sum_\rvec \tilde{s}_\rvec 
\end{align}
is then the ($Z_2$) symmetry-breaking order parameter because its expectation $\mathbb{E}m$ is proportional to the gradient of the free energy with respect to the symmetry-breaking field $h$. %, where the expectation $\mathbb{E}f := \int f(\xvec) \pi(\xvec) d\xvec$ of some observable $f$ is that predicted by the Boltzmann model $\pi(\xvec) \propto e^{-\beta U(\xvec)}$, with $U$ the interaction potential and $\beta > 0$ the inverse temperature.  %Throughout this review, we also define $\metric(\rvec, \rvec')$ as the shortest distance between points $\rvec$ and $\rvec'$ on any topologically toroidal surface.

\begin{figure}[t]
\includegraphics[width=\linewidth]{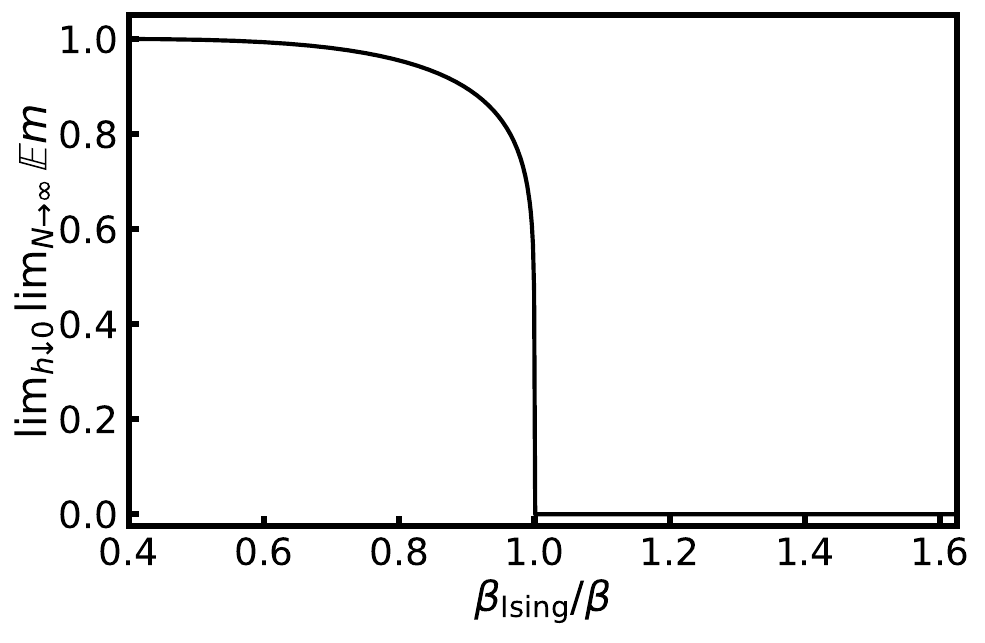}
\caption{Spontaneous magnetization [defined in equation~\eqref{eq:2dOnsagerYangSolution}] of the 2D Ising model vs reduced temperature $\beta_{\rm Ising} / \beta$.  This predicts a phase transition at $\beta = \beta_{\rm Ising}, h = 0$.  The nonzero low-temperature values reflect the probability of the zero-field magnetization changing sign (under single-spin-flip Metropolis dynamics) going to zero in the thermodynamic limit.}
\label{fig:2dIsingSpontaneousMagnetisation}
\end{figure}

\begin{figure*}
\includegraphics[width=\linewidth]{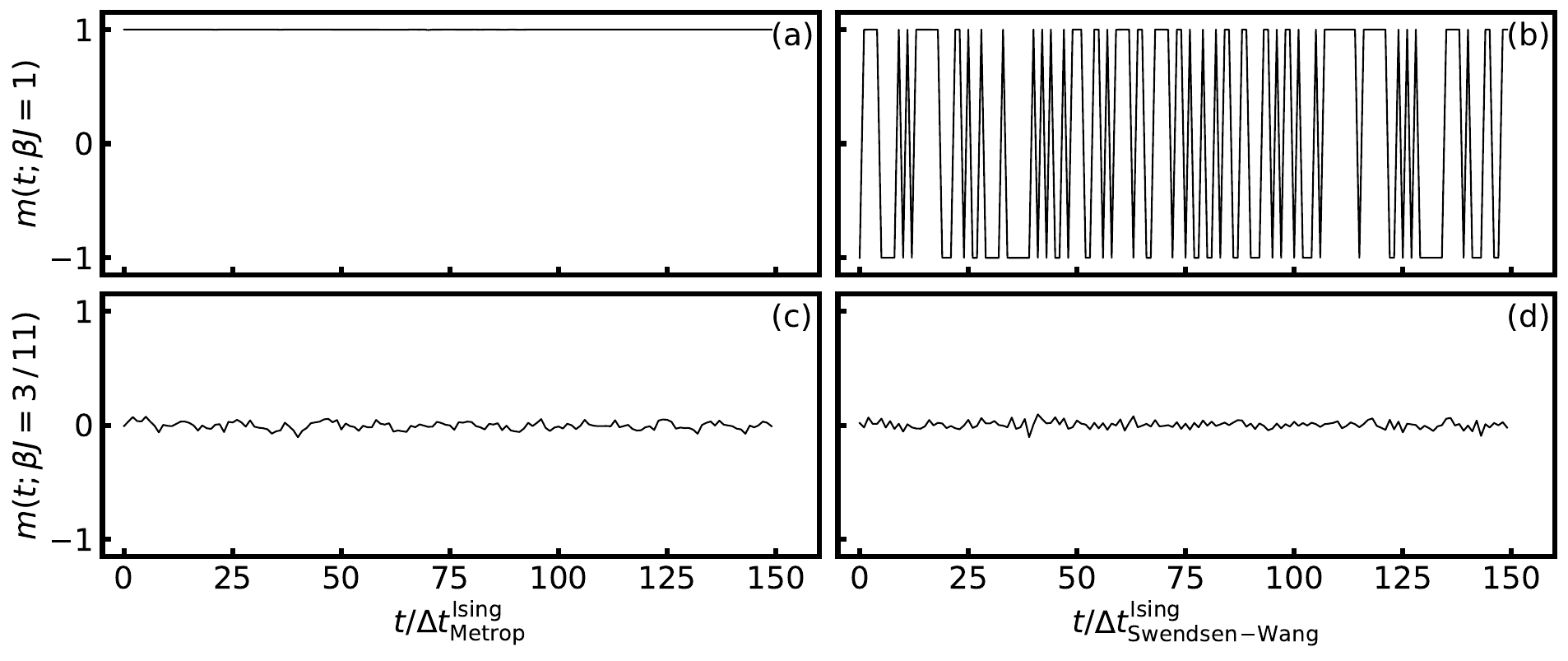}
\caption{Zero-field Ising magnetization $m := \sum_\rvec \tilde{s}_\rvec / N$ versus time $t$ for $N = 64 \! \times \! 64$-spin simulations at low (top) and high (bottom) temperature using Metropolis (left) and Swendsen--Wang (right) dynamics.  $\Delta t_{\rm Metrop}^{\rm Ising}$  and $\Delta t_{\rm Swendsen--Wang}^{\rm Ising}$ are the time steps of the Metropolis and Swendsen--Wang algorithms, respectively.  The former is defined as the elapsed time between $N$ attempted single-spin flips.  The latter is defined as the elapsed time between each partitioning of the lattice into clusters of aligned spins.  Both high-temperature simulations are symmetric about $m = 0$, but only the Swendsen--Wang dynamics are symmetric at low temperature.  Indeed, the low-temperature Metropolis simulation stays in the $m = 1$ state throughout the simulation.  This reflects the probability of it changing sign (under single-spin-flip Metropolis dynamics) vanishing in the thermodynamic limit, leading to a divergent (or asymptotically long) directional mixing timescale.  $10^4$ equilibration samples were discarded from each simulation.}
\label{fig:2dIsingMagDensityVsTime}
\end{figure*}

%Building on the initial work of Kramers~\cite{Kramers1941StatisticsOfTwoDimFerromagnetI,Kramers1941StatisticsOfTwoDimFerromagnetII}, Onsager~\cite{Onsager1944CrystalStatistics} showed that the thermodynamic limit of the expected zero-field ($h = 0$) specific heat per particle is 
%\begin{align}
%    \lim_{N \to \infty} \left( 
%    \frac{1}{N} \check{C}_V(\beta, J, h = 0, N) \right) = \beta^2 \partial_{\beta}^2 \gamma(\beta) ,
%\label{eq:2dIsingExpectedSpecHeat}
%\end{align}
%where 
%\begin{align}
%\gamma(\beta) := \ln \left( 2\cosh (2\beta J) \right) + \frac{1}{\pi} \int_0^{\pi / 2} \ln \left[ \frac{1}{2} \glb 1 + \sqrt{1 - \frac{4 \sinh^2 (2\beta J) \sin^2w}{\cosh^4 (2\beta J)}} \grb \right] dw .
%\end{align}
%The function in equation \eqref{eq:2dIsingExpectedSpecHeat} diverges logarithmically at the inverse critical temperature $\beta_{\rm Ising} := \ln (1 + \sqrt{2}) / (2J)$.  This predicts a phase transition at $\beta = \beta_{\rm Ising}, h = 0$, as supported by the black curve in Figure \ref{fig:2dIsingSpecHeatAndAbsMagDensityAnalytical}.  In addition, the magnetization 
Onsager~\cite{Onsager1949Discussion} and Yang~\cite{Yang1952SpontaneousMagnetization} demonstrated a non-differentiability in the \emph{spontaneous magnetization} 
\begin{align}
    %\lim_{h \downarrow \uparrow 0} \lim_{N \to \infty} \mathbb{E}m (\beta, h, N) = \pm
    \lim_{h \downarrow 0} \lim_{N \to \infty} \mathbb{E}m (\beta, h, N) = \qquad\qquad\qquad\qquad\qquad\quad & \nonumber\\
    \begin{cases}
        \left( 1 - [\sinh (2 \beta J)]^{-4} \right)^{1 / 8} & {\rm for} \,\, \beta > \beta_{\rm Ising} , \\
        0 & {\rm for} \,\, \beta < \beta_{\rm Ising} ,
    \end{cases}
\label{eq:2dOnsagerYangSolution}
\end{align}
with $\beta_{\rm Ising} := \ln (1 + \sqrt{2}) / (2J)$ the inverse critical temperature.  This predicted a phase transition at $\beta = \beta_{\rm Ising}$, $h = 0$ between a low-temperature ($\beta > \beta_{\rm Ising}$) long-range-ordered phase and a high-temperature ($\beta < \beta_{\rm Ising}$) disordered one, as supported by the curve in figure~\ref{fig:2dIsingSpontaneousMagnetisation}.  %With $\metric(\rvec, \rvec') > 0$ defined as the shortest distance between points $\rvec$ and $\rvec'$ on any topologically toroidal surface, the 
The long-range order at low temperature is characterized by a spin-spin correlation function $\sum_{\rvec \ne \rvec'} \mathbb{I} \glc r = \metric(\rvec, \rvec') \grc \mathbb{E} \glc \tilde{s}_\rvec \tilde{s}_{\rvec'} \grc / 2$ that scales with a constant prefactor on long distances $r > 0$, while exponentially decaying spin-spin correlations reflect the high-temperature disorder.  Equation~\eqref{eq:2dOnsagerYangSolution} also implies spontaneously broken symmetry in the low-temperature phase, as this phenomenon is defined by $\lim_{h \uparrow \downarrow 0} \lim_{N \to \infty} \mathbb{E}m (\beta, h, N) \ne 0$.  This is, however, a consequence of the more foundational viewpoint of broken symmetry -- a dynamical phenomenon corresponding to asymptotically slow mixing between equilibrium states of equal probability mass -- which we explore below.

%The choice of system dynamics is central to zero-field observations of symmetry breaking on experimental timescales.  %Such observations are seen in the magnetization $m$, which is the symmetry-breaking order parameter because its expectation is proportional to the gradient of the free energy with respect to the symmetry-breaking field $h$.  
The choice of system dynamics is central to experimental/simulation observations of the dynamical phenomenon of broken symmetry.  At $h = 0$, the Boltzmann distribution is symmetric with respect to global spin flips (i.e., $\tilde{s}_\rvec \mapsto -\tilde{s}_\rvec$ for all $\rvec$) for any finite $\beta, J, N$.  %The magnetization $m$ therefore ergodically explores the set $\{-1, -1 + 1 / N, \dots, +1\}$ on some finite symmetry-restoring timescale at high temperature.  Under certain dynamics, however, the long-range order at low temperature causes the system to spontaneously choose either $m = -1$ or $m = +1$ and stay close to this chosen $\beta \to \infty$ state on a timescale that diverges with $N$.  
Under ergodic dynamics, the magnetization $m$ then ergodically explores the set $\{-1, -1 + 1 / N, \dots, +1\}$ on some finite \emph{directional mixing timescale}: the mean magnetization converges to its expected value $\mathbb{E}m = 0$ on this timescale (defined by its fluctuations also converging to their expected value $\sqrt{\mathbb{E}m^2} > 0$ within some small $\varepsilon > 0$) where it is independent of global spin flips.  %This holds on non-divergent timescales in the high-temperature phase under a broad choice of dynamics.  
Under certain dynamics, however, the directional mixing timescale diverges with system size at low temperature.  This is because the long-range order causes the system to spontaneously choose either %$m = -1$ or $m = +1$ and stay close to this chosen $\beta \to \infty$ state 
$m > 0$ or $m < 0$ and keep this sign on a timescale that diverges with system size.  In contrast, more sophisticated dynamics can be constructed that ergodically explore configuration space on short non-divergent timescales at all nonzero temperatures.  For example, in figure~\ref{fig:2dIsingMagDensityVsTime} we compare simulation results generated using the local Metropolis~\cite{Metropolis1953EquationOfState} and Swendsen--Wang~\cite{Swendsen1987Nonuniversal} algorithms.  As outlined in \cite{Faulkner2024SamplingAlgorithms}, the former proposes single spin flips at each algorithm iteration, while the latter employs a dynamics that flips entire clusters of aligned spins at each iteration~\footnote{Each iteration of the Swendsen--Wang algorithm partitions the lattice into clusters of aligned and connected spins.  Each cluster starts as a single spin and is grown by adding neighboring aligned spins with probability $1 - \exp (-2\beta J)$.  For each cluster, all component spins are then flipped with probability $1/2$.}.  %As a result, the low-temperature symmetric-mixing timescale of the Metropolis algorithm diverges with system size $N$ \blue{[ref]}, while Swendsen--Wang dynamics ensure $Z_2$ symmetry on non-divergent timescales \blue{[ref]} (and also overcome critical slowing down near the transition \blue{[ref]}).  
At low temperature, this results in asymptotically slow directional mixing of the $Z_2$ order parameter ($m$) under single-spin-flip Metropolis dynamics (i.e., a divergent directional mixing timescale) while Swendsen--Wang dynamics ensure $Z_2$-symmetric mixing on non-divergent timescales at all nonzero temperatures [and also overcome critical slowing down near the transition -- see, e.g. \cite{Faulkner2024SamplingAlgorithms}].  This is reflected in the low-temperature zero-field magnetization trace plots in figures~\ref{fig:2dIsingMagDensityVsTime}(a) and (b).  %On the presented simulation timescale, the 
The Metropolis simulation starts at $m = 1$ and stays in this state, while the Swendsen--Wang simulation mixes between $m = 1$ and $m = -1$.  In contrast, the high-temperature outputs in figures~\ref{fig:2dIsingMagDensityVsTime}(c) and (d) reflect both dynamics ergodically exploring the support of the magnetization on short timescales.  The Swendsen--Wang simulations therefore display $Z_2$-symmetric mixing at both low and high temperature, but we observe discrepancies between the low-temperature Metropolis simulations and predictions of the Boltzmann model in zero field -- as a result of the low-temperature spin order.  Indeed, at low temperature and under single-spin-flip Metropolis dynamics, the magnetization changes sign (on any finite timescale) with probability zero in the thermodynamic limit. % we state this in final para of section

The above zero-field discrepancies between experimental/simulation observations and predictions of the Boltzmann model are the essence of broken symmetry.  Such discrepancies are typically accompanied by some singular limit, as seen in this case upon characterizing the above results in terms of the \emph{long-time directional stability} %(of the order parameter) 
\begin{align}
\gamma_{\rm Ising}(\beta) := \lim_{\tau \to \infty} \lim_{N \to \infty} g_{\rm Ising}(\beta, \tau, N) ,
\label{eq:DirectionalStabilityIsingDef}
\end{align}
where the \emph{finite directional stability}
\begin{align}
    g_{\rm Ising}(\beta, \tau, N) := 1 - \sqrt{\frac{\langle s_m^2(\beta, \tau, N, h = 0) \rangle}{{\rm Var}[m](\beta, N, h = 0)}} 
\label{eq:IsingVarianceRatioFn}
\end{align}
of some simulation method measures discrepancies between the \emph{directional fluctuations} $\sqrt{\langle s_m^2(\beta, \tau, N, h = 0) \rangle}$ and their expected value $\sqrt{{\rm Var}[m](\beta, N, h = 0)}$ ($\tau$ is the simulation timescale). 
%Here, $s_m^2(\beta, \tau, N, h = 0)$ is the (unbiased) simulation variance of some $N$-spin zero-field simulation of timescale $\tau$ and inverse temperature $\beta$. %, and $\langle \cdot \rangle$ denotes a mean over an infinite number of independent simulations at fixed $\beta, \tau, N, h$.  
For single-spin-flip Metropolis dynamics, it then follows that 
\begin{align}
\gamma_{\rm Ising}^{\rm Metrop}(\beta) = 
\begin{cases}
    1 & {\rm for} \,\, \beta > \beta_{\rm Ising} , \\
    0 & {\rm for} \,\, \beta < \beta_{\rm Ising} .
\end{cases}
\label{eq:DirectionalStabilityIsingMetrop}
\end{align}
The nonzero value at low temperature results from vanishing directional fluctuations in the thermodynamic limit, as reflected by the absence of directional mixing in figure~\ref{fig:2dIsingMagDensityVsTime}(a).  %This thermodynamic limit is singular because exchanging the order of the limits in equation~\eqref{eq:DirectionalStabilityIsingDef} returns zero at all nonzero temperatures [with Faulkner~\cite{Faulkner2024SymmetryBreakingBKT} noting that singular semiclassical limits analogously involving long times are commonplace in quantum chaos~\cite{Berry2001Chaos}].  This is because all unbiased estimators of statistical expectations -- necessarily finite-size quantities -- eventually converge to predictions of the Boltzmann model on some finite timescale (the directional mixing timescale in this case).  %, resulting in $Z_2$-symmetric zero-field simulations on long enough timescales.  
This thermodynamic limit is singular because exchanging the order of the limits in equation~\eqref{eq:DirectionalStabilityIsingDef} returns zero at all nonzero temperatures -- since all unbiased estimators of statistical expectations eventually converge to predictions of the Boltzmann model on some finite timescale (the directional mixing timescale in this case).  It was noted~\cite{Faulkner2024SymmetryBreakingBKT} that singular semiclassical limits analogously involving long times are commonplace in quantum chaos~\cite{Berry2001Chaos}, and we will see below in Section~\ref{sec:TopologicalNonergodicity} that functions of the form of the long-time directional stability in equation~\eqref{eq:DirectionalStabilityIsingDef} can be used to characterize broken ergodicity (under some dynamics) more generally.  In contrast with equation~\eqref{eq:DirectionalStabilityIsingMetrop}, the long-time directional stability of the Swendsen--Wang algorithm is zero for all $\beta < \infty$.    

%Moreover, the \emph{static} spontaneous magnetization in equation~\eqref{eq:2dOnsagerYangSolution} elegantly characterizes the \emph{dynamical} zero-field phenomenon described by equation~\eqref{eq:DirectionalStabilityIsing} in terms of thermodynamic expectations.  %and reflected in the data in figure~\ref{fig:2dIsingSpontaneousMagnetisation}(b). 
%where $m_0(\beta) \ge 0$ is the spontaneous magnetic density defined in \eqref{eq:2dOnsagerYangSolution}. 

\begin{figure}
\includegraphics[width=\linewidth]{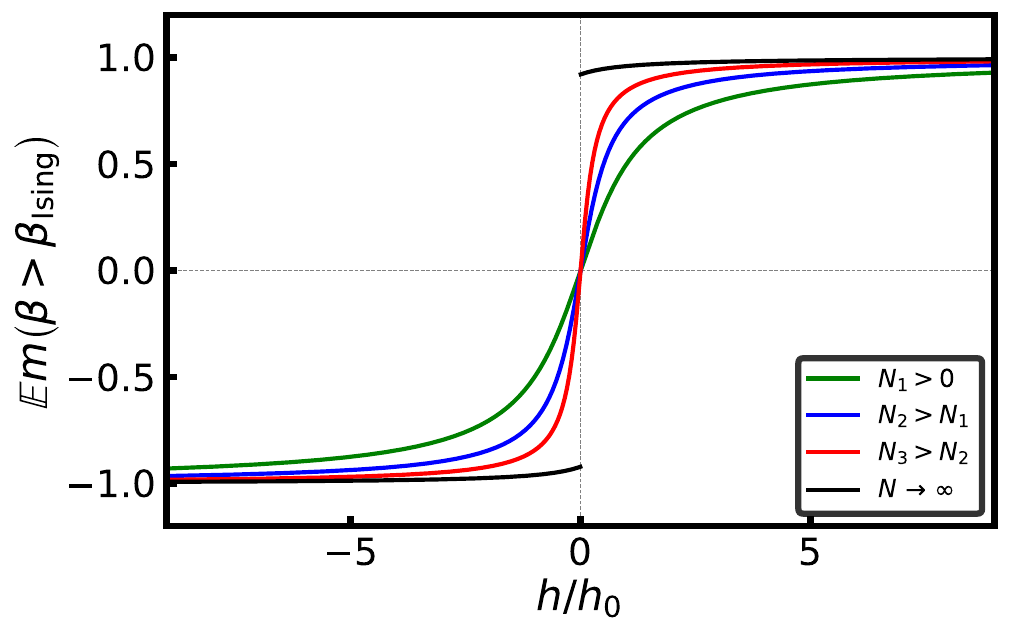}
\caption{Schematic of the expected Ising magnetization as a function of field $h$ and system size $N$ at low temperature, with $h_0 > 0$ some base field strength.  The discontinuity that develops at $h = 0$ as $N \to \infty$ reflects the divergent (or asymptotically long) directional mixing timescale of low-temperature single-spin-flip Metropolis simulations.  This discontinuity is not a feature of the high-temperature phase as the directional mixing timescale is non-divergent there, even under single-spin-flip Metropolis dynamics.}
\label{fig:SchematicMagnetisationVsField}
\end{figure}

Moreover, the \emph{dynamical} zero-field phenomenon described by equation~\eqref{eq:DirectionalStabilityIsingMetrop} was characterized by the \emph{static} spontaneous magnetization in equation~\eqref{eq:2dOnsagerYangSolution}.  As stated there, this formulation in terms of thermodynamic expectations defines the framework of spontaneous symmetry breaking.  This framework is analogously [to equation~\eqref{eq:DirectionalStabilityIsingMetrop}] encapsulated by a singular thermodynamic limit because exchanging the order of the $|h| \to 0$ and $N \to \infty$ limits in equation~\eqref{eq:2dOnsagerYangSolution} returns zero at all nonzero temperatures.  This is reflected by the schematic plots of the expected low-temperature magnetization as a function of symmetry-breaking field $h$ for various system sizes in figure~\ref{fig:SchematicMagnetisationVsField}.  The expected magnetization goes to zero smoothly with $|h|$ for any finite system size $N$ (the non-black curves) reflecting even single-spin-flip Metropolis simulations being $Z_2$-symmetric on long enough timescales in zero field [symmetric simulations are defined by a directional simulation variance that has converged to its expected value ($s_m^2$ and ${\rm Var}[m]$ in this case) within some small $\varepsilon > 0$].  In contrast, taking the limit $N \to \infty$ before the zero-field limit (the black curve) leads to a discontinuity at $h = 0$.  While the spontaneous magnetization is independent of algorithm dynamics, this discontinuity reflects the divergent directional mixing timescale of low-temperature single-spin-flip Metropolis simulations in zero field.  %, with the probability of the magnetization changing sign under such dynamics going to zero in the thermodynamic limit.  % stated above instead
The singular nature of the low-temperature thermodynamic limit in equation~\eqref{eq:2dOnsagerYangSolution} therefore analogously [to equation~\eqref{eq:DirectionalStabilityIsingMetrop}] reflects measurable zero-field discrepancies between experimental observations and predictions of the Boltzmann model, but instead in terms of a static thermodynamic expectation formed in nonzero field.  %The long-time directional stability defined by equation~\eqref{eq:DirectionalStabilityIsingDef} is therefore not required to described broken symmetry in this model, though it is still a useful additional metric, as it characterizes the symmetric nature of Swendsen--Wang dynamics.  % no longer agree that equation~\eqref{eq:DirectionalStabilityIsingDef} is not reqd, given its utility in describing whether some dynamics are symmetric 
For completeness, we note that the $h = 0$ discontinuity is not a feature of the high-temperature phase as the directional mixing timescale is non-divergent here, even under single-spin-flip Metropolis dynamics.  The thermodynamic limit in equation~\eqref{eq:2dOnsagerYangSolution} is therefore non-singular at high temperature.

%Moreover, the local Metropolis dynamics can be supplemented with a global dynamics that ensure $Z_2$ symmetry at all nonzero temperatures.  These global dynamics flip all spins at the end of each sweep of local spin flips, resulting in the zero-field magnetization trace plots in figure~Y(a).  \blue{INSERT}  These results are $Z_2$-symmetric and similar to the low-temperature Swendsen--Wang trace plots in figure~\ref{fig:2dIsingMagDensityVsTime}(b).  Indeed, the breaking of some global symmetry is typically associated with an accompanying global symmetry-restoring dynamics.  %We add that the supplemental global dynamics result in the high-temperature trace plot in figure~Y(b), with similar form to that of the high-temperature Swendsen--Wang trace plots in figure~\ref{fig:2dIsingMagDensityVsTime}(d).  The global dynamics do not overcome, however, the issue of critical slowing down near the transition, though this falls outside the scope of this review.
%We add that the supplemental global dynamics result in the high-temperature trace plot in figure~Y(b), which differs in form from both the Metropolis and Swendsen--Wang trace plots at high temperature in figures~\ref{fig:2dIsingMagDensityVsTime}(c) and (d).  This reflects, in part, the fact that the global dynamics do not overcome the issue of critical slowing down near the transition, though this falls outside the scope of this review.

\section{Two-dimensional XY model and general symmetry breaking}
\label{sec:XYModelAndGSB}

We now turn to the 2DXY model of magnetism and explore its symmetry-breaking properties.  This prototypical model of BKT physics is the set of unit-length $U(1)$ spins fixed at the $N$ sites of a topologically toroidal, square lattice with interaction potential  
\begin{align}
U_{\rm XY} := - J \sum_{\langle \rvec, \rvec' \rangle} \cos \left( \varphi_\rvec - \varphi_{\rvec'} \right) - \hvec \cdot \sum_\rvec \begin{pmatrix} \cos{\varphi_\rvec} \\ \sin{\varphi_\rvec} \end{pmatrix} .
\label{eq:2DXYModelDefinition}
\end{align}
Here $\hvec \in \mathbb{R}^2$ is the 2D symmetry-breaking field and $\varphi_\rvec \in [-\pi, \pi)$ is the spin phase at lattice site $\rvec \in D'$.  We also define the (XY) \emph{magnetization} 
\begin{align}
    \mvec := \frac{1}{N}\sum_\rvec \begin{pmatrix} \cos{\varphi_\rvec} \\ \sin{\varphi_\rvec} \end{pmatrix} ,
    \label{eq:MagnetizationXY}
\end{align}
which is the [$U(1)$] symmetry-breaking order parameter because its expectation is proportional to the gradient of the free energy with respect to the symmetry-breaking field $\hvec$, in analogy with the Ising magnetization $m$.  Writing $\mvec = (\| \mvec \|, \phi_\mvec)$ in polar coordinates then defines the \emph{global $U(1)$ phase} $\phi_\mvec \in [-\pi, \pi)$.  Unless otherwise stated, we set $\hvec = 0$ throughout.  For simplicity, we redefine some quantities used in Section~\ref{sec:IsingModelAndSSB}, e.g., the (XY) magnetization in equation~\eqref{eq:MagnetizationXY}.

\subsection{Mermin--Wagner--Hohenberg theorem and finite systems}
\label{sec:MWHandBHtheories}

%Upon redefining the spin-spin correlation function 
%\begin{align}
%    g(r, N) := \frac{1}{2} \sum_{i, j} \mathbb{I} \glc r = \metric(\rvec_i, \rvec_j) \grc \mathbb{E} \glc e^{-i \varphi_i} e^{i \varphi_j} \grc ,
%\label{eq:XYCorrelationFunction}
%\end{align}
%algebraic correlations on long distances $r > 0$ indicate quasi-long-range order in the low-temperature phase, while exponential decay on long distances $r$ again implies disorder at high temperature~\cite{Berezinskii1973DestructionLongRangeOrder,Kosterlitz1973OrderingMetastability}.  
In zero field, the low-temperature phase is characterized by quasi-long-range order due to the spin-spin correlation function 
\begin{align}
    \frac{1}{2} \sum_{\rvec \ne \rvec'} \mathbb{I} \glc r = \metric(\rvec, \rvec') \grc \mathbb{E} \glc e^{-i \varphi_\rvec} e^{i \varphi_{\rvec'}} \grc 
\label{eq:XYCorrelationFunction}
\end{align}
displaying algebraic correlations on long distances $r$ for all finite $\beta > \beta_{\rm BKT}^{\rm XY}$ (with $\beta_{\rm BKT}^{\rm XY}$ the BKT phase transition of the 2DXY model), while exponential decay $\exp (- r / \xi)$ on long distances $r$ again implies disorder at high temperature (with $\xi > 0$ the correlation length)~\cite{Berezinskii1973DestructionLongRangeOrder,Kosterlitz1973OrderingMetastability}.  The Mermin--Wagner--Hohenberg theorem~\cite{Mermin1966AbsenceFerromagnetism,Hohenberg1967ExistenceOfLong-RangeOrder} states that the quasi-long-range order precludes the spontaneous symmetry breaking associated with true long-range order at any nonzero temperature, i.e.,
\begin{align}
    \lim_{\| \hvec \| \to 0} \lim_{N \to \infty} \mathbb{E} \mvec (\beta, \hvec, N) = 0 \,\, \textrm{for all} \,\, \beta < \infty .
\label{eq:MWHTheorem}
\end{align}
This presented, however, the paradox between theory and experiment described above, as measurements consistent with system-spanning symmetry-broken spin-phase coherence have been measured across a broad and diverse array of experimental BKT systems~\cite{Baity2016EffectiveTwoDimensionalThickness,Shi2016EvidenceCorrelatedDynamics,Bishop1978StudySuperfluid,Bramwell2015PhaseOrder,Wolf1981TwoDimensionalPhaseTransition,Resnick1981KosterlitzThouless,Bramwell1993Magnetization,Huang1994MagnetismFewMonolayersLimit,Elmers1996CriticalPhenomenaTwoDimensionalMagnet,BedoyaPinto2021Intrinsic2DXYFerromagnetism,Hadzibabic2006KosterlitzThouless,Fletcher2015ConnectingBKTAndBEC,Christodoulou2021Observation} (n.b., `spin-phase coherence' describes positional coherence of the local spin phases, which spans the system when symmetry is broken).  %Moreover, the discovery of a transition to quasi-long-range order at low temperature was highly non-conventional at the time.  Its mechanism was explained, however, by the KT description of the phase transition in terms of topological defects, which we explore below in detail in Section~\ref{sec:EmergentElectrolyte}.
%As regards the symmetry-breaking paradox, however, 
This was partially resolved in large finite systems by the expected low-temperature norm of the $U(1)$ order parameter going to zero very slowly and at the same rate as its %fluctuations~\cite{Tobochnik1979MonteCarlo,Archambault1997MagneticFluctuations,Bramwell1998Universality}.  Tobochnik and Chester~\cite{Tobochnik1979MonteCarlo} and then Archambault, Bramwell \& Holdsworth~\cite{Archambault1997MagneticFluctuations} 
fluctuations~\cite{Archambault1997MagneticFluctuations,Bramwell1998Universality}.  Archambault, Bramwell \& Holdsworth~\cite{Archambault1997MagneticFluctuations} 
demonstrated that the expected norm $\mathbb{E} \| \mvec \|$ and its fluctuations $\sigma_{\| \mvec \|}$ both scale like $N^{-1 / (8\pi \beta J)}$ at large $N$ and as $\beta \to \infty$~\footnote{Tobochnik and Chester also presented the scaling of the fluctuations~\cite{Tobochnik1979MonteCarlo}.}.  This demonstrates that no finite system size exists at which the expected norm can be considered to have reached its thermodynamic value of zero, as the relative fluctuations $\sigma_{\| \mvec \|} / \mathbb{E} \| \mvec \|$ are system-size independent and therefore cannot be made arbitrarily small with increasing system size.  This is a consequence of the law of large numbers not applying to sums of correlated random variables, which is the case here due to the spin-phase fluctuations being correlated on all length scales (quasi-long-range order) so that correlations are cut off on long length scales in finite systems.  It follows that i) the thermodynamic limit of the expected norm is not reached at arbitrarily large system size, and ii) the probability density function (PDF) of the fluctuation-normalized order parameter $\mvec / \sigma_{\| \mvec \|}$ has a well-defined sombrero form in the thermodynamic limit~\cite{Faulkner2024SymmetryBreakingBKT}.

In response to this, Bramwell \& Holdsworth~\cite{Bramwell1993Magnetization,Bramwell1994Magnetization} developed a comprehensive theory of finite-size transition temperatures and a symmetry-breaking critical exponent $\widetilde{\beta} = 3 \pi^2 / 128$ in large macroscopic systems, where
\begin{align}
    \widetilde{\beta} := \lim_{N \to \infty} \left. \frac{\partial \ln \mathbb{E} \| \mvec \|}{\partial \ln t(N)}  \right|_{\beta \uparrow \beta^*(N)} 
\end{align}
is in fact defined in the thermodynamic limit.  Here, the lower finite-size transition temperature $1 / \beta^*(N)$ is defined to coincide with $\mathbb{E}\| \mvec \| = (cN)^{-1 / 16}$ ($c = 1.8456$) and $t(N)$ is defined as $\beta_0 / \beta_{\rm c}(N) - \beta_0 / \beta$ for all $\beta \in ( \beta_{\rm c}(N), \beta^*(N) ]$, where the upper finite-size transition temperature $1 / \beta_{\rm c}(N)$ is defined as the highest temperature at which the correlation length $\xi \sim \sqrt{N}$, and $\beta_0 > 0$ is some reference inverse temperature.  Both inverse finite-size transition temperatures converge to $\beta_{\rm BKT}^{\rm XY}$ as $N \to \infty$, and these results were confirmed by Chung~\cite{Chung1999EssentialFiniteSizeEffect} using a transfer-matrix method.  We note also that 
the definition of $\beta^*(N)$ is equivalent to renormalizing to a spin-wave model with spin stiffness $K$ to give $\mathbb{E}\| \mvec \| = (cN)^{-1 / (8\pi K)}$, and then assuming $K = 2 / \pi$ at the finite-size transition -- the spin stiffness predicted by Jos\'{e} {\it et al}.~\cite{Jose1977Renormalization} and Nelson \& Kosterlitz~\cite{Nelson1977UniversalJump} at the phase transition (see Section~\ref{sec:SpinStiffness}).

%Bramwell--Holdsworth theory 
This finite-system framework 
was extremely successful in describing experimental measurements related to the expected norm~\cite{Bramwell1993Magnetization,Huang1994MagnetismFewMonolayersLimit,Elmers1996CriticalPhenomenaTwoDimensionalMagnet,Chung1999EssentialFiniteSizeEffect,Bramwell2015PhaseOrder,BedoyaPinto2021Intrinsic2DXYFerromagnetism,Venus2022RenormalizationGroup}, but a rigorous dynamical framework for the phase fluctuations in both finite and thermodynamic systems was still lacking.  This gap in the literature was crucial, as broken symmetry is necessarily an asymptotically slow mixing between equilibrium regions of equal (measure and) probability mass -- corresponding to asymptotically slow mixing of the directional global $U(1)$ phase in this case.  Experimentally, this posed particular problems to the superconducting film and Josephson-junction array, as the electrical resistance is a directly measurable quantity that is conjugate to the directional condensate phases.  %This meant that the strongly nonergodic/autocorrelated electrical resistance measured at the transition in the superconducting films~\cite{Shi2016EvidenceCorrelatedDynamics} could not be described by increasingly large symmetry-broken regions persisting on significant nonergodic timescales as the system temperature is reduced towards the transition -- because this effect (which leads to critical slowing down) is a necessary precursor to the system-spanning symmetry-broken condensate-phase coherence that induces asymptotically slow directional mixing.  Similarly, the low-temperature magnetization vector in BKT magnetic films and the orientational order parameter in the hexatic phase of colloidal films should provide direct experimental evidence of system-spanning symmetry-broken spin-phase coherence, but no theory existed to predict the phenomenon.
The strongly autocorrelated electrical resistance measured at the transition in the superconducting films~\cite{Shi2016EvidenceCorrelatedDynamics} could not therefore be explained.  Similarly, the low-temperature magnetization vector in XY magnetic films and the orientational order parameter in the hexatic phase of colloidal films should provide direct experimental evidence of the asymptotically slow directional mixing, but no theory existed to predict the phenomenon.

%Rather than long-range order with respect to the order parameter, the BKT transition in fact induces topological order in the low-temperature phase -- described by a nonergodic suppression of topological-sector fluctuations~\cite{Faulkner2015TSFandErgodicityBreaking}.  The symmetry-breaking paradox was only fully resolved, however, once it was shown that this ergodicity-breaking mechanism induces a more general symmetry form of symmetry breaking to that defined by spontaneous symmetry breaking.  

\subsection{Order-parameter dynamics and singular limit}
\label{sec:SingularLimitGSB}

The symmetry-breaking paradox was fully resolved by the introduction of a more general framework for characterizing broken symmetry than that defined by spontaneous symmetry breaking~\cite{Faulkner2024SymmetryBreakingBKT}.  Some global symmetry is generally broken if the directional fluctuations of the symmetry-breaking order parameter go to zero in the thermodynamic limit, provided they are asymptotically smaller than the expected norm of the order parameter.  This mathematically reflects the order parameter arbitrarily choosing some well-defined direction in the thermodynamic limit, providing a theoretical framework for all cases of negligible directional fluctuations compared to the expected norm in arbitrarily large experimental systems.  This is clearly fulfilled by the case of spontaneous symmetry breaking in the 2D Ising model with local Metropolis dynamics.  This is because the simulation variance of the zero-field magnetization goes to zero at low temperature in the thermodynamic limit, while the expected absolute value of the zero-field magnetization goes to zero (as $N \to \infty$) only at $\beta < \beta_{\rm Ising}$, and is system-size independent at sufficiently large $\beta < \infty$.  %As stated above, however, this broader framework and its accompanying singular limit in equation~\eqref{eq:DirectionalStabilityIsingMetrop} are not required for the 2D Ising model, as its symmetry breaking is well described by the spontaneous symmetry breaking of equation~\eqref{eq:2dOnsagerYangSolution}.  
This broader framework is not required, however, to describe broken symmetry in the 2D Ising model, as this is well described by the spontaneous symmetry breaking of equation~\eqref{eq:2dOnsagerYangSolution}.
Indeed, this is due to the expected absolute value of the zero-field magnetization going to zero only at $\beta < \beta_{\rm Ising}$.  

For the case of the 2DXY model, however, things are not so clear, as the expected zero-field norm $\mathbb{E}\|\mvec\|$ goes to zero as $N \to \infty$ at all nonzero temperatures.  It was therefore necessary to show that the low-temperature directional fluctuations are asymptotically smaller than the expected norm at large $N$ and under local Metropolis/Brownian spin dynamics~\cite{Faulkner2024SymmetryBreakingBKT}.  To draw analogies with the broken symmetry of the 2D Ising model, this work additionally benchmarked the diffusive local Metropolis dynamics against the ballistic-style dynamics of the event-chain Monte Carlo algorithm~\cite{Michel2015EventChain}.  This is because the latter induces global rotations and therefore $U(1)$ symmetry on short timescales, in analogy with Swendsen--Wang simulations of the 2D Ising model achieving $Z_2$ symmetry on short timescales.  In this subsection, we outline the Metropolis and event-chain algorithms for the 2DXY model before reviewing the symmetry-breaking results of \cite{Faulkner2024SymmetryBreakingBKT}.

\begin{figure}[t]
    \includegraphics[width=0.85\linewidth]{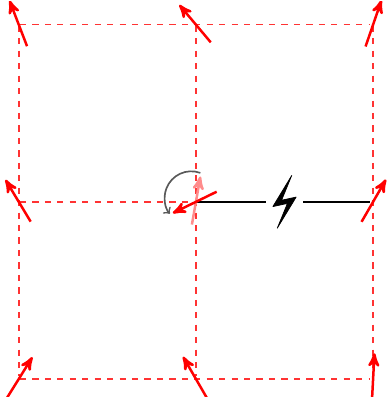}
  \caption{Example of the event-chain algorithm continuously advancing some active spin between particle events, using the superposition of independent (two-spin) Poisson processes.  The central spin advances with constant anticlockwise velocity $v > 0$ until an event is triggered by one of its four nearest-neighboring spins.  One particle-event time is sampled per two-spin nearest-neighbor potential and the soonest of these defines the next event, with the corresponding vetoing spin becoming active following the event [assuming no boundary events occur -- see the text below equation~\eqref{eq:MasterEquationECMC}].  The light red arrow represents the active spin at the time of the initial particle event.  The dark red arrows represent the spins at the time of the next particle event.  The gray arrow represents the direction of motion of the active spin between events.  The red-dashed and black lines represent the edges of the periodic spin lattice $D'$.  The lightning bolt separates the active and vetoing spins.}
  \label{fig:DynamicsECMC}
\end{figure}

The local Metropolis algorithm (for the 2DXY model) with Gaussian noise was used in \cite{Faulkner2024SymmetryBreakingBKT}.  With $\sigma_{\rm noise}^2$ the variance of the noise distribution $\mathcal{N}(0, \sigma_{\rm noise}^2)$, each Metropolis proposal attempts to perturb a single spin phase by an amount $\delta \sim \mathcal{N}(0, \sigma_{\rm noise}^2)$.  This is accepted with probability $\min [ 1, \exp(-\beta \Delta U) ]$ by comparing this probability with a random number $\tilde{\Upsilon} \sim \mathcal{U}(0, 1)$ ($\Delta U$ is the potential difference between the proposed and current configurations).  Samples are drawn after each (2DXY) Metropolis Monte Carlo time step $\Delta t_{\rm Metrop}^{\rm XY}$ (defined as the elapsed simulation time between $N$ attempted single-spin moves) and $\sigma_{\rm noise}$ is tuned such that the Metropolis acceptance rate $a_{\rm Metrop} \simeq 0.6$.  This reflects Brownian spin dynamics, as Brownian dynamics with an $N$-independent %physical time step $\Delta t_{\rm phys} := a_{\rm Metrop} \sigma_{\rm noise}^2 \Delta t_{\rm Metrop}^{\rm XY} / 2$ 
diffusion constant $a_{\rm Metrop} \sigma_{\rm noise}^2 / 2$ were proven to be the (thermodynamic) limiting behavior of Metropolis dynamics for a set of linearly coupled harmonic oscillators~\cite{Roberts1997WeakConvergence,Neal2006OptimalScalingForPartially}.  Informal proofs~\cite{Kikuchi1991Metropolis,Sanz2010DynamicMCVsBrownianDynamics} and simulation data from a broad range of physical systems~\cite{Kikuchi1991Metropolis,Sanz2010DynamicMCVsBrownianDynamics,Kaiser2013OnsagersWienEffect,Jaubert2009SignatureOfMagneticMonopole,Kaiser2015ACWienEffectInSpinIce} have also been presented (for uniform noise) by the physics community.  

\begin{figure*}[t]
  \includegraphics[width=\linewidth]{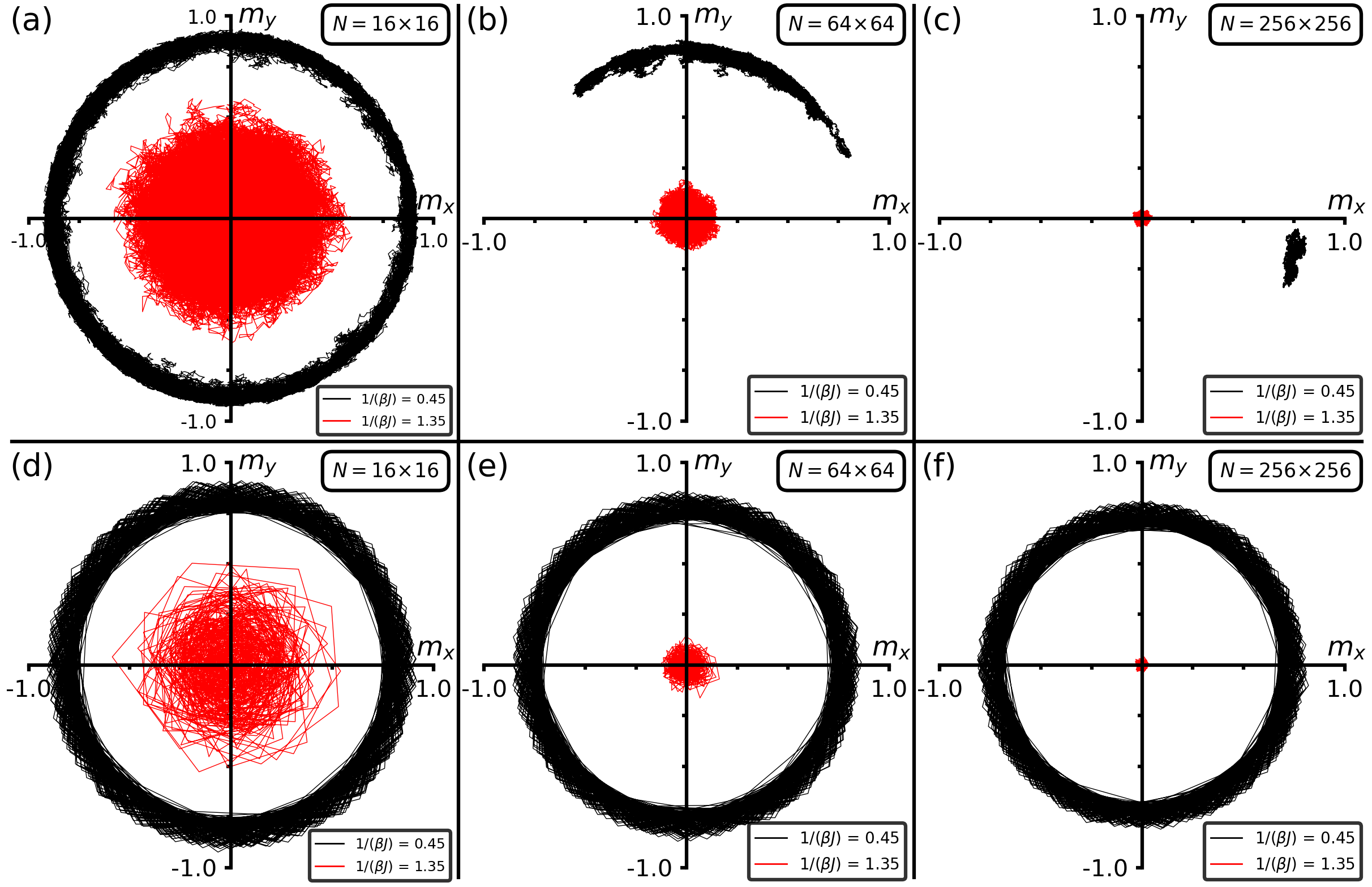}
  \caption{Evolutions of the $U(1)$ order parameter $\mvec$ [defined in Eq.\eqref{eq:MagnetizationXY}] over the course of single Metropolis [(a)-(c)] and event-chain [(d)-(f)] simulations are/are not consistent with local Metropolis/event-chain dynamics being characterized by the singular limit described below equation~\eqref{eq:DirectionalStabilityXYMetrop}.  The Metropolis results [(a)-(c)] suggest a low/high-temperature directional mixing timescale that does/does not diverge with $N$ at finite $\beta$.  The event-chain results [(d)-(f)], by contrast, are consistent with a directional mixing timescale that does not diverge with $N$ at finite $\beta$.  Metropolis simulations comprise $10^5$ samples with acceptance rate $a_{\rm Metrop} \simeq 0.6$.  Event-chain simulations comprise $10^3$ samples.  Straight lines connect samples between time steps.  $10^4$ / $10^5$ equilibration samples were discarded from each event-chain / Metropolis simulation.  Figure is taken from \cite{Faulkner2024SymmetryBreakingBKT}.}
  \label{fig:MagnetisationEvolution}
\end{figure*}

In contrast with these diffusive dynamics, the event-chain Monte Carlo algorithm operates by effectively inverting the local Metropolis algorithm: rather than proposing a single discrete spin-phase perturbation before drawing a random number $\tilde{\Upsilon} \sim \mathcal{U}(0, 1)$ to decide whether to accept the move, event-chain Monte Carlo draws the random number and then continuously advances some \emph{active spin} (in spin-phase space) at some fixed anticlockwise \emph{auxiliary velocity} $v > 0$ until the first time at which a Metropolis rejection would have occurred, given the random number (below we compare this method with $v \in \mathbb{R}$, possibly for multiple active spins~\footnote{Indeed, one can choose to vary the velocity (via a variety of methods) and/or advance multiple active spins.}).  This time defines the next \emph{particle-event time} $t_\eta > 0$ and the process essentially amounts to performing a sequence of infinitesimally small Metropolis moves in a fixed direction until $t_\eta$.  To sample $t_\eta$ we therefore consider a sequence of $m$ proposed Metropolis translations of length $\delta > 0$ in the positive (i.e., anticlockwise) spin-phase direction.  Beginning from some initial time $t_0 \ge 0$, defining $\Delta U_{t_0,i} := U_{\rm XY} \left[ \varphi_1(t_0), \dots , \varphi_a(t_0) + i \delta, \dots , \varphi_N(t_0) \right] - U_{\rm XY} \left[ \varphi_1(t_0), \dots , \varphi_a(t_0) + (i - 1) \delta, \dots , \varphi_N(t_0) \right]$ and assuming $\varphi_a(t_0) + m \delta < \pi$ (as $\varphi_k \in [-\pi, \pi )$ for all spins $k$), the probability of accepting all proposals and translating the active spin $a$ through the total distance $\eta := m \delta$ is 
\begin{align}
    p_{t_0, \eta} = & \prod_{i = 1}^{\eta / \delta} \min \left[ 1, \exp (-\beta \Delta U_{t_0, i}) \right] \nonumber\\
    = & \exp \left[ -\beta \sum_{i = 1}^{\eta / \delta} \max \left( 0, \Delta U_{t_0, i} \right) \right] . 
\end{align}
Defining $\boldsymbol{\varphi} := (\varphi_1, \dots , \varphi_N)^T$, it follows that 
\begin{align}
    p_{t_0, \eta} \to \exp \left[ -\beta \int_{\varphi_a(t_0)}^{\varphi_a(t_0) + \eta} \max \left[ 0, \partial_a U_{\rm XY}(\boldsymbol{\varphi}') \right] d\varphi_a' \right] 
\end{align}
as $\delta \to 0$ with $\eta$ fixed, i.e., in the continuous-time limit.  We therefore draw some random number $\tilde{\Upsilon} \sim \mathcal{U}(0, 1)$ at $t_0$ and if 
\begin{align}
    -\log \tilde{\Upsilon} < \beta \int_{\varphi_a(t_0)}^\pi \max \left[ 0, \partial_a U_{\rm XY}(\boldsymbol{\varphi}') \right] d\varphi_a' ,
    \label{eq:BoundaryEventECMC}
\end{align}
we solve 
\begin{align}
    -\log \tilde{\Upsilon} = \beta \int_{t_0}^{t_\eta} \max \left[ 0, v \, \partial_a U_{\rm XY}(\boldsymbol{\varphi}(t)) \right] dt 
    \label{eq:MasterEquationECMC}
\end{align}
to sample the next particle-event time $t_\eta = t_0 + \eta / v$.  This defines a \emph{particle event} and one of the four (nearest) neighboring spins $k$ then becomes active with probability $\propto \max \left[ 0, - v \, \partial_k U_{\rm XY}(\boldsymbol{\varphi}(t = t_\eta)) \right]$.  If equation~\eqref{eq:BoundaryEventECMC} does not hold, however, then $\varphi_a(t = t_{\rm b}) = \pi$ at the \emph{boundary-event time} $t_{\rm b} := t_0 + [\pi - \varphi_a(t_0)] / v$ and the active spin is instantaneously translated to $-\pi$ (then remains active).  This can be viewed as a \emph{teleportation portal}~\cite{Bierkens2023PDMPsWithBoundaries}.  %One may recycle $\tilde{\Upsilon}$ following a boundary event.

It follows from equation~\eqref{eq:MasterEquationECMC} that each next particle-event time is sampled from a non-homogeneous Poisson process with intensity function (or \emph{event rate}) $\beta \max \left[ 0, v \, \partial_a U_{\rm XY}(\boldsymbol{\varphi}(t)) \right]$.  It is typically challenging, however, to solve equation~\eqref{eq:MasterEquationECMC} in practice.  The present simulations instead sample one particle/boundary-event time per two-spin potential and define the next event time as the soonest of these.  If this corresponds to a particle event, the other spin of the corresponding two-spin potential becomes active, as demonstrated in figure~\ref{fig:DynamicsECMC}.  This follows from the superposition of independent Poisson processes.  %Observations were drawn at every $N^{\rm th}$ particle event, with the event-chain Monte Carlo time step defined as $\Delta t_{\rm ECMC} := \tau / n$.  This Monte Carlo time step then results in directional mixing timescales in units of CPU time, to allow comparison with the directional mixing timescales of the Metropolis simulations.  When estimating quantities related to the norm of the magnetisation $\| {\bf m} \|$, however, one should draw observations every $N$ units of event-chain time or via some homogeneous Poisson process with intensity function proportional to $N$ (this generates a sample of $\| {\bf m} \|$ that reflects the time spent at all values of $\| {\bf m} \|$, rather than a sample of $\| {\bf m} \|$ at the event times). 
Samples were drawn every $N$ units of event-chain time~\footnote{It was erroneously stated in \cite{Faulkner2024SymmetryBreakingBKT} that samples were drawn at every $N^{\rm th}$ particle event (which would have resulted in a biased sample of $\| {\bf m} \|$ but not of $\phi_{\bf m}$).}.  The event-chain Monte Carlo time step is defined as $\Delta t_{\rm ECMC} := \tau / n$ with $\tau$ the simulation timescale and $n$ the sample size.

Restricting the auxiliary velocity $v$ to positive values is equivalent to breaking symmetry on the velocity space of symmetric-velocity event-chain Monte Carlo.  The symmetric framework augments the state (or configuration) space by introducing an auxiliary velocity vector $\vvec \in \mathbb{R}^N$ drawn from some symmetric distribution $\nu$.  Convergence on the joint distribution $\mu(\rvec, \vvec) = \pi(\rvec)\nu(\vvec)$ of the augmented space $[-\pi, \pi)^N \times \mathbb{R}^N$ is then ensured by constructing a process that generates a non-biased sample of $\vvec$ (i.e., the sample is symmetric on velocity space).  %Correctness of this symmetric-velocity process has been proven for similar piecewise deterministic Markov processes defined on Euclidean state spaces in Bayesian computation [ref].  
For a broad class of particle models, $\pi$-invariance has been demonstrated (via $\mu$-invariance) for the symmetric-velocity~\cite{Andrieu2020GeneralPerspective,Andrieu2021PeskunTierneyOrdering} and asymmetric-velocity~\cite{Faulkner2024SamplingAlgorithms} processes, while irreducibility on the original state space has only been demonstrated for the symmetric-velocity process~\cite{Monemvassitis2023PDMPCharacterisation}.  Elsewhere, both $\mu$-invariance and irreducibility (i.e., correctness with aperiodicity) have been shown for a similar piecewise deterministic Markov process (from Bayesian computation) with symmetric-velocity refreshment~\cite{Bierkens2019Ergodicity}.  Breaking symmetry on velocity space appears, however, to accelerate mixing for translationally symmetric models defined on state spaces with toroidal topology (as are common in statistical physics)~\cite{Bernard2009EventChain}.  Moreover, numerical results suggest correctness for a variety of such models~\cite{Bernard2009EventChain,Bernard2011TwoStepMelting,Michel2015EventChain,Faulkner2018AllAtomComputations}, and irreducibility on the original state space  is trivial for the 2DXY model.  It will be interesting to explore whether correctness of the asymmetric-velocity process holds rigorously for a broad class of such models, perhaps due to the toroidal topology and translational symmetry (i.e., whether symmetry can be broken on velocity space thanks to symmetries on the original state space).  %Moreover, it will be interesting to explore whether the asymmetric-velocity process presented here leads to faster directional mixing (in this 2DXY case) than its symmetric-velocity counterpart.

%Christophe and Sam showed pi invariance (or simply invariant) of ECMC with symmetric-velocity refreshment~\cite{Andrieu2021PeskunTierneyOrdering} then Sam (with me-ish)~\cite{Faulkner2024SamplingAlgorithms} showed pi invariance for asymmetric(xy)-velocity refreshment

%Manon and others showed ergodicity (pi invariance + irreducibility) of ECMC with symmetric-velocity refreshment~\cite{Monemvassitis2023PDMPCharacterisation}

%Gareth and Joris showed ergodicity (pi invariance + irreducibility) of zig zag with symmetric-velocity refreshment~\cite{Bierkens2019Ergodicity}

Figure~\ref{fig:MagnetisationEvolution} shows evolutions of the order parameter $\mvec$ at various systems sizes and temperatures, where each evolution is over the course of a single simulation using either local Metropolis or (asymmetric-velocity) event-chain dynamics.  The low-temperature simulations reflect the PDF with sombrero form described in Section~\ref{sec:MWHandBHtheories}, while those at high temperature display the single well (centred at $\mvec = 0$) associated with a $U(1)$-symmetric thermodynamic phase [$1 / \beta_{\rm BKT}^{\rm XY} \simeq 0.887 J$~\cite{Weber1988MonteCarloDetermination}].  %While the radius of the low-temperature sombrero potential tends to zero in the thermodynamic limit, we stress again that $\mathbb{E} \| \widetilde{m} \|$ is system-size independent.  
We note that all analysis of the global $U(1)$ phase $\phi_\mvec$ holds to the thermodynamic limit because $\mathbb{E} \| \mvec \|$ does not reach its thermodynamic value of zero at arbitrarily large system size.  This can also be argued from the equivalence of the directional phases of $\mvec$ and $\mvec / \sigma_{\| \mvec \|}$.

\begin{figure*}[t]
  \includegraphics[width=\linewidth]{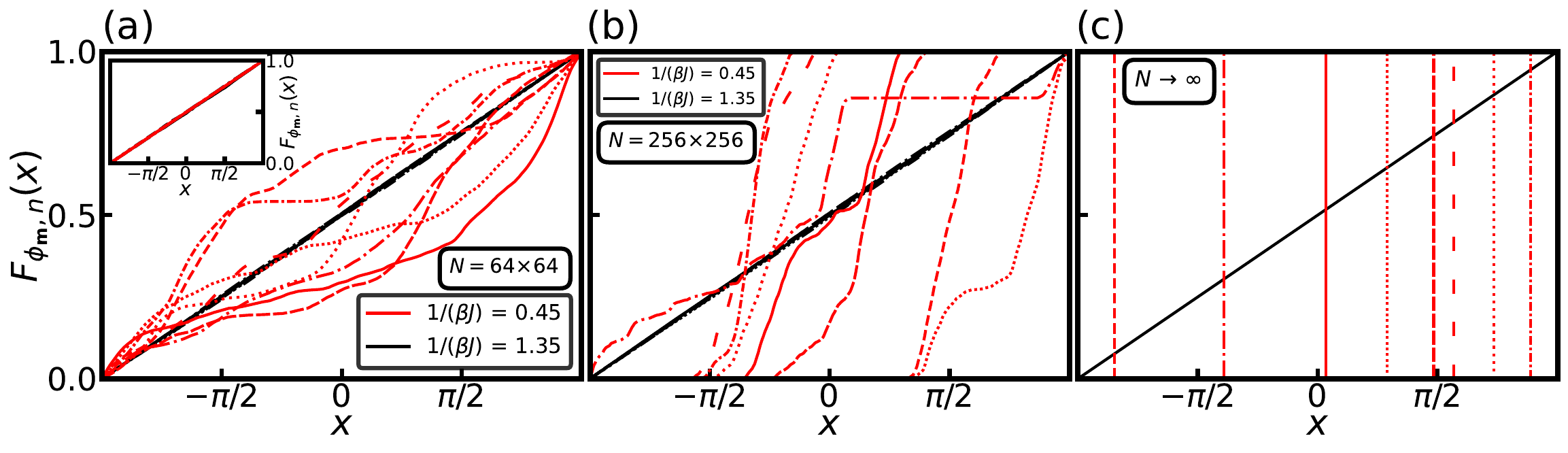}
  \caption{ECDFs [defined in equation~\eqref{eq:ECDF}] of the global $U(1)$ phase $\phi_\mvec$ [defined below equation~\eqref{eq:MagnetizationXY}] for local Metropolis simulations [(a)-(b)] reflect the symmetry properties of the low/high-temperature phase -- which are demonstrated by the red/black schematics in (c).  Different line styles represent different realizations.  (a)-(b)  Eight local Metropolis simulations of $n = 10^6$ samples at low and high temperature, with acceptance rate $a_{\rm Metrop} \simeq 0.6$  Inset: Four event-chain simulations of $n = 10^4$ samples at low and high temperature.  $10^4$ / $10^5$ equilibration samples were discarded from each event-chain / Metropolis simulation.  Figure is taken from \cite{Faulkner2024SymmetryBreakingBKT}.}
  \label{fig:ECDFs}
\end{figure*}

In analogy with Section~\ref{sec:IsingModelAndSSB}, symmetric simulations are defined by a directional simulation variance $s_{\phi_\mvec}^2$ that has converged to its expected value ${\rm Var}[\phi_\mvec]$ (within some small $\varepsilon > 0$) and the timescale on which this is first achieved is the directional mixing timescale $\tau_{\rm mix}$ [n.b., ${\rm Var}[\phi_\mvec] = \pi^2 / 3$ as $\phi_\mvec \sim \mathcal{U}(-\pi, \pi)$].  
%Symmetric simulations are analogously (to those in Section~\ref{sec:IsingModelAndSSB}) defined by a directional simulation variance $s_{\phi_\mvec}^2$ that has converged to its expected value ${\rm Var}[\phi_\mvec] = \pi^2 / 3$ [the variance of the uniform distribution $\mathcal{U}(-\pi, \pi)$] within some small $\varepsilon > 0$.  The timescale on which this is first achieved again defines the directional mixing timescale.  
Figures~\ref{fig:MagnetisationEvolution}(a)-(c) use local Metropolis dynamics.  For $N \ge 64 \! \times \! 64$, the low-temperature outputs are evidently asymmetric on the simulation timescale, while the high-temperature outputs suggest $U(1)$-symmetric simulations for a broad range of system sizes.  Moreover, the Metropolis results are consistent with the low-temperature directional mixing timescale increasing with system size, suggesting a singular limit analogous to that in equation~\eqref{eq:DirectionalStabilityIsingMetrop}.  To see this, we define the (XY) \emph{finite directional stability} 
\begin{align}
    g_{\rm XY}(\beta, \tau, N) := 1 - \sqrt{\frac{\langle s_{\phi_\mvec}^2(\beta, \tau, N) \rangle}{{\rm Var}[\phi_\mvec]}} ,
\label{eq:XYVarianceRatioFn}
\end{align}
%where the expected value of the squared phase fluctuations $\langle s_{\phi_\mvec}^2(\beta, \tau, N) \rangle$ is ${\rm Var}[\phi_\mvec] = \pi^2 / 3$ -- the variance of the uniform distribution $\mathcal{U}(-\pi, \pi)$.  
where $\sqrt{\langle s_{\phi_\mvec}^2(\beta, \tau, N) \rangle}$ are the (XY) \emph{directional (phase) fluctuations}. %, whose expected value is given by the standard deviation ($\pi / \sqrt{3}$) of the uniform distribution $\mathcal{U}(-\pi, \pi)$.  
Defining the (XY) \emph{long-time directional stability} 
\begin{align}
\gamma_{\rm XY}(\beta) := \lim_{\tau \to \infty} \lim_{N \to \infty} g_{\rm XY}(\beta, \tau, N),
\label{eq:DirectionalStabilityXYDef}
\end{align}
the Metropolis results are then consistent with 
\begin{align}
\gamma_{\rm XY}^{\rm Metrop}(\beta) = 
\begin{cases}
    1 & {\rm for} \,\, \beta > \beta_{\rm BKT}^{\rm XY} , \\
    0 & {\rm for} \,\, \beta < \beta_{\rm BKT}^{\rm XY} .
\end{cases}
\label{eq:DirectionalStabilityXYMetrop}
\end{align}
In analogy with the $Z_2$-symmetric Swendsen--Wang simulations of the 2D Ising model, by contrast, all event-chain results in figures~\ref{fig:MagnetisationEvolution}(d)-(f) suggest directional mixing timescales that do not diverge with $N$, consistent with zero long-time directional stability for all finite $\beta$.  As reviewed in detail in Section~\ref{sec:SymmetryBreakingGSB} below, it was shown~\cite{Faulkner2024SymmetryBreakingBKT} that these observations do indeed hold -- with the Metropolis case due to vanishing low-temperature phase fluctuations in the thermodynamic limit.  In analogy with the Ising case [described below equation~\eqref{eq:DirectionalStabilityIsingMetrop}] this thermodynamic limit is singular for all finite $\beta > \beta_{\rm BKT}^{\rm XY}$ because exchanging the order of the limits in equation~\eqref{eq:DirectionalStabilityXYDef} returns zero at all nonzero temperatures.  We note again that functions of the form of the long-time directional stabilities in equations~\eqref{eq:DirectionalStabilityIsingDef} and \eqref{eq:DirectionalStabilityXYDef} characterize broken ergodicity more generally (see Section~\ref{sec:TopologicalNonergodicity}).  In the remainder, we assume local Metropolis/Brownian spin dynamics unless otherwise stated. 

It is also important to note that the low-temperature Metropolis outputs in figures~\ref{fig:MagnetisationEvolution}(a)-(b) display small fluctuations towards $\mvec = 0$ at random values of the global $U(1)$ phase $\phi_\mvec$, while the event-chain outputs in figures~\ref{fig:MagnetisationEvolution}(d)-(f) appear to reflect well-converged simulation variances in $\| \mvec \|$.  This is likely to be due to the asymmetry of the low-temperature $\| \mvec \|$ distributions (see figure 2 of \cite{Archambault1997MagneticFluctuations}) as Metropolis dynamics tend to mix poorly into heavy-tailed regions like that seen towards $\mvec = 0$ -- but this hypothesis is left to a future article.

\subsection{Symmetry breaking}
\label{sec:SymmetryBreakingGSB}

\begin{figure}[t]
  \includegraphics[width=\linewidth]{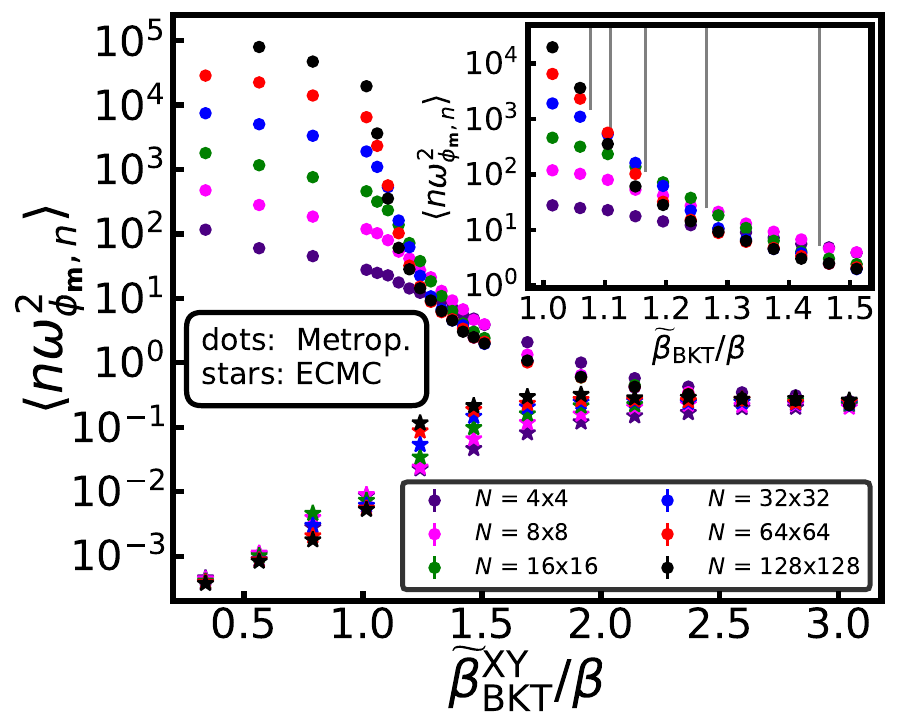}
  \caption{Local Metropolis dynamics break symmetry throughout the low-temperature phase, in contrast with event-chain dynamics.  Estimates of the Cram\'{e}r-von Mises statistic [defined below equation~\eqref{eq:CvM}] are presented against reduced temperature $\tilde{\beta}_{\rm BKT}^{\rm XY} / \beta$ and system size $N$ for local Metropolis (circles) and event-chain (stars) simulations [$\tilde{\beta}_{\rm BKT}^{\rm XY} := 1 / (0.887 J)$ is the approximate inverse 2DXY transition temperature].  Results indicate a directional mixing timescale $\tau_{\rm mix} \sim N^{z / 2}$ with $z = 2$ and $z = 0$ for local Metropolis and event-chain dynamics at low temperature (and $z = 0$ at high temperature for both).  Metropolis outputs intersect near the transition, marked by vertical gray lines in the inset.  It was demonstrated~\cite{Faulkner2024SymmetryBreakingBKT} that i) the intercept temperatures extrapolate to the phase transition as $N \to \infty$, and ii) the intercept values display approximate $\sim N$ scaling.  Outputs are averaged over $560$ simulations with $n = 10^6$ at $\tilde{\beta}_{\rm BKT}^{\rm XY} / \beta > 1.2$ and $n = 3 \! \times \! 10^7$/$n = 10^7$ at $\tilde{\beta}_{\rm BKT}^{\rm XY} / \beta < 1.2$ for $N \gtrless 40 \! \times \! 40$.  Local Metropolis acceptance rates $a_{\rm Metrop.} \simeq 0.6$.  $10^4$ / $10^5$ equilibration samples were discarded from each event-chain / Metropolis simulation.  Estimates are improved by supplemental global-twist dynamics (defined in Section~\ref{sec:GlobalTwists}).  Figure is adapted from \cite{Faulkner2024SymmetryBreakingBKT}.}
  \label{fig:SymmetryBreaking}
\end{figure}

Here we review the low-temperature broken symmetry~\cite{Faulkner2024SymmetryBreakingBKT} suggested by the Metropolis outputs in figure~\ref{fig:MagnetisationEvolution}.  The simulation variance $s_{\phi_\mvec}^2$ will eventually converge to its expected value of ${\rm Var}[\phi_\mvec] = \pi^2 / 3$ on the directional mixing timescale $\tau_{\rm mix} > 0$.  Assuming that the relationship $\langle s_{\phi_\mvec}^2 \rangle \propto \tau / \tau_{\rm mix}$ (for large enough simulation timescale $\tau < \tau_{\rm mix}$) describes this convergence under the diffusive Metropolis dynamics, it follows that $\tau_{\rm mix}$ provides a measure of the scaling of the fixed-timescale phase fluctuations with system size.  %The directional mixing timescale associated with some chosen dynamics 
For any chosen dynamics, this timescale can be estimated by measuring the empirical cumulative distribution functions (ECDFs) 
\begin{align}
    F_{\phi_\mvec,n}(x) := \frac{1}{n} \sum_{i = 1}^n \mathbb{I} \left[ \phi_\mvec(t_i) < x \right] 
    \label{eq:ECDF}
\end{align}
of multiple realizations of the dynamics.  Here, $t_i$ is the Monte Carlo time at observation $i$ and the indicator function $\mathbb{I}(A)$ is one/zero if $A$ does/does not hold.  Each ECDF then measures the number of samples with $\phi_\mvec < x \in [-\pi, \pi)$, meaning that symmetric simulations display small deviations from the target CDF $F(x) := \mathbb{P}(\phi_\mvec < x)$ of the uniform distribution $\mathcal{U}(-\pi, \pi)$.  

Figures~\ref{fig:ECDFs}(a) and (b) present the ECDFs of multiple realizations of both Metropolis and event-chain dynamics.  The ECDFs of the low-temperature Metropolis simulations deviate further from the target CDF $F(x)$ than those performed at high temperature.  Moreover, at low temperature, the mean of the deviations increases with system size, and each realization generates a non-reproducible distribution -- a fundamental characteristic of broken symmetry.  This is consistent with broken symmetry in the low-temperature phase.  In contrast, the event-chain results in the inset are consistent with symmetric convergence at all nonzero temperatures.  Finally, the two possible forms of the ECDF in the thermodynamic limit are demonstrated by the schematic in figure~\ref{fig:ECDFs}(c): the Heaviside-step functions associated with the symmetry-broken thermodynamic phase and the target CDF $F(x) := \mathbb{P}(\phi_\mvec < x)$ of the uniform distribution $\mathcal{U}(-\pi, \pi)$ observed at high temperature.  Indeed, the Metropolis results in figures~\ref{fig:ECDFs}(a) and (b) suggest that any sequence of low-temperature Metropolis ECDFs (at fixed temperature and increasing $N$) tends to some Heaviside step function in the large-$N$ limit.  

\begin{figure*}[t]
  \hspace{-0.6em}
  \includegraphics[width=0.424\linewidth]{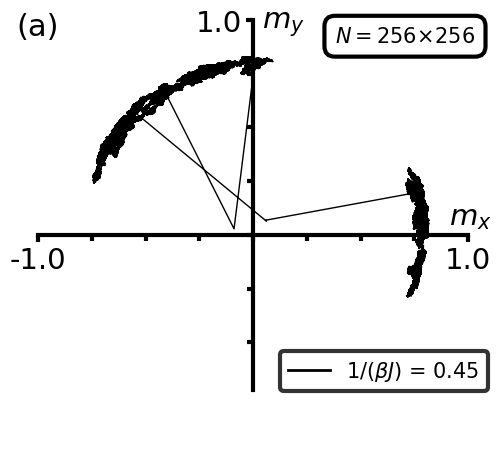}
  \includegraphics[width=0.574\linewidth]{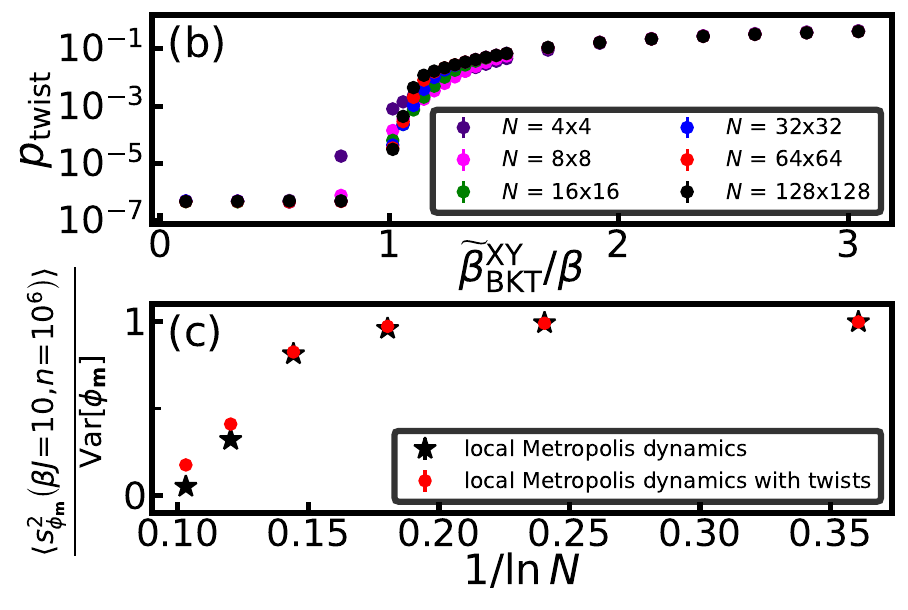}
  \vspace{-2.0em}
  \caption{Supplemental global-twist dynamics [defined below equation~\eqref{eq:DefGlobalSpinTwists}] ensure $U(1)$-symmetric 2DXY simulations on non-divergent timescales for all $\beta < \infty$.  Global-twist events occur in pairs that can be viewed as compound tunnelling events through the sombrero potential, with two such tunnelling events seen in (a).  Global-twist events occur with system-size-independent probability sufficiently far from the transition [(b)] and result in $U(1)$-symmetric simulations on non-divergent timescales for all $\beta < \infty$, characterized by zero long-time directional stability [defined in equation~\eqref{eq:DirectionalStabilityXYDef}].  The phase fluctuations $\sqrt{\langle s_{\phi_\mvec}^2 \rangle}$ of local Metropolis simulations with supplemental global-twist dynamics are then greater than those without supplemental global-twist dynamics, with the difference increasing with system size [(c)].  Simulations use local Metropolis dynamics with acceptance rate $a_{\rm Metrop} \simeq 0.6$.  (a) Evolution of the $U(1)$ order parameter ${\bf m}$ [defined in Eq.\eqref{eq:MagnetizationXY}] over the course of a single Metropolis simulation of an $N = 256 \! \times \! 256$ 2DXY system.  This simulation uses supplemental global-twist dynamics and consists of $n = 5 \! \times \! 10^5$ Monte Carlo time steps.  (b) Probability of global-twist events vs reduced temperature $\tilde{\beta}_{\rm BKT}^{\rm XY} / \beta$ and system size $N$ for $560 n$ attempted global twists, with $n = 10^6$ at $\tilde{\beta}_{\rm BKT}^{\rm XY} / \beta > 1.2$ and $n = 3 \! \times \! 10^7$/$n = 10^7$ at $\tilde{\beta}_{\rm BKT}^{\rm XY} / \beta < 1.2$ for $N \gtrless 40 \! \times \! 40$ [$\tilde{\beta}_{\rm BKT}^{\rm XY} := 1 / (0.887 J)$ is the approximate inverse 2DXY transition temperature].  (c) Low-temperature ($\beta J = 10$) squared phase fluctuations vs $1 / \ln{N}$ with and without supplemental global-twist dynamics.  $\tau = 10^6 \Delta t_{\rm Metrop}$ and the estimate is based on $5600$ simulations.  $10^5$ equilibration samples were discarded from each simulation.  Figure is taken from \cite{Faulkner2024SymmetryBreakingBKT}.}
  \label{fig:GlobalTwists}
\end{figure*}

The Cram\'{e}r-von Mises mean square distance~\cite{Cramer1928OnTheComposition,vonMises1928Wahrscheinlichkeit} 
\begin{align}
    \omega_{\phi_\mvec,n}^2 := \int \left[ F_{\phi_\mvec,n}(x) - F(x) \right]^2 dF(x) 
    \label{eq:CvM}
\end{align}
between $F_{\phi_\mvec,n}(x)$ and the target CDF $F(x)$ is a natural measure of the deviations between the two functions.  It was demonstrated~\cite{Faulkner2024SymmetryBreakingBKT} that the Cram\'{e}r-von Mises statistic $\langle n \omega_{\phi_\mvec,n}^2 \rangle \sim N^{z / 2}$, with $z = 2$ and $z = 0$ (respectively) for low- and high-temperature Metropolis dynamics, and $z = 0$ for event-chain dynamics at all nonzero temperatures.  This is reflected in figure~\ref{fig:SymmetryBreaking}, which shows estimates of $\langle n \omega_{\phi_{\bf m},n}^2 \rangle $ as a function of temperature and system size for local Metropolis and event-chain dynamics.  Given that $\langle \tau \omega_{\phi_\mvec,n}^2 \rangle$ converges on the directional mixing timescale $\tau_{\rm mix}$ and $\langle \omega_{\phi_\mvec,n}^2 \rangle \to 0$ as $\tau \to \infty$, it follows that $\langle \omega_{\phi_\mvec,n}^2 \rangle \propto \tau_{\rm mix} / \tau$ for all $\tau > \tau_{\rm mix}$, implying that $\tau_{\rm mix} \sim N^{z / 2}$.  For local Metropolis/Brownian spin dynamics and large enough $\tau < \tau_{\rm mix}$, the assumption $\langle s_{\phi_\mvec}^2 \rangle \propto \tau / \tau_{\rm mix}$ then implies that $\langle s_{\phi_\mvec}^2 \rangle \sim N^{-1}$ at low temperature.  Moreover, the Metropolis data sets intersect near the transition, marked by the vertical gray lines in the inset.  It was demonstrated~\cite{Faulkner2024SymmetryBreakingBKT} that i) the intercept temperatures extrapolate to the phase transition in the thermodynamic limit, and ii) the intercept values display approximate $\sim N$ scaling.  This confirmed equation~\eqref{eq:DirectionalStabilityXYMetrop} and the singular thermodynamic limit of the low-temperature phase fluctuations, due to nonzero long-time directional stability.  In addition, the phase fluctuations are asymptotically smaller than the expected norm for all $\beta > \beta_{\rm BKT}^{\rm XY}$, where $1 / \mathbb{E}\| \mvec \|$ is $\mathcal{O}\left(N^{1 / 16}\right)$~\cite{Bramwell1993Magnetization}. 

At low temperature, the algebraic correlations therefore combine with the diffusive Brownian dynamics to provoke a divergence (with system size) of the directional mixing timescale. %, i.e., the timescale on which the directional phase of the $U(1)$ order parameter ergodically explores $[-\pi, \pi)$.  
This corresponds to the low-temperature $U(1)$ phase fluctuations going to zero in the thermodynamic limit while being asymptotically smaller than the expected norm of the order parameter.  This constitutes broken symmetry (throughout the low-temperature BKT phase) within the framework of general symmetry breaking described in the first paragraph of Section~\ref{sec:SingularLimitGSB}.  This case is distinct from spontaneous symmetry breaking as it cannot be identified via a singular limit of the expected $U(1)$ order parameter $\mathbb{E} \mvec$.  We emphasize again that spontaneous symmetry breaking is a special case of general symmetry breaking.   

Finally, it will be interesting to explore in future whether the result $z = 0$ for event-chain dynamics (at all nonzero temperatures) is unique to the asymmetric-velocity process presented here, i.e., whether symmetry on the original state space is ensured (on non-divergent timescales) thanks to broken symmetry on the auxiliary velocity space.

\subsection{Supplemental global-twist dynamics}
\label{sec:GlobalTwists}

It was also shown~\cite{Faulkner2024SymmetryBreakingBKT} that local 2DXY dynamics can be supplemented with a global dynamics that guarantee zero long-time directional stability at all nonzero temperatures.  These dynamics consist of proposing externally applied global spin twists 
\begin{align}
    \varphi_\rvec \mapsto \varphi_\rvec + \frac{2\pi}{L} q_{x / y} r_{x/y} 
    \label{eq:DefGlobalSpinTwists}
\end{align}
for all lattice sites $\rvec \in D'$ along the $x/y$ dimension, where $\qvec \in \mathbb{Z}^2$.  With $q_{x/y} = \pm 1$, the dynamics are defined by one such Metropolis proposal along each Cartesian dimension at each Monte Carlo time step.  Global-twist events then occur in pairs that can be viewed as compound tunnelling events.  Figure~\ref{fig:GlobalTwists}(a) displays two such compound events, each due to a global-twist event to small $\| \mvec \|$ before another global-twist event back to the well of the sombrero potential.  The tunnelling events constitute rapid transitions to different positions in the well of the sombrero potential, typically corresponding to a significantly different global $U(1)$ phase $\phi_\mvec$.  %This ensures $U(1)$-symmetric simulations on non-divergent timescales because the probability of 2DXY global-twist events is system-size independent sufficiently far from the transition and non-negligible at all system sizes and temperatures~\cite{Faulkner2024SymmetryBreakingBKT} -- reflecting this probability scaling like $\exp(-2 \pi^2 \beta J)$ as $N \rightarrow \infty$ in the absence of other excitations.  

As reflected in figure~\ref{fig:GlobalTwists}(b), the probability of global-twist events is system-size independent sufficiently far from the transition and non-negligible at all system sizes and temperatures.  This reflects it scaling like $\exp(-2 \pi^2 \beta J)$ as $N \rightarrow \infty$ in the absence of other excitations.  It follows that, for any fixed simulation timescale, any sequence of histograms (with suitably chosen bin size) of the global $U(1)$ phase $\phi_\mvec$ tends to some normalized sum over randomly distributed (around the well of $\widetilde{\mvec}$ sombrero potential) Dirac distributions in the large-$N$ limit, with the expected number of Dirac distributions increasing with (fixed) simulation timescale.  This implies $U(1)$-symmetric simulations on non-divergent timescales at all nonzero temperatures, corresponding to zero long-time directional stability at all nonzero temperatures.  This is reflected in the estimates of the fixed-timescale squared phase fluctuations vs $1 / \ln N$ in figure~\ref{fig:GlobalTwists}(c).  The results reflect the phase fluctuations going to zero in the thermodynamic limit under local Metropolis dynamics, with supplemental global-twist dynamics leading to increased phase fluctuations as an increasing function of system size.  We will see below that supplemental global-twist dynamics also ensure topological ergodicity on non-divergent timescales at all nonzero temperatures, providing the link between topological order and broken $U(1)$ symmetry.

%The algebraic correlations therefore combine with the diffusive Brownian dynamics at low temperature to provoke a divergence (with system size) of the symmetric-mixing timescale on which the directional phase of the $U(1)$ order parameter ergodically explores $[-\pi, \pi)$.  It follows that the global $U(1)$ symmetry is broken at the BKT transition, but symmetry can be restored by non-physical global-twist dynamics that tunnel through the $U(1)$ sombrero %potential, in analogy with global spin-flip dynamics restoring the global $Z_2$ symmetry in the 2D Ising model.  
%potential. % and restore topological ergodicity.  
%The divergent timescale corresponds to the low-temperature $U(1)$ phase fluctuations going to zero in the thermodynamic limit while being asymptotically smaller than the expected norm of the order parameter.  This falls within the framework of general symmetry breaking set out above, but is distinct from spontaneous symmetry breaking as it cannot be identified via a singular limit of the expected $U(1)$ order parameter. %~\footnote{However, as general symmetry breaking implies that the spontaneous $U(1)$ order parameter (left-hand side of equation~\eqref{eq:MWHTheorem}) has a well-defined direction, one may argue that spontaneous symmetry breaking is indeed satisfied.}  

%\section{Emergent electrolyte and topological order/nonergodicity}
\section{Emergent electrolyte and topological order}
\label{sec:EmergentElectrolyte}

\begin{figure*}[t]
  \hspace{-0.6em}
  \includegraphics[width=\linewidth]{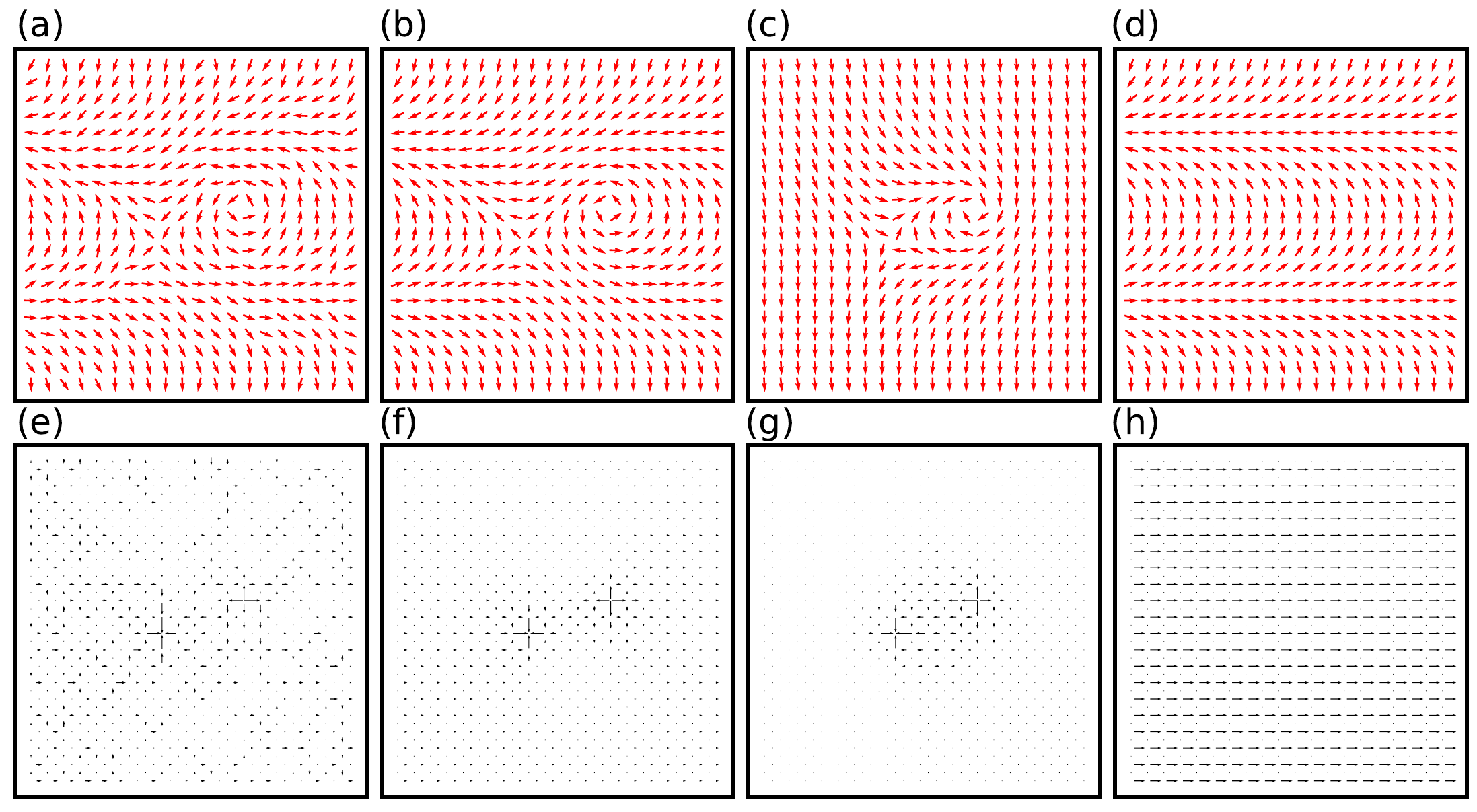}
  \caption{(a)-(d) Any spin configuration decomposes into contributions from three principal excitations: local topological defects (spin vortices), global topological defects (internal global spin twists) and continuous fluctuations around the topological defects (spin waves).  Red arrows represent spins.  (a) Typical 2DHXY configuration at $\beta J = 1.5$ with fixed topological defects [i.e., for illustrative purposes, two local defects (and one global defect) are fixed in position and spin waves are allowed to fluctuate around this constrained configuration -- this is distinct from the non-constrained simulations presented elsewhere in the paper].  (b) Zero-temperature annealed minimization of (a), i.e., the spin waves have been annealed away.  The corresponding spin-difference field is the defect field defined below equation~\eqref{eq:HelmholtzDecompPhaseDifferenceField}.  (c)-(d) Configuration in (b) split into its vortex [(c)] and global-twist [(d)] components.  (e)-(h) The emergent-field representation of (a)-(d).  Black arrows represent the emergent (electric) flux (due to the relevant spin-difference field, i.e., either the total spin-difference field or one of its decomposed components) flowing between emergent charge sites, with the length of the arrow representing the relative magnitude.  (e) The total emergent field defined in equation~\eqref{eq:ElectricFieldHXY}.  (f) The irrotational Coulomb field of equation~\eqref{eq:TopologicalDefectEquivalence}.  (g)  The low-energy Coulomb field defined above equation~\eqref{eq:TSGlobalTwistRelationship}.  (h) $2\pi \wvec / L$ where $\wvec$ is the topological sector defined below equation~\eqref{eq:OriginIndependentPolarisation} (the magnitude has been increased by a scale factor of four to improve visibility).  (a)-(d) are taken from \cite{Faulkner2024SymmetryBreakingBKT}.}
  \label{fig:TopDefects}
\end{figure*}

%The topological nature of the BKT transition suggests the breaking of some form of topological order in the high-temperature phase.  Kosterlitz and Thouless defined topological order in terms of the helicity modulus~\cite{Kosterlitz1973OrderingMetastability}.  The onset of order typically induces, however, an ergodicity breaking~\cite{Palmer1982}, which was not identified by the helicity modulus.  An alternative framework resolved this issue by defining topological order in terms of the global-twist dynamics presented in Section~\ref{sec:GlobalTwists}~\cite{Faulkner2015TSFandErgodicityBreaking,Faulkner2017AnElectricFieldRepresentation}.  In the remainder of this paper, we present the emergent electrostatic-field theory that both defines this topological order and connects it to the symmetry breaking described in Section~\ref{sec:XYModelAndGSB}.  We begin in this section by decomposing the 2DXY model into its three principal excitations, two of which define its topological defects.
The BKT transition induces a topological ordering in the low-temperature phase.  Jos\'{e} {\it et al}.~\cite{Jose1977Renormalization} and Nelson \& Kosterlitz~\cite{Nelson1977UniversalJump} followed the work of Salzberg \& Prager~\cite{Salzberg1963EquationOfStateTwoDimensional} and BKT~\cite{Berezinskii1973DestructionLongRangeOrder,Kosterlitz1973OrderingMetastability} by characterizing the topological order in terms of the \emph{spin stiffness} (defined below in Section~\ref{sec:SpinStiffness}).  This was fundamental to modelling the superfluid stiffness in the experiments of Bishop \& Reppy~\cite{Bishop1978StudySuperfluid}, but its connection with the symmetry breaking described in Section~\ref{sec:XYModelAndGSB} is not clear.  In addition, the onset of order typically induces a large-$N$ nonergodicity under local Brownian dynamics~\cite{Palmer1982}, but the spin stiffness cannot discern whether the topological ordering does indeed break some form of topological ergodicity.  An alternative framework~\cite{Faulkner2015TSFandErgodicityBreaking,Faulkner2017AnElectricFieldRepresentation} reformulated the model system as an emergent electrostatic-field theory on the torus.  This resolved both issues by reframing topological order in terms of a topological nonergodicity that can be restored by the symmetry-restoring global-twist dynamics presented in Section~\ref{sec:GlobalTwists}.  %This framework defined topological order as the nonergodic freezing of global topological defects and connected this topological order/nonergodicity to the symmetry breaking described in Section~\ref{sec:XYModelAndGSB}.  
%Indeed, the notion of topological order is not present in the model of long-range interacting Coulomb charges analysed by Salzberg \& Prager, so where has the topological order gone in the Salzberg--Prager theory, and precisely how do the long- and short-range models exhibit the same phase transition?  %This also resolves the additional paradox of the long-range Salzberg--Prager~\cite{Salzberg1963EquationOfStateTwoDimensional} and short-range BKT~\cite{Berezinskii1973DestructionLongRangeOrder,Kosterlitz1973OrderingMetastability} pictures exhibiting the same phase transition.  
This also answered the closely related question as to how the \emph{short-range} interactions of the 2DXY model map to the \emph{long-range} interactions of the electrolyte: electric fields are the local fundamental objects of electrostatics, and the long-range interacting electrostatic charges of the Salzberg--Prager model are local topological defects in the short-range interacting electric field.  %This elucidated the fact that neither the Salzberg--Prager nor BKT pictures included global topological defects in the (emergent) electric field -- the degrees of freedom required to describe the topological order/nonergodicity. 
This section reviews and expands on these recent advances.

%Throughout the remainder of the paper, we employ the framework for discrete vector calculus presented by Chew~\cite{Chew1994ElectromagneticTheoryLattice}.  All functions are defined to be the discrete counterparts of smooth vector fields.  Any two-dimensional lattice vector field $\Fvec$ is defined as 
%\begin{align}
%    \Fvec(\rvec) = F_x \glb \rvec + \frac{a}{2} \UnitX \grb + F_y \glb \rvec + \frac{a}{2} \UnitY \grb ,
%\end{align}
%where the Cartesian components $F_x, F_y$ are defined equidistant from neighboring lattice sites, and with three-dimensional lattice vector fields defined analogously.  The operators $\widetilde{\boldsymbol{\nabla}}$ and $\hat{\boldsymbol{\nabla}}$ are the forwards and backwards finite-difference operators on a lattice, with the lattice Laplacian defined as $\boldsymbol{\nabla}^2 := \hat{\boldsymbol{\nabla}} \cdot \widetilde{\boldsymbol{\nabla}}$.

%Following the pioneering work of Salzberg, Prager, BKT and others, Vallat \& Beck~\cite{Vallat1994CoulombGas} reformulated the above theories on the torus, which laid the foundations for our emergent electrostatic-field theory~\cite{Faulkner2017AnElectricFieldRepresentation}

\subsection{Harmonic XY model and principal excitations}
\label{sec:HXYModel}

We begin by decomposing the 2DXY model into its three principal excitations, two of which define its topological defects.  To proceed, we turn off the non-linear cosine couplings by transforming to the 2D harmonic XY (HXY) model~\cite{Vallat1992ClassicalFrustratedXY,Bramwell1994Magnetization}.  This is a piecewise-parabolic analogue of the 2DXY model that retains the local $2\pi$-modular XY symmetry while additionally mapping to the 2D lattice-field electrolyte~\cite{Faulkner2017AnElectricFieldRepresentation}.  It is defined by the same 2DXY spins and spin lattice $D'$ but with zero-field potential %(we set $h = 0$ throughout the remainder of the paper) 
\begin{align}
U_{\rm HXY} = \frac{J}{2} \sum_{{\bf r} \in D'} \| \DeltaPhi({\bf r}) \|^2 . % - h \cdot \sum_i \begin{pmatrix} \cos{\varphi_i} \\ \sin{\varphi_i} \end{pmatrix} .
\label{eq:HXYHamiltonian}
\end{align}
Here, %the \emph{spin lattice} $D' := \{ (1, 1), (2, 1), \dots , (\sqrt{N}, \sqrt{N}) \}$ is the set of all spin sites and 
$\DeltaPhi : D' \to [-\pi, \pi)^2$ is the (modular) \emph{phase-difference field}, defined via 
\begin{align}
\glc \DeltaPhi \grc_\mu \glb \rvec + \frac{a}{2} \mathbf{e}_\mu \grb  := \left[ \delta \varphi(\rvec, \mu) + \pi \right] \!\!\!\!\!\! \mod \! (2\pi) - \pi
\label{eq:ModularPhaseDiff}
\end{align}
for each Cartesian component $\mu \in \{ x, y \}$, with $\delta \varphi(\rvec, \mu) := \varphi(\rvec \oplus a \mathbf{e}_\mu) - \varphi(\rvec)$ the \emph{absolute phase difference} along the $\mathbf{e}_\mu$ direction.  Each Cartesian component of the phase-difference field then takes values on the interval $\glc -\pi, \pi \grb$, due to the modular arithmetic in equation~\eqref{eq:ModularPhaseDiff}.  %This induces topological defects via an equivalent mechanism to the cosine function of the 2DXY model.  
As both 2DXY symmetries are retained, this model also experiences a BKT phase transition between a low-temperature phase of algebraic spin-spin correlations and a high-temperature disordered phase -- but at a lower inverse temperature $\beta_{\rm BKT}^{\rm elec}$ [$1 / \beta_{\rm BKT}^{\rm elec} \simeq 1.351 J$~\cite{Bramwell1994Magnetization}] denoted with a superscript ``elec'' due to the mapping to the 2D lattice-field electrolyte.  We note that the 2DHXY model may be considered a realistic lattice model of the 2D condensate composed of a large number of bosons, as such large numbers result in negligible fluctuations in the  amplitude of the condensate wavefunction~\cite{Nelson1977UniversalJump}. 

\begin{figure*}
    \includegraphics[width=\linewidth]{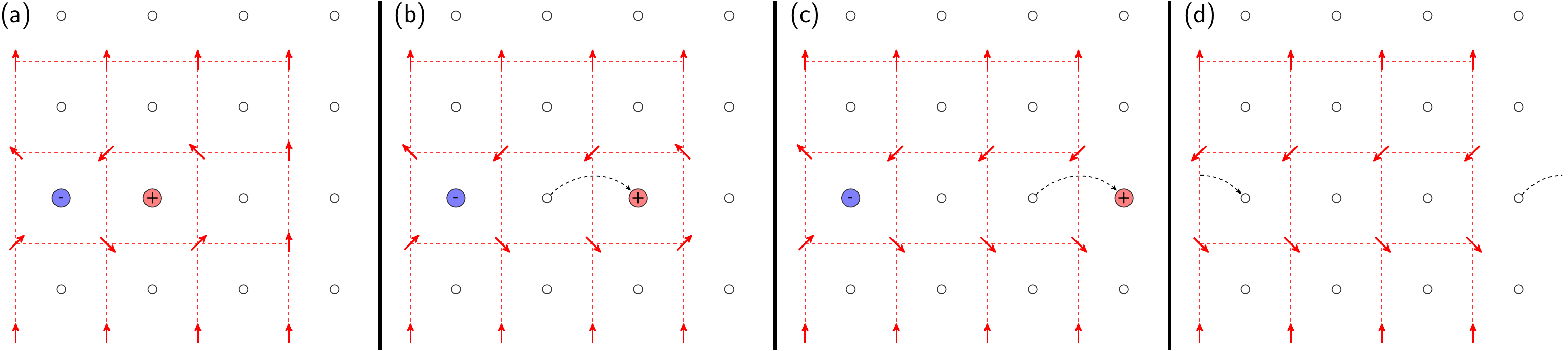}
\caption{Global topological defects can be generated by deconfined local topological defects.  The red arrows are the spins.  The red dashed lines represent the edges of the periodic spin lattice $D'$ (though we omit the edges at the periodic `boundaries').  Each red/blue/white circle represents positive/negative/zero emergent charge.  The curly dashed black arrows represent vortex hops.  (a) A 2DHXY configuration containing a charge-neutral pair of emergent charges.  (b) and (c) The positive emergent charge hops in the positive $x$ direction.  (d) The positive emergent charge annihilates the negative emergent charge by making its final hop (in the positive $x$ direction) around the toroidal system.  Figures are adapted from \cite{Faulkner2017AnElectricFieldRepresentation}. %(\textcopyright IOP Publishing. Reproduced with  permission. All rights reserved.)
}
  \label{fig:GlobalTwistGeneration}
\end{figure*}

Both two-dimensional XY models contain two symmetries.  One is the global $U(1)$ symmetry associated with a global rotation of all spins, while the other is the local $2\pi$-modular symmetry [with respect to each absolute phase difference $\delta \varphi(\rvec, \mu)$] that allows the system to admit local topological defects in the phase-difference field $\DeltaPhi$.  The local topological defects are defined by the \emph{emergent charge density} 
\begin{align}
\rho(\rvec) := \frac{1}{a^2} \sum_{\rvec' \in \partial \Gamma_{\rvec}} \DeltaPhi
  (\rvec') \cdot \lvec_\rvec (\rvec') 
\label{eq:DefinitionVortex}
\end{align}
at some node $\rvec$ of the \emph{charge lattice} $D$ (the centres of the plaquettes of the spin lattice $D'$, i.e., its conjugate).  Here, $\Gamma_{\rvec}$ is the plaquette of $D'$ that contains $\rvec \in D$, $\partial \Gamma_{\rvec}$ is the anticlockwise closed contour in $D'$ around this plaquette and the \emph{directing vector field} $\lvec_\rvec \in \{ 0, -\UnitX, \UnitY, \UnitX - \UnitY \}$ directs the sum around this contour.  A positive/negative vortex is then a point $\rvec \in D$ at which $\rho(\rvec) = \pm 2 \pi / a^2$.  We note that $| \rho(\rvec) | > 2\pi / a^2$ is not geometrically possible.  When the system is placed on the torus, these spin vortices are supplemented with additional global topological defects in the phase-difference field -- the internal global spin twists defined below.  In addition, \emph{spin waves} are continuous field deformations around the defects and represent the third principal excitation.  

The 2DHXY model elucidates the decomposition of the spin field into the three principal excitations.  A typical 2DHXY configuration at $\beta J = 1.5$ and with a fixed topological-defect configuration is shown in figure~\ref{fig:TopDefects}(a) [i.e., for illustrative purposes, two local defects (and one global defect) are fixed in position and spin waves are allowed to fluctuate around this constrained configuration -- this is distinct from the non-constrained simulations presented elsewhere in the paper].  The local topological defect on the right/left is a positive/negative vortex, about which the spins rotate by $\pm 2\pi$.  %We therefore define the emergent charge at some point $\rvec$ on the charge lattice $D$ (the centres of the plaquettes of the spin lattice $D'$) by 
%\begin{align}
%\rho(\rvec) := \sum_{\rvec' \in \partial \Gamma_{\rvec}} \DeltaPhi (\rvec') \cdot \lvec (\rvec') .
%\label{eq:DefinitionVortex}
%\end{align}
%Here, $\Gamma_{\rvec}$ is the plaquette of $D'$ that contains $\rvec \in D$, and $\partial \Gamma_{\rvec}$ is the anticlockwise closed contour in $D'$ around this plaquette.  A positive/negative vortex is then a point $\rvec \in D$ at which $\rho(\rvec) = \pm 2 \pi$.  We note that $| \rho(\rvec) | > 2\pi$ is not geometrically possible.  
Figure~\ref{fig:TopDefects}(b) is the zero-temperature %2DHXY 
annealed minimization of this configuration: topological defects are fixed and spin waves are annealed away.  Figure~\ref{fig:TopDefects}(c) depicts figure~\ref{fig:TopDefects}(b) with global spin twists applied along each Cartesian dimension until the potential is minimized by some $\qvec \in \mathbb{Z}^2$ [see equation~\eqref{eq:DefGlobalSpinTwists}].  This leaves behind the vortex field which contains only local topological defects.  $\qvec$ is identified with the \emph{global twist-relaxation field} $\tilde{\tvec}$, which removed the \emph{internal global spin twist} $\tvec = (0, 1)^T$ depicted in figure~\ref{fig:TopDefects}(d) ($\tvec := - \tilde{\tvec}$).  It is important to note that $\tilde{\tvec}$ can be energy degenerate for certain spin configurations (e.g., certain cases of a single charge-neutral pair of vortices with minimal separation distance $L / 2$) with $\tilde{t}_\mu = \omega$ and $\tilde{t}_\mu = \omega + 1$ resulting in the same minimized potential for some $\omega \in \ZZ$ and $\mu \in \{ x, y \}$.  To align with the emergent-field convention presented in Section~\ref{sec:HarmonicMode}, we choose $\tilde{t}_x = \max \{ \omega, \omega + 1 \}$ and $\tilde{t}_y = \min \{ \omega, \omega + 1 \}$ when such cases arise.  Both this degeneracy and the subtleties of 2DXY internal global spin twists are discussed in more detail in Section~\ref{sec:HarmonicMode}.      %In the emergent electrostatic-field representation (presented below in Section~\ref{sec:EmergentElectricField}) in which the positive/negative vortex maps to a positive/negative emergent charge, the vortex, spin-wave and internal global-twist components map to (respectively)~\cite{Faulkner2017AnElectricFieldRepresentation} the Poisson solution to the Gauss law for the emergent charges, the purely rotational auxiliary gauge field of the 2D lattice electrolyte~\cite{Maggs2002LocalSimulationAlgorithms,Faulkner2015TSFandErgodicityBreaking} and the topological sector of the 2D electrolyte~\cite{Faulkner2015TSFandErgodicityBreaking}.  The same decomposition recipe defines the global twist-relaxation field $\tilde{\tvec}$ in the 2DXY model, but this mapping to the electrolyte field components is then only approximate.  One may circumvent this by using the 2DHXY potential in the 2DXY decomposition recipe.  This defines the local and global topological defects, but the global topological defects will not always correspond to the global twist-relaxation field.  % This is because the non-linear cosine couplings blur the emergent charges, enforcing $\sum_{{\rm around} {\rm plaquette}} \sin (\Delta \theta(\rvec)) = \rho(\rvec)$ (as a functional with Lagrange multipliers can then be written) rather than the Gauss law, i.e., $\sum_{{\rm around} {\rm plaquette}} \Delta \theta(\rvec) = \rho(\rvec)$.  

\begin{figure*}[htb]
\includegraphics[width=\linewidth]{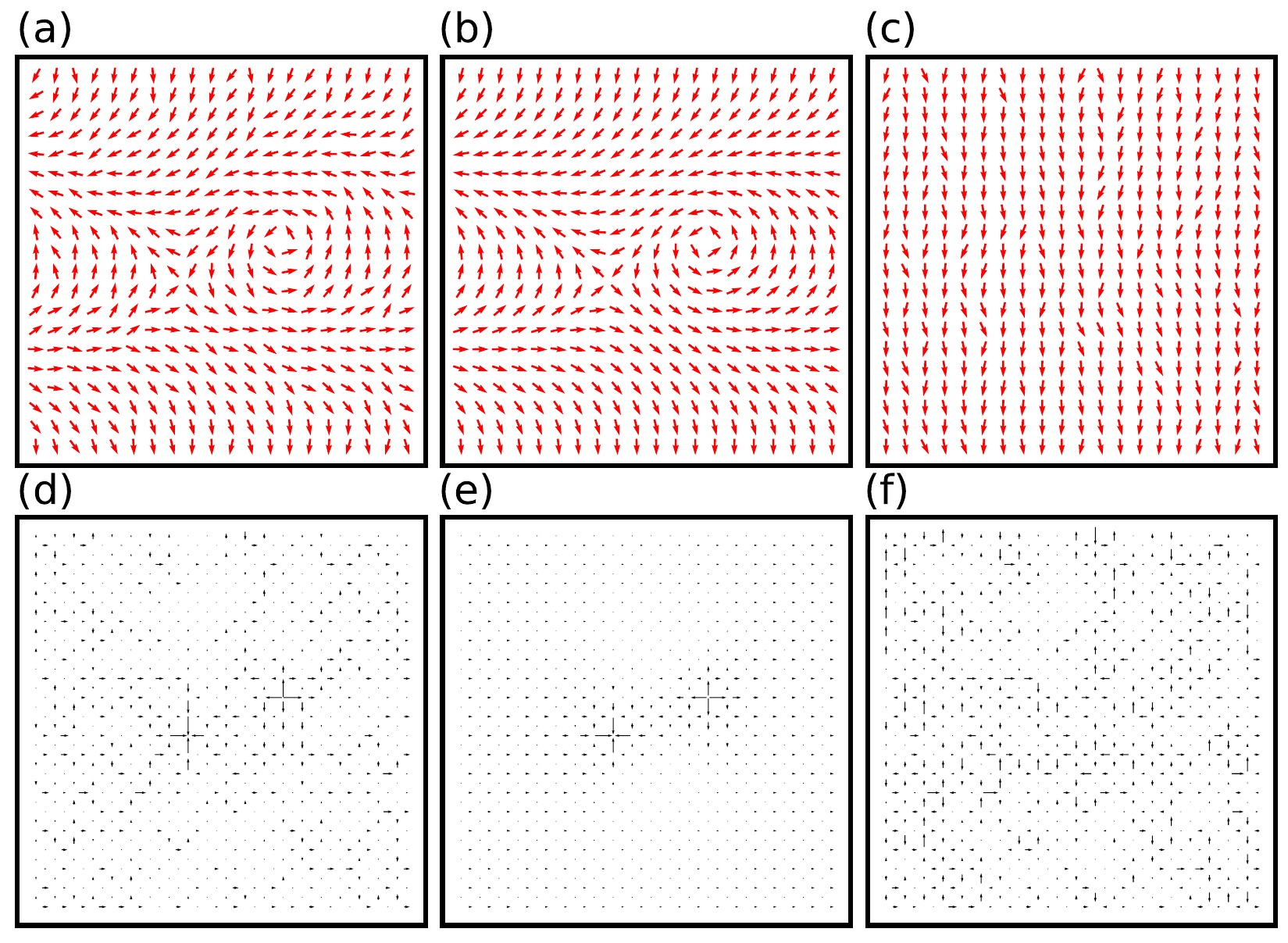}
\caption{(a)-(c) To complement figure~\ref{fig:TopDefects}, we decompose the spin-difference field of the 2DHXY configuration of figure~\ref{fig:TopDefects}(a) into its defect $\DeltaPhiHat$ and spin-wave $\DeltaPsi$ fields, as defined below equation~\eqref{eq:HelmholtzDecompPhaseDifferenceField} [up to an arbitrary choice of global $U(1)$ gauge].  Red arrows represent spins.  (a) Typical 2DHXY configuration at $\beta J = 1.5$ with fixed topological defects [i.e., for illustrative purposes, two local defects (and one global defect) are fixed in position and spin waves are allowed to fluctuate around this constrained configuration -- this is distinct from the non-constrained simulations presented elsewhere in the paper].  This is equivalent to figure~\ref{fig:TopDefects}(a).  (b) Zero-temperature annealed minimization of (a), i.e., the spin waves have been annealed away.  This is equivalent to figure~\ref{fig:TopDefects}(b) and the corresponding spin-difference field is the defect field defined below equation~\eqref{eq:HelmholtzDecompPhaseDifferenceField}.  (c)  The spin waves that were annealed away from (b), up to an arbitrary choice of global $U(1)$ gauge.  The corresponding spin-difference field is the spin-wave field $\DeltaPsi$.  (d)-(f) The emergent-field representation of (a)-(c).  Black arrows represent the emergent (electric) flux (due to the relevant spin-difference field, i.e., either the total spin-difference field or one of its decomposed components) flowing between emergent charge sites, with the length of the arrow representing the relative magnitude.  (d) The total emergent field defined in equation~\eqref{eq:ElectricFieldHXY}.  (e) The irrotational Coulomb field of equation~\eqref{eq:TopologicalDefectEquivalence}.  (f)  The auxiliary gauge field of equation~\eqref{eq:ContinuousDeformationsEquivalence} (the magnitude has been multiplied by two to improve visibility).  (a)-(b) are taken from \cite{Faulkner2024SymmetryBreakingBKT}.}
\label{fig:HelmholtzHodgePhaseConfigures}
\end{figure*}

Applying the supplemental global-twist dynamics of Section~\ref{sec:GlobalTwists} to a fully aligned 2DHXY (or 2DXY) spin configuration results in configurations such as those in figure~\ref{fig:TopDefects}(d).  Moreover, internal global spin twists can be generated by deconfined vortices~\cite{Faulkner2017AnElectricFieldRepresentation} as demonstrated by the motion of the neutral vortex pair in figure~\ref{fig:GlobalTwistGeneration}.  When unbinding is possible, the positive vortex can trace a closed path around the torus until it annihilates the negative vortex.  This path in the $\UnitX$ direction leaves an internal global twist in the spin field along the $\UnitY$ direction -- perpendicular to the net path followed by the vortex, where an internal global spin twist along the $\UnitX$ direction can be formed by an analogous vortex path through the $- \UnitY$ direction.  %These objects are global topological defects because they cannot be removed by continuous deformation of the phase-difference field.
We note that vortices passing through the boundaries of a window of size $L^2$ (within an infinite-size simply connected system) would generate the same global objects -- where a neutral pair of vortices disappearing through opposite boundaries of the window is analogous to them annihilating one another after tracing a net closed path around the torus of volume $L^2$.  We explore in detail the dynamics of the global topological defects (and a non-annealed analogue of the global twist-relaxation field) in Section~\ref{sec:TSF}.

Helmholtz-Hodge decomposition of the phase-difference field $\DeltaPhi$ splits it into divergence-free $\DeltaPhiHat$ and divergence-full $\DeltaPsi$ components:
\begin{align}
\DeltaPhi (\rvec) = \DeltaPhiHat (\rvec) + \DeltaPsi (\rvec) .
\label{eq:HelmholtzDecompPhaseDifferenceField}
\end{align}
The topological defects are described by the \emph{defect field} $\DeltaPhiHat$, while the \emph{spin-wave field} $\DeltaPsi$ describes the spin waves.  To complement figure~\ref{fig:TopDefects}, figure~\ref{fig:HelmholtzHodgePhaseConfigures} decomposes the 2DHXY configuration of figure~\ref{fig:TopDefects}(a) into its defect and spin-wave fields.  Figures~\ref{fig:HelmholtzHodgePhaseConfigures}(a) and (b) are respectively identical to figures~\ref{fig:TopDefects}(a) and (b) (representing the total phase-difference $\DeltaPhi$ and defect $\DeltaPhiHat$ fields) while figure~\ref{fig:HelmholtzHodgePhaseConfigures}(c) is the spin-wave field $\DeltaPsi$ [up to an arbitrary choice of global $U(1)$ gauge].

%\subsection{Helmholtz-Hodge decomposition and generalized lattice electric field}
%$\subsection{Helmholtz-Hodge decomposition and emergent electric field}
%\label{sec:HHDecomposition}
\subsection{Emergent electric field}
\label{sec:EmergentElectricField}

\begin{figure*}[t]
\includegraphics[width=\linewidth]{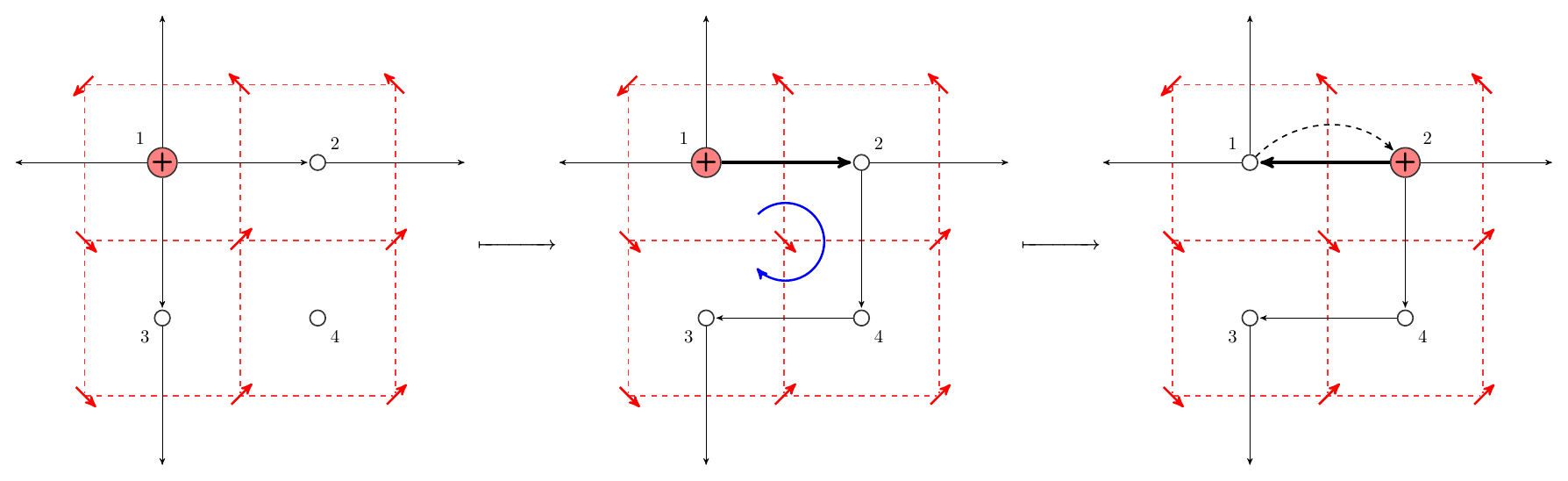}
\caption{Single 2DHXY/2DXY spin rotations can lead to emergent-charge hops.  The red arrows are spins.  The red dashed lines represent the periodic spin lattice $D'$.  Each red/white circle represents positive/zero emergent charge.  The curly dashed black arrow represents an emergent-charge hop.  The black arrows represent the emergent (electric) flux $E_{ab}$ flowing directly from emergent-charge site $a$ to emergent-charge site $b$ (the thickness represents the relative magnitude).  The blue arrow represents the direction of the spin rotation.  Left-hand to central panel: The value of the central spin phase decreases by an amount $\pi / 2 + \varepsilon$, for some small $\varepsilon > 0$.  Central to right-hand panel: The $2\pi$-modular XY symmetry [enforced by equation~\eqref{eq:ModularPhaseDiff}/the two-particle cosine potential] then enforces $E_{12} + \pi J / 2 + \varepsilon J \, \mapsto \, E_{12} + \pi J / 2 + \varepsilon J - 2\pi J$.  This figure is taken from \cite{Faulkner2017AnElectricFieldRepresentation}.  %(\textcopyright IOP Publishing. Reproduced with  permission. All rights reserved.)
}
\label{fig:ChargeHopHXY}
\end{figure*}

%The Helmholtz-Hodge decomposition in equation~\eqref{eq:HelmholtzDecompPhaseDifferenceField} better elucidates the picture of topological defects and continuous spin waves, but the divergence-full and divergence-free components are coupled to each other in even the 2DHXY model.  The internal global-twist component cannot therefore be isolated and uniquely manipulated by the global-twist dynamics, as such dynamics may also alter the vortex and spin-wave configurations.  In order to isolate and uniquely manipulate the global topological defects, we therefore release the coupling by transforming to the real lattice electrolyte in Section~\ref{sec:GeneralisedLatticeElectrolyte}.  %We begin here, however, by presenting the emergent 2DHXY electrolyte.  
%To understand this decoupling, however, we must first present the 2DHXY model as an emergent electrolyte.  To proceed, we define the lattice electric field 
The Helmholtz-Hodge decomposition in equation~\eqref{eq:HelmholtzDecompPhaseDifferenceField} mathematically expresses the picture of topological defects and continuous spin waves, but %a reformulation of the model as an emergent lattice electrolyte is required to elucidate the connection between global-defect excitations and the phase transition.  To proceed, we define the lattice electric field 
reformulating the model as an emergent lattice electrolyte will allow us to split the topological-defect field into its local and global components in Section~\ref{sec:HarmonicMode} -- to then express the phase transition in terms of the global-defect excitations.  To proceed, we define the emergent (electric) field 
\begin{align}
\Evec(\rvec) := \frac{J}{a} 
\begin{pmatrix}
    \glc \DeltaPhi \grc_y \glb \rvec + a \UnitX / 2 \grb  \\
    \\
    - \glc \DeltaPhi \grc_x \glb \rvec + a \UnitY / 2 \grb
\end{pmatrix} %= \frac{J}{a} \DeltaPhi (\rvec) \times \UnitZ 
\label{eq:ElectricFieldHXY}
\end{align}
for all $\rvec \in D$.  This is very closely related to $\DeltaPhi(\rvec') \times \UnitZ$ ($\rvec' \in D'$) and transforms the 2DHXY potential [equation~\eqref{eq:HXYHamiltonian}] to the form of that of a lattice-field electrolyte:
\begin{align}
U_{\rm HXY} = \frac{a^2}{2J} \sum_{{\bf r} \in D} \| \Evec({\bf r}) \|^2 . 
\label{eq:ElectricFieldPotentialHXY}
\end{align}
Figures~\ref{fig:TopDefects}(e) and \ref{fig:HelmholtzHodgePhaseConfigures}(d) depict the spin configuration of figures~\ref{fig:TopDefects}(a) and \ref{fig:HelmholtzHodgePhaseConfigures}(a) in this emergent-field representation.  The vortices now become emergent Coulomb charges, in part due to the combination of equations~\eqref{eq:DefinitionVortex} and \eqref{eq:ElectricFieldHXY} resulting in the lattice Gauss law of two-dimensional electrostatics: 
\begin{align}
\NablaHat \cdot \Evec (\rvec) = J \rho (\rvec) .
\label{eq:GaussLawHXY}
\end{align}
In Sections~\ref{sec:ConstrainedElectrolytePDF} and \ref{sec:GeneralisedLatticeElectrolyte}, we elucidate fully this emergent-Coulomb behaviour by expressing the 2DHXY Boltzmann distribution as that of a constrained lattice-field electrolyte.  We note that charge neutrality is enforced by the periodic boundaries: $\sum_\rvec \rho(\rvec) = \NablaHat \cdot \sum_\rvec \Evec (\rvec) / J = 0$.  

The most general solution to the lattice Gauss law of equation~\eqref{eq:GaussLawHXY} is the Helmholtz-Hodge decomposition of the emergent field $\Evec$: %into its Poisson $-\NablaTildePhi$, harmonic $\Ebar$, and auxiliary-gauge-field $\boldsymbol{\nabla} \times \Qvec$ components:
\begin{align}
\Evec (\rvec) = - \NablaTildePhi (\rvec) + \Ebar + \hat{\boldsymbol{\nabla}} \times \Qvec (\rvec) .
\label{eq:ElectricFieldHelmholtzHodge}
\end{align}
Here the two-dimensional \emph{auxiliary gauge field} $\hat{\boldsymbol{\nabla}} \times \Qvec$ is purely rotational, the $x$ and $y$ components of the three-dimensional vector potential $\Qvec(\rvec) \equiv [0, 0, Q(\rvec)]^T$ are zero with $Q : D \to \RR$ the \emph{auxiliary gauge potential}, the \emph{harmonic mode} 
\begin{align}
\Ebar := \frac{1}{N} \sum_{\rvec \in D} \Evec (\rvec) 
\label{eq:EbarDefinition}
\end{align}
is the zero Fourier mode and the \emph{Poisson field} $-\NablaTildePhi$ is divergence-full, where the \emph{scalar potential} $\phi$ is a solution of the Poisson equation of lattice electrostatics (on the torus) 
\begin{align}
\NablaBF^2 \phi (\rvec) = - J \rho (\rvec) .
\label{eq:PoissonLawHXY}
\end{align}
We now combine Eqs \eqref{eq:HelmholtzDecompPhaseDifferenceField}, \eqref{eq:ElectricFieldHXY} and \eqref{eq:ElectricFieldHelmholtzHodge} to map the divergence-free $\DeltaPhiHat$ and divergence-full $\DeltaPsi$ fields to the irrotational $-\NablaTilde \phi + \Ebar$ and purely rotational $\hat{\boldsymbol{\nabla}} \times \Qvec$ components of the emergent field:
\begin{align}
-\NablaTilde \phi (\rvec) + \Ebar = \frac{J}{a} 
\begin{pmatrix}
\glc \DeltaPhiHat \grc_y \glb \rvec + a \UnitX / 2 \grb \\
\\
- \glc \DeltaPhiHat \grc_x \glb \rvec + a\UnitY / 2 \grb
\end{pmatrix} ;
\label{eq:TopologicalDefectEquivalence}\\
\hat{\boldsymbol{\nabla}} \times \Qvec (\rvec) = \frac{J}{a} 
\begin{pmatrix}
\glc \DeltaPsi \grc_y \glb \rvec + a \UnitX / 2 \grb \\
\\
- \glc \DeltaPsi \grc_x \glb \rvec + a \UnitY / 2 \grb
\end{pmatrix} .
\label{eq:ContinuousDeformationsEquivalence}
\end{align}
The irrotational \emph{Coulomb field} in equation~\eqref{eq:TopologicalDefectEquivalence} describes both the local and global topological defects, while the rotational auxiliary gauge field in equation~\eqref{eq:ContinuousDeformationsEquivalence} describes the continuous spin waves.  This mapping is reflected in figures~\ref{fig:HelmholtzHodgePhaseConfigures}(b), (c), (e) and (f). 
%figures~\ref{fig:HelmholtzHodgePhaseConfigures}(e) and (f) depict the spin configurations of figures~\ref{fig:HelmholtzHodgePhaseConfigures}(b) and (c) in their emergent-field representations.  %This reflects the correspondence between local topological defects and charges, and between linear phase fluctuations and the auxiliary field. 

In addition, the 2DHXY potential of equations~\eqref{eq:HXYHamiltonian} and \eqref{eq:ElectricFieldPotentialHXY} can now be expressed in terms of $\rho(\rvec)$: 
\begin{align}
U_{\rm HXY} = & - \mu \tilde{n} + \frac{a^4 J}{2} \sum_{\rvec \ne \rvec'} \rho(\rvec) G(\rvec, \rvec') \rho(\rvec') \nonumber\\
& + \frac{L^2}{2J} \| \Ebar \|^2 + \frac{a^2}{2J} \sum_{\rvec \in D} \| \hat{\boldsymbol{\nabla}} \times \Qvec(\rvec) \|^2 , \label{eq:DecomposedHXYPotential}
\end{align}
where 
\begin{align}
G(\rvec, \rvec') = \frac{1}{2N} \sum_{\kvec \ne 0} \frac{e^{i\kvec \cdot (\rvec - \rvec')}}{2 - \cos (k_x a) - \cos (k_y a)} 
\label{eq:LatticeGreensFunctionSolution}
\end{align}
is the lattice Green's function between any two sites $\rvec, \rvec'$ on the charge lattice, $\mu := - 2 \pi^2 J G(0)$ is the chemical potential for the introduction of an emergent charge, $\tilde{n}$ is the number of emergent charges and $G(0) := G(\rvec, \rvec)$ is the same-site lattice Green's function.  In Appendix~\ref{sec:GreensFunction}, we show that equation~\eqref{eq:LatticeGreensFunctionSolution} is a solution of 
\begin{align}
a^2 \boldsymbol{\nabla}_\rvec^2 G(\rvec, \rvec') = - \mathbb{I} \left[ \rvec = \rvec' \right] \,\, \forall \, \rvec, \rvec' \in D ,
\label{eq:LatticeGreensFunctionDefinition}
\end{align}
which itself comes from the Poisson equation of equation~\eqref{eq:PoissonLawHXY}.  
%Note that we set the solution of the $\kvec = 0$ mode (in equation~\eqref{eq:KspaceLatticeGreensFunctionSoln}) to zero as it does not appear in the potential in equation~\eqref{eq:DecomposedHXYPotential} for a charge-nfeutral system~\footnote{It was erroneously stated in ref.~\cite{Faulkner2015TSFandErgodicityBreaking} that $\tilde{G}_{\rvec'}(\kvec = 0) = 0$ was chosen due to a (incorrect) relationship between $\tilde{G}_{\rvec'}(\kvec = 0)$ and the zero mode of $\Evec$.}.
Note that we chose to set the $\kvec = 0$ mode of the lattice Green's function to zero but could choose any real number as this $\kvec = 0$ mode does not appear in the potential for a charge-neutral system~\footnote{It was erroneously stated in \cite{Faulkner2015TSFandErgodicityBreaking} that the zero-valued $\kvec = 0$ mode was chosen due to an incorrectly stated relationship with $\Ebar$ (the zero mode of $\Evec$).  Note also that the indicator function in equation~\eqref{eq:LatticeGreensFunctionDefinition} is equivalent to the Kronecker delta function.}.

Moreover, the emergent-field representation elucidates how the coupling between the topological defects and spin waves leads to the emergent charge-hop dynamics depicted in figure~\ref{fig:ChargeHopHXY}.  A clockwise spin rotation by an amount $\Delta$ rotates (also in the clockwise direction) the immediately surrounding emergent (electric) flux by an amount $\Delta J$.  The emergent-charge configuration then changes if an absolute phase difference $\delta \varphi(\rvec, x / y)$ either leaves or enters the interval $\glc -\pi, \pi \grb$, as in figure~\ref{fig:ChargeHopHXY} where the continuous spin rotation is followed by a discrete change to the emergent field and an accompanying emergent-charge hop to an adjacent charge site.  This second stage is due to the local $2\pi$-modular XY symmetry enforced by the modular arithmetic in equation~\eqref{eq:ModularPhaseDiff}.  The subtleties of the 2DXY model are discussed at the end of Section~\ref{sec:HarmonicMode} and we will see below in Section~\ref{sec:GeneralisedLatticeElectrolyte} the close connection between these dynamics and those of the generalized lattice-field electrolyte.

\subsection{Harmonic mode}
\label{sec:HarmonicMode}

The irrotational Coulomb field describes the local and global topological defects, but we only require its harmonic component to describe the global defects.  Due to the toroidal topology of the charge lattice $D$, the uniform harmonic component is the mean electric field defined in equation~\eqref{eq:EbarDefinition}.  In Appendix~\ref{sec:Polarisation}, we demonstrate that this zero Fourier mode is given by 
\begin{align}
\Ebar = - J \Pvec + \frac{2\pi J}{L} \wvec_0 ,
\label{eq:EbarPolarisationTopSector}
\end{align}
where $\Pvec := \sum_{\rvec \in D} \rvec \rho (\rvec) / N$ is the \emph{origin-dependent polarization} and $w_{0, x} : = a \sum_{y = a / 2}^{L - a / 2} E_x (L, y) / (2 \pi J) \in \ZZ$ is the $x$ component of the \emph{origin-dependent winding field}, which changes when an emergent charge traces a closed path around the torus (the $y$ component is defined analogously)~\cite{Vallat1994CoulombGas,Faulkner2015TSFandErgodicityBreaking}.  Equation~\eqref{eq:EbarPolarisationTopSector} is a general expression for any electrolyte on the torus with elementary charge $q = 2\pi$ [in the case of the 2DHXY model, the (emergent) charge density is restricted to $\rho(\rvec) \in \{ 0, \pm 2 \pi / a^2 \}$ but equation~\eqref{eq:EbarPolarisationTopSector} also holds for electrolytes whose charge density $\rho (\rvec) \in 2 \pi \ZZ / a^2$].  The left-hand side is origin-independent, while each component on the right-hand side is dependent on the choice of origin -- due to the concepts `close together' and `far apart' being ambiguous on the torus.  We must therefore adopt a toroidal model for the polarization.  For an electrolyte of elementary charges [$\rho(\rvec) \in \{ 0, \pm 2 \pi / a^2 \}$], we may express the harmonic mode $\Ebar$ as
\begin{align}
\Ebar = \Ebar_{\rm p} + \frac{2 \pi J}{L} \wvec ,
\label{eq:EbarToroidalModel}
\end{align}
%where $\Ebar_{\rm p}$ is the origin-independent polarization component of the harmonic mode and $\wvec \in \ZZ^2$ is the topological sector of the system.  The topological sector is fixed by the condition
%\begin{align}
%\bar{E}_{{\rm p}, x/y} \in \glb - \frac{\pi J}{L}, \frac{\pi J}{L} \grc
%\end{align}
%and changes when a charge traces a closed path around the torus.  For any given charge configuration $\gld \rho(\rvec) \grd$, this condition corresponds to 
%\begin{align}
%\bar{E}_{\rm p, \mu} = \left( - \frac{J}{N} \sum_{\mathbf{r}\in D} \mu \rho(\mathbf{r}) + \frac{\pi J}{L} \right) \!\!\!\!\! \mod \left( \frac{2\pi J}{L} \right) - \frac{\pi J}{L} \,\,\, \textrm{for all} \,\,\, \mu \in \{ x, y \} .
%\label{eq:OriginIndependentPolarisation}
%\end{align}
where 
\begin{align}
\bar{E}_{\rm p, \mu} := \left( - \frac{J}{N} \sum_{\mathbf{r}\in D} \rvec_\mu \rho(\mathbf{r}) + \frac{\pi J}{L} \right) \!\!\!\!\! \mod \left( \frac{2\pi J}{L} \right) - \frac{\pi J}{L} 
\label{eq:OriginIndependentPolarisation}
\end{align}
is the $\mu \in \{ x, y \}$ component of the (origin-independent) \emph{polarization field} and $\wvec \in \ZZ^2$ is the (origin-independent) \emph{topological sector}.  The polarization field $\Ebar_{\rm p}$ is the low-energy harmonic mode of some charge distribution $\{ \rho(\mathbf{r}) : \rvec \in D \}$, while nonzero values of the topological sector describe higher energy solutions.  Indeed, for some harmonic mode $\Ebar$, equations~\eqref{eq:EbarToroidalModel} and \eqref{eq:OriginIndependentPolarisation} are equivalent to 
\begin{align}
\bar{E}_{{\rm p}, x / y} = \left( \bar{E}_{x / y} + \frac{\pi J}{L} \right) \!\!\!\!\! \mod \left( \frac{2\pi J}{L} \right) - \frac{\pi J}{L} 
\label{eq:OriginIndependentPolarisationFromEbar}
\end{align}
and 
\begin{align}
w_{x / y} = \left\lfloor \frac{L\bar{E}_{x / y} + \pi J}{2\pi J} \right\rfloor ,
\label{eq:OriginIndependentTopSectorFromEbar}
\end{align}
where the floor function $\left\lfloor \cdot \right\rfloor$ returns the greatest integer less than or equal to the input real number.  Equations~\eqref{eq:OriginIndependentPolarisationFromEbar} and \eqref{eq:OriginIndependentTopSectorFromEbar} follow from the potential difference 
\begin{align}
    \Delta U = 2\pi L \omega \left( \bar{E}_{x/y} + \frac{\pi J}{L} \omega \right) 
\label{eq:LowEnergyHarmonicMode}
\end{align}
that results from adding $\omega \in \ZZ$ Cartesian topological sectors to the $x / y$ component of the harmonic mode $\Ebar$ [i.e., applying $\bar{E}_{x/y} \mapsto \bar{E}_{x/y} + 2\pi J \omega / L$ to equation~\eqref{eq:DecomposedHXYPotential}]~\footnote{In practice, this transformation must be applied to the 2DHXY model in the absence of spin-wave excitations.  This is not required for the generalized lattice-field electrolyte of Section~\ref{sec:GeneralisedLatticeElectrolyte}.}.  For $\bar{E}_{x/y} = \bar{E}_{{\rm p}, x/y}$, the constraint 
\begin{align}
\bar{E}_{{\rm p}, x/y} \in \glc - \frac{\pi J}{L}, \frac{\pi J}{L} \grb
\label{eq:OriginIndependentPolarisationConstraint}
\end{align}
induced by equation~\eqref{eq:OriginIndependentPolarisation} then means that $\omega = 0$ is the low-energy solution of equation~\eqref{eq:LowEnergyHarmonicMode}, as required.  We note that equation~\eqref{eq:LowEnergyHarmonicMode} is degenerate for $\bar{E}_{x/y} = (2k + 1) \pi J / L$ ($k \in \ZZ$) since the low-energy solutions $\omega = -k$ and $\omega = -(k + 1)$ both result in the potential difference $\Delta U = - 2\pi^2 J k (k + 1)$.  Examples include certain cases of a single charge-neutral pair of local topological defects with minimal separation distance $L/2$.  The degeneracy is resolved by the interval in Eq.\eqref{eq:OriginIndependentPolarisationConstraint} being half-open and is analogous to that discussed in Section~\ref{sec:HXYModel} in the context of the global twist-relaxation field $\tilde{\tvec}$ [the interval switches from right-open to left-open upon making the transformation $\pm (2)\pi J / L \mapsto \mp (2)\pi J / L$ in equation~\eqref{eq:OriginIndependentPolarisation}, with the left-open convention used in \cite{Faulkner2015TSFandErgodicityBreaking,Faulkner2017AnElectricFieldRepresentation}].  The above origin-independent framework will become useful when defining topological order in terms of topological-sector fluctuations in Section~\ref{sec:TopologicalNonergodicity}.

The 2DHXY model admits only elementary emergent charges, meaning that we may adopt the toroidal polarization model of equation~\eqref{eq:EbarToroidalModel}. For any given emergent-charge configuration, the polarization field $\Ebar_{\rm p}$ is then the low-energy solution of the harmonic mode of the emergent field.  Its sum with the Poisson field then gives the \emph{low-energy Coulomb field} $-\NablaTildePhi (\rvec) + \Ebar_{\rm p}$.  This describes the emergent field at $\rvec \in D$ due to the local topological defects, while the global topological defects correspond to nonzero values of each Cartesian component of the topological sector $\wvec$.  Multiple topological sectors therefore describe any given emergent-charge configuration, mapping to the internal global spin twists via
\begin{align}
\wvec = (t_y, - t_x)^T ,
\label{eq:TSGlobalTwistRelationship}
\end{align}
where $\tvec \in \mathbb{Z}^2$ is the internal global-twist field described in Section~\ref{sec:HXYModel}.  %In \sect{sec:Nonergodicity}, we present an analysis of topological-sector fluctuations in the real (rather than the emergent) lattice electrolyte. An absence of topological-sector fluctuations (described by the additional degrees of freedom $\Ebar$) defines topological order and corresponds to the ergodic freezing of global condensate-phase twists. This identification of nonergodicity in the condensate-phase field displays the power of Helmholtz-Hodge decomposition of the electric field of the emergent electrolyte.

The zero-temperature annealing described in Section~\ref{sec:HXYModel} fixes the topological defects in position before continuously removing the spin waves/auxiliary gauge field.  This leaves behind the defect field $\DeltaPhiHat$ which splits into its local and global components (respectively mapping to $- \NablaTildePhi + \Ebar_{\rm p}$ and $2 \pi J \wvec / L$) with the spin/emergent-field representations depicted in figures~\ref{fig:TopDefects}(c)/(g) and (d)/(h).  More precisely, the emergent-field representation maps~\cite{Faulkner2017AnElectricFieldRepresentation} the vortex, spin-wave and internal global-twist components to (respectively) the low-energy solution to the Gauss law for the emergent charges $- \NablaTildePhi + \Ebar_{\rm p}$, the purely rotational auxiliary gauge field $\hat{\boldsymbol{\nabla}} \times \Qvec$ and the topological-sector component of the emergent field $2 \pi J \wvec / L$.  

The subtleties of the 2DXY model are now elucidated by considering the recipe that decomposed the 2DHXY model into its three principal excitations in Section~\ref{sec:HXYModel}.  Applying this same decomposition recipe to the 2DXY model also defines its global twist-relaxation field $\tilde{\tvec}$, but the mapping to the electrolyte-field components is then only approximate.  One may circumvent this by using the 2DHXY potential in the 2DXY decomposition recipe.  This defines the 2DXY local and global topological defects -- along with their corresponding topological sector $\wvec$ -- but the global topological defects will not always correspond to the global twist-relaxation field.  % -- and global-defect fluctuations will not then turn on at the transition, as seen below in Sections~\ref{sec:TopologicalNonergodicity} and \ref{sec:TSF}.  
This is essentially because the non-linear cosine couplings soften the emergent charges, leading to $\sum_{\rvec' \in \partial \Gamma_\rvec} \sum_{\mu \in \{ x, y \}} \sin \left( \glc \DeltaPhi \grc_\mu (\rvec' + a\mathbf{e}_\mu / 2) \right) \mathbf{e}_\mu \cdot \lvec_\rvec(\rvec')$ in place of $\sum_{\rvec' \in \partial \Gamma_\rvec} \DeltaPhi (\rvec') \cdot \lvec_\rvec(\rvec')$ in equation~\eqref{eq:DefinitionVortex} -- as a result of minimizing the 2DXY potential with Lagrange multipliers.  Nevertheless, the emergent-charge dynamics described at the end of Section~\ref{sec:EmergentElectricField} also applies to the 2DXY model with respect to the local topological defects defined here.  
%We note that Ref.~\cite{Faulkner2017AnElectricFieldRepresentation} was correct but not as clear on these subtleties.  
This subsection additionally clarifies that the supplemental global-twist dynamics of Section~\ref{sec:GlobalTwists} pass through high-energy global topological defects when tunnelling through the $U(1)$ sombrero potential [as in figure~\ref{fig:GlobalTwists}(a)] to guarantee $U(1)$ symmetry on non-divergent timescales.

\begin{figure*}[t]
\includegraphics[width=\linewidth]{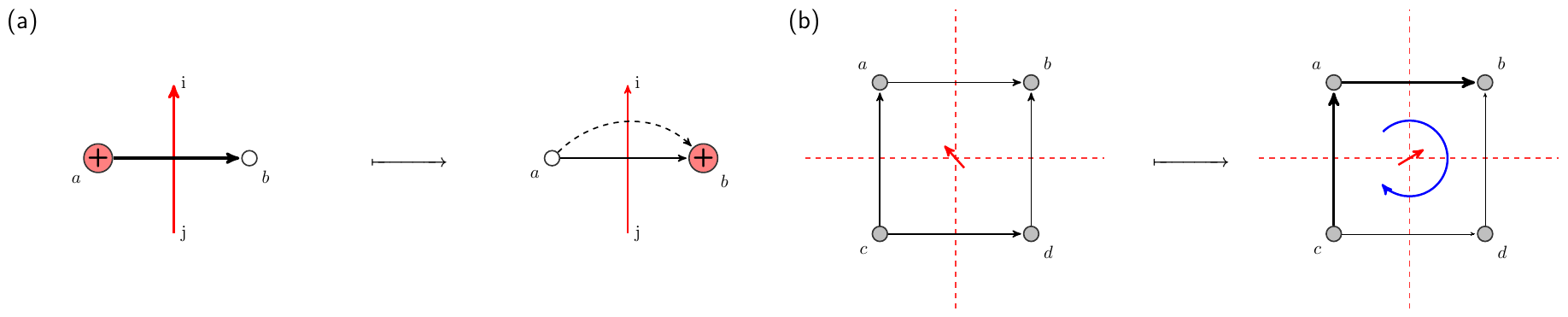}
\caption{Electric-field dynamics in the generalized lattice-field electrolyte.  Each red/white/gray circle represents positive/zero/arbitrary charge.  The solid black arrows represent the electric flux $E_{ab}$ flowing directly from charge site $a$ to charge site $b$ (the thickness represents the relative magnitude).  The curly dashed black arrow represents a charge hop.  The blue arrow represents the direction of the electric-field rotation.  We additionally denote the spin-like fields defined in Appendix~\ref{sec:MaggsToVillain}.  The long red arrows in (a) represent the $\hat{s}(\rvec_i, \rvec_j)$ field from spin site $j$ to spin site $i$ (the thickness represents the relative magnitude).  The short red arrows represent the $\varphi$ field at the spin site contained within the charge plaquette.  The red dashed lines represent the edges of the periodic spin lattice $D'$ (which is conjugate to the periodic charge lattice $D$).  (a) The movement of a positive charge $q$ from charge site $a$ to charge site $b$ is induced by a $q / \epsilon_0$ decrease in the electric flux $E_{ab}$ flowing from site $a$ to site $b$ [equivalently, $\hat{s}(\rvec_i, \rvec_j)$ decreases by $1$].  (b) A clockwise electric-flux rotation of $\Delta / \epsilon_0$ increases $E_{ab}$ and $E_{ca}$ by $\Delta / \epsilon_0$ and decreases $E_{cd}$ and $E_{db}$ by $\Delta / \epsilon_0$, altering only the auxiliary gauge field.  Equivalently, the $\varphi$ component at $\rvec_0 \in D'$ (the centre of the charge plaquette) decreases by $\Delta$.  As described in Appendix~\ref{sec:MaggsToVillain}, to ensure that $\varphi(\rvec_0)$ remains within $[-\pi, \pi)$, the operation $\varphi(\rvec_0) \mapsto \left[ \varphi(\rvec_0) + \pi \right] \!\!\! \mod (2\pi) - \pi$ is then applied; if this operation has an effect [i.e., if $\varphi(\rvec_0) \notin [-\pi, \pi)$ between the $\varphi$ rotation and the operation] each of the four $\hat{s}$ field components associated with $\rvec_0$ are updated by $\pm 1$ such that the new charge configuration equals the charge configuration before the $\varphi$ rotation.  Figures are adapted from \cite{Faulkner2015TSFandErgodicityBreaking}.}
\label{fig:MaggsRossettoDynamics}
\end{figure*}

\subsection{Constrained lattice-field electrolyte}
\label{sec:ConstrainedElectrolytePDF}

We are now able to express the 2DHXY model as a constrained lattice-field electrolyte.  Upon adopting the toroidal polarization model of equation~\eqref{eq:EbarToroidalModel}, the 2DHXY Boltzmann distribution may be written in terms of the emergent charges $\rho(\rvec)$, auxiliary gauge potential $Q(\rvec)$ and topological sector $\wvec$: 
\begin{widetext}
\begin{align}
\pi_{\rm HXY} \left( \{ \rho(\rvec), Q(\rvec) \}, \wvec \right) \propto & \int \mathcal{D} \mathbf{E} \prod_{\mathbf{r}\in D} \left[ \delta \left( \boldsymbol{\hat{\nabla}} \cdot \mathbf{E}(\mathbf{r}) \! - \! J \rho(\mathbf{r}) \right) \delta \left(\widetilde{\boldsymbol{\nabla}} \! \times \! \Evec(\rvec) \! + \! \boldsymbol{\nabla}^2 Q(\rvec)\UnitZ \right) \!\!\!\! \prod_{\mu \in \{ x, y \}} \!\!\!\! \Theta \left(\frac{\pi J}{a} \! - \! \left| E_\mu \left(\mathbf{r} \! + \! \frac{a}{2} \mathbf{e}_\mu \right) \right| \right) \right] \nonumber\\
& \times \prod_{\mu \in \{ x, y \}} \mathbb{I} \left[ \frac{\bar{E}_{{\rm p}, \mu}}{J} = \left( - \sum_{\mathbf{r} \in D} \frac{\rvec_\mu \rho(\mathbf{r})}{N} + \frac{\pi}{L} \right) \!\!\!\!\! \mod \left( \frac{2\pi}{L} \right) - \frac{\pi}{L} \right] \nonumber\\
& \times \delta \left( \frac{1}{N} \sum_{\mathbf{r}\in D} \mathbf{E}(\mathbf{r}) - \left( \Ebar_{\rm p} + \frac{2\pi J}{L} \wvec \right) \right) \exp \glb - \frac{\beta a^2}{2 J} \sum_{\mathbf{r}\in D} \|\mathbf{E}(\mathbf{r})\|^2 \grb ,
\label{eq:ConstrainedElectrolytePDF1}
\end{align}
\end{widetext}
where the Heaviside step function $\Theta : \RR \to \{ 0, 1\}$ couples the topological defects to the auxiliary gauge field.  
%Lifting this constraint leaves us with a generalized lattice-field electrolyte of elementary electrical charges (with charge value $2\pi$), as presented in ref.~\cite{Faulkner2015TSFandErgodicityBreaking} and based on the model of Maggs \& Rossetto~\cite{Maggs2002LocalSimulationAlgorithms}.  Below we lift the constraint in order to study the ergodic properties of the global topological defects in the real lattice-field electrolyte -- as this allows us to isolate and uniquely manipulate the global topological defects.  
We will see below in Section~\ref{sec:GeneralisedLatticeElectrolyte} that this Heaviside function is the only distinction between this Boltzmann distribution and that of the (real) generalized lattice-field electrolyte.  In equation~\eqref{eq:ConstrainedElectrolytePDF1}, the first Dirac object $\delta(\cdot)$ imposes the Gauss law, the second identifies the rotational component of the emergent field with the auxiliary gauge field $\hat{\boldsymbol{\nabla}} \times \Qvec$, and the third identifies the harmonic mode $\Ebar$ with the sum of its polarization $\Ebar_{\rm p}$ and topological-sector $2\pi J \wvec / L$ components [see equation~\eqref{eq:EbarToroidalModel}], with the polarization field $\Ebar_{\rm p}$ defined by the indicator function.  In Appendix~\ref{sec:DiscreteVectorCalculus}, we derive the relation $\widetilde{\boldsymbol{\nabla}} \times \Evec(\rvec) = - \boldsymbol{\nabla}^2 Q(\rvec)\UnitZ$ (enforced by the second Dirac object) for $\Qvec(\rvec) = [0, 0, Q(\rvec)]^T$ on a 2D lattice, and we also show that starting from $\widetilde{\boldsymbol{\nabla}} \times \Evec(\rvec) = - \boldsymbol{\nabla}^2 Q(\rvec)\UnitZ$ (on a 2D lattice) generates the necessary properties in the rotational component of the emergent field.  Note that, in a slight abuse of notation, $\delta(x) dx$ is the Dirac measure $\delta(dx)$ for $x \in \RR$ and $\delta(\Fvec) \equiv \delta(F_x) \delta(F_y)$ for $\Fvec \in \RR^2$ (as outlined in Section~\ref{sec:Definitions}).  Another alternative representation of the Dirac measures is to formulate the problem in terms of Riemann-Stieltjes integrals over $d\Theta(F_x) d\Theta(F_y)$.

We can in fact go further and express the 2DHXY Boltzmann distribution in terms of the Helmholtz-Hodge-decomposed potential of equation~\eqref{eq:DecomposedHXYPotential}.  For any 2D vector field $\Evec(\rvec)$ with rotational component $\hat{\boldsymbol{\nabla}} \times \Qvec'(\rvec)$ described by some 3D vector field $\Qvec'(\rvec)$ on a 2D lattice, we show in Appendix~\ref{sec:DiscreteVectorCalculus} that 
$\widetilde{\boldsymbol{\nabla}} \times \Evec(\rvec) = - \boldsymbol{\nabla}^2 Q(\rvec)\UnitZ$ enforces $\left[ \hat{\boldsymbol{\nabla}} \times \Qvec'(\rvec) \right]_{x / y} = \pm \hat{\boldsymbol{\nabla}}_{y / x} Q(\rvec)$ and $\exp \left( - \| \hat{\boldsymbol{\nabla}} \times \Qvec'(\rvec) \|^2 \right) \propto \exp \left( - \| \hat{\boldsymbol{\nabla}} Q(\rvec) \|^2 \right)$.  Combining this with equations~\eqref{eq:DecomposedHXYPotential} and \eqref{eq:ConstrainedElectrolytePDF1} leads to 
\begin{widetext}
\begin{align}
\pi_{\rm HXY} \glb \gld \rho(\rvec), Q(\rvec) \grd , \wvec \grb & \propto \mathbb{I} \glc \sum_{\rvec \in D} \rho(\rvec) = 0 \grc \prod_{\mu \in \{ x, y \}} \mathbb{I} \left[ \frac{\bar{E}_{{\rm p}, \mu}}{J} = \left( - \sum_{\mathbf{r} \in D} \frac{\rvec_\mu \rho(\mathbf{r})}{N} + \frac{\pi}{L} \right) \!\!\!\!\! \mod \left( \frac{2\pi}{L} \right) - \frac{\pi}{L} \right] \label{eq:ConstrainedElectrolytePDF2} \\
\times \prod_{\mathbf{r}\in D} & \prod_{\{ x, y \}} \Theta \left( \frac{\pi J}{a} - \left| - Ja^2 \NablaTilde_{x / y} \sum_{\rvec' \in D} G(\rvec, \rvec') \rho(\rvec') + \bar{E}_{{\rm p}, x / y} + \frac{2\pi J}{L} w_{x / y} \pm \hat{\boldsymbol{\nabla}}_{y / x} Q(\rvec) \right| \right) e^{\beta \mu a^2 \sum_{\rvec \in D} \left| \rho(\rvec) \right|} \nonumber\\
\times \exp & \glc -\frac{\beta a^4 J}{2} \sum_{\rvec_i \ne \rvec_j} \rho(\rvec_i)G(\rvec_i, \rvec_j)\rho(\rvec_j) \grc \exp \glb - \frac{\beta L^2}{2 J} \| \Ebar_{\rm p} + \frac{2\pi J}{L} \wvec \|^2 \grb   \exp \glc - \frac{\beta a^2}{2 J} \sum_{\rvec \in D} \| \hat{\boldsymbol{\nabla}} Q(\rvec) \|^2 \grc . \nonumber 
\end{align}
\end{widetext}
%where $\mu := - J G(0)$ is the emergent chemical potential for the introduction of charge-neutral pair of emergent charges and $n$ is the number of such pairs.  
Again, the Heaviside function in equation~\eqref{eq:ConstrainedElectrolytePDF2} couples the topological defects to the auxiliary gauge field.  In this case, the first indicator function is required to impose charge neutrality and we have used the fact that the number of charges $\tilde{n} = a^2 \sum_{\rvec \in D} \left| \rho(\rvec) \right|$.   %Below we lift this constraint in order to study the ergodic properties of the global topological defects in the real lattice-field electrolyte -- as this allows us to isolate and uniquely manipulate the global topological defects.  

\subsection{Generalized lattice-field electrolyte}
\label{sec:GeneralisedLatticeElectrolyte}

%To fully appreciate the emergent electrostatics, we now release the coupling between the irrotational and purely rotational components of the emergent electric field.  This transforms the system to a two-dimensional, grand-canonical analogue~\cite{Faulkner2015TSFandErgodicityBreaking,Faulkner2017AnElectricFieldRepresentation} of the generalized lattice-field electrolyte first introduced by Maggs \& Rossetto~\cite{Maggs2002LocalSimulationAlgorithms,Rossetto2002Thesis} for a lattice-field algorithm of three-dimensional electrostatics.  The algorithm generalizes to any spatial dimension greater than one, and operates very similarly to the Metropolis algorithm of the 2DHXY model.  The key difference is that the charge hops are proposed independent of rotations of the auxiliary gauge field, as depicted in figure~\ref{fig:MaggsRossettoDynamics}, where one of the two is proposed at each iteration of the algorithm.  
To elucidate fully the emergent electrostatics, we now compare the 2DHXY model with the generalized lattice-field electrolyte defined by the potential 
\begin{align}
U_{\rm Maggs} = \frac{\epsilon_0 a^2}{2} \sum_{{\bf r} \in D} \| \Evec({\bf r}) \|^2 
\end{align}
for lattice electric fields $\Evec : D \to \RR^2$ with units of $1 / [\beta L]$ and subject to the lattice Gauss law $\NablaHat \cdot \Evec (\rvec) = \rho (\rvec) / \epsilon_0$.  Here, $\rho : D \to q \ZZ / a^2$ is the \emph{charge density}, $\epsilon_0 > 0$ is the \emph{vacuum permittivity} (with units of $[\beta]$) and $q > 0$ is the dimensionless \emph{elementary charge}.  This is a two-dimensional grand-canonical analogue of the model first introduced by Maggs \&  Rossetto~\cite{Maggs2002LocalSimulationAlgorithms,Rossetto2002Thesis} for a lattice-field algorithm of three-dimensional electrostatics.  The algorithm generalizes to any spatial dimension greater than one, and 
exploits the Helmholtz-Hodge decomposition of equation~\eqref{eq:ElectricFieldHelmholtzHodge} to operate very similarly to the Metropolis algorithm of the 2DHXY model.  The key difference is that the charge hops are proposed independently of rotations of the auxiliary gauge field, resulting in topological defects that are independent of the auxiliary gauge field.  The two local Metropolis moves are depicted in figure~\ref{fig:MaggsRossettoDynamics}, where one of the two is proposed at each iteration of the algorithm.  An elementary charge hop from some charge site $a$ to a neighboring site $b$ [as presented in figure~\ref{fig:MaggsRossettoDynamics}(a)] is proposed with probability $p \in (0, 1)$ by proposing an increase in the electric-field flux $E_{ab}$ flowing from site $a$ to site $b$ by $\pm q / \epsilon_0$.  Accepted charge hops alter, however, all Helmholtz-Hodge-decomposed components of the electric field.  To relax the auxiliary gauge field, electric-field rotations are therefore proposed with probability $1 - p$.  This is achieved by proposing electric-field rotations around single plaquettes, as depicted in figure~\ref{fig:MaggsRossettoDynamics}(b).  %These moves are identical to 2DHXY spin rotations that do not alter the emergent-charge configuration.  
One Monte Carlo time step is then defined as the elapsed simulation time between $N$ algorithm iterations on an $N$-site lattice.  

The generalized lattice-field electrolyte therefore augments the Poisson field $-\NablaTildePhi$ of the electrostatic problem to include both the harmonic mode $\Ebar$ and a purely rotational auxiliary gauge field $\Etilde$, which locally propagates the long-range Coulomb interactions throughout the electrolyte (as in the 2DHXY case above, we choose the $x$ and $y$ components of the three-dimensional vector potential $\Qvec(\rvec) \equiv [0, 0, Q(\rvec)]^T$ to be zero with $Q : D \to \RR$ the auxiliary gauge potential).  This framework is exactly as described in Sections~\ref{sec:EmergentElectricField}--\ref{sec:ConstrainedElectrolytePDF}, but with no constraint $| E_{x / y} (\mathbf{r}) | \le q / (2a\epsilon_0) \, \forall \mathbf{r} \in D$ on the electric-field components, and therefore no coupling between the topological defects and auxiliary gauge field.  Restricting to an electrolyte of elementary charges (i.e., local topological defects with charge value $\pm q$), the Maggs-Rossetto generalized lattice-field electrolyte is therefore an array of Cartesian lattice electric-field components $E_{x/y}$ connecting the fixed points of a regular, topologically toroidal and square lattice with Boltzmann distribution 
\begin{widetext}
\begin{align}
\pi_{\rm Maggs} \left( \{ \rho(\rvec), Q(\rvec) \}, \wvec \right) \propto & \int \mathcal{D} \mathbf{E} \,\,  \prod_{\mathbf{r}\in D} \left[ \delta \left( \boldsymbol{\hat{\nabla}} \cdot \mathbf{E}(\mathbf{r}) - \frac{\rho(\mathbf{r})}{\epsilon_0} \right) \delta \left(\widetilde{\boldsymbol{\nabla}} \times \Evec(\rvec) + \boldsymbol{\nabla}^2 Q(\rvec)\UnitZ \right) \mathbb{I} \left[ |\rho\left(\mathbf{r}\right)| \le \frac{q}{a^2} \right] \right] \nonumber\\
& \times \prod_{\mu \in \{ x, y \}} \mathbb{I} \left[ \epsilon_0 \bar{E}_{{\rm p}, \mu} = \left( - \sum_{\mathbf{r} \in D} \frac{\rvec_\mu \rho(\mathbf{r})}{N} + \frac{q}{2L} \right) \!\!\!\!\! \mod \left( \frac{q}{L} \right) - \frac{q}{2L} \right] \nonumber\\
& \times \delta \left( \frac{1}{N} \sum_{\mathbf{r}\in D} \mathbf{E}(\mathbf{r}) - \left( \Ebar_{\rm p} + \frac{q}{L\epsilon_0} \wvec \right) \right) \exp \glb - \frac{\beta \epsilon_0 a^2}{2} \sum_{\mathbf{r}\in D} \|\mathbf{E}(\mathbf{r})\|^2 \grb ,
\label{eq:MaggsRossettoPDF}
\end{align}
\end{widetext}
where the indicator function $\mathbb{I} \left[ |\rho\left(\mathbf{r}\right)| \le q / a^2 \right]$ restricts to elementary charges.  Upon setting the elementary charge $q = 2\pi$ and vacuum permittivity $\epsilon_0 = 1 / J$, the only difference with the 2DHXY model is then that the Heaviside function(s) $\Theta \left(\pi J / a - \left| E_{x / y} (\mathbf{r} + a\mathbf{e}_{x / y} / 2) \right| \right)$ of equation~\eqref{eq:ConstrainedElectrolytePDF1} does not appear in this Boltzmann distribution [n.b., this Heaviside function(s) enforces the constraint $| \rho(\rvec) | \leq 2\pi / a^2 \, \forall \rvec \in D$, hence the indicator function(s) $\mathbb{I} \left[ |\rho\left(\mathbf{r}\right)| \le 2\pi / a^2 \right]$ is redundant in the 2DHXY case].  Indeed, the absence of $\Theta \left(\pi J / a - \left| E_{x / y} (\mathbf{r}) \right| \right)$ means that equation~\eqref{eq:MaggsRossettoPDF} factorizes into its Coulomb and auxiliary marginals:
\begin{align}
\pi_{\rm Maggs} \left(\{\rho(\rvec), Q(\rvec)\}, \wvec \right) = \pi_{\rm c} \left(\{\rho(\rvec)\}, \wvec \right) \pi_{\rm aux} \left(\{Q(\rvec)\}\right) ,
\label{eq:RealElectrolyteFactorisedPDF}
\end{align}
where
\begin{widetext}
\begin{align}
\pi_{\rm c} \glb \gld \rho(\rvec) \grd , \wvec \grb \propto \,\, & \mathbb{I} \glc \sum_{\rvec \in D} \rho(\rvec) = 0 \grc \prod_{\mu \in \{ x, y \}} \mathbb{I} \left[ \frac{\bar{E}_{{\rm p}, \mu}}{J} = \left( - \sum_{\mathbf{r} \in D} \frac{\rvec_\mu \rho(\mathbf{r})}{N} + \frac{\pi}{L} \right) \!\!\!\!\! \mod \left( \frac{2\pi}{L} \right) - \frac{\pi}{L} \right] \prod_{\mathbf{r}\in D} \mathbb{I} \left[ |\rho\left(\mathbf{r}\right)| \le \frac{2\pi}{a^2} \right] \nonumber\\
& \times \exp \glc -\frac{\beta a^4 J}{2} \sum_{\rvec_i \ne \rvec_j} \rho(\rvec_i)G(\rvec_i, \rvec_j)\rho(\rvec_j) \grc \exp \glb - \frac{\beta L^2}{2 J} \| \Ebar_{\rm p} + \frac{2\pi J}{L} \wvec \|^2 \grb e^{\beta \mu a^2 \sum_{\rvec \in D} \left| \rho(\rvec) \right|}
\label{eq:RealElectrolyteCoulombPDF}
\end{align}
\end{widetext}
is the \emph{Coulomb marginal} and 
\begin{align}
\pi_{\rm aux} \left(\{Q(\rvec)\}\right) \propto \exp \glc - \frac{\beta a^2}{2 J} \sum_{\rvec \in D} \| \hat{\boldsymbol{\nabla}} Q(\rvec) \|^2 \grc
\label{eq:RealElectrolyteAuxiliaryPDF}
\end{align}
is the \emph{auxiliary marginal}.  It is interesting to note the similarity between the auxiliary marginal and the Boltzmann factor of the Poisson field $-\NablaTildePhi$. %, which is explained by discontinuities in the electric field (due to the topological defects) being described by the harmonic mode $\Ebar$.  
This reflects the similar algebraic structure of the spin-spin~\cite{Mermin1966AbsenceFerromagnetism} and charge-charge~\cite{Alastuey1997PartI} correlations at low temperature.  To tune the chemical potentials in the generalized lattice-field electrolyte of multi-valued charges, \cite{Faulkner2015TSFandErgodicityBreaking,Faulkner2017AnElectricFieldRepresentation} added a core-energy term $a^4\sum_{\mathbf{r}\in D}\epsilon_{\rm core}\left(m(\rvec)\right)\rho(\mathbf{r})^2 / 2$ to the potential, where $\epsilon_{\rm core}(m)$ is the core-energy constant of each charge $mq$ ($m \in \ZZ$) with $\epsilon_{\rm core}(m) = \epsilon_{\rm core}(-m)$ as charges are excited in neutral pairs.  %The corresponding Boltzmann distribution also factorizes into Coulomb and auxiliary marginals.

The factorization of the Boltzmann distribution into the product of two marginals reflects the independence of the topological defects and auxiliary gauge field.  This is the only distinction between the generalized lattice-field electrolyte (of elementary charges) and the 2DHXY model.  In both cases, long-range interacting Coulomb charges emerge as local topological defects in the short-range interacting (`real' or emergent) electric field -- the local fundamental object of electrostatics -- but while the lattice-Coulomb physics is exact in the generalized lattice-field electrolyte, it is constrained in the 2DHXY model by the coupling between the topological defects and the spin waves.  It follows that the local 2DHXY spin dynamics (demonstrated in figure~\ref{fig:ChargeHopHXY}) propagate the emergent (but constrained) long-range Coulomb interactions throughout the 2DHXY model via the spin-wave fluctuations.  The second stage of the composite 2DHXY charge hop in figure~\ref{fig:ChargeHopHXY} is equivalent to a charge hop in the generalized lattice-field electrolyte [figure~\ref{fig:MaggsRossettoDynamics}(a)] while a 2DHXY spin rotation that does not change the emergent-charge configuration is equivalent to the electric-field rotation depicted in figure~\ref{fig:MaggsRossettoDynamics}(b).  In conjunction with equations~\eqref{eq:ConstrainedElectrolytePDF1}--\eqref{eq:RealElectrolyteAuxiliaryPDF}, this elucidates fully the emergent-Coulomb behaviour of the 2DHXY vortices [see the text around equation~\eqref{eq:GaussLawHXY}] and resolves the question as to how the \emph{long-range} interactions of the Salzberg--Prager model emerge from the \emph{short-range} interactions of the BKT picture.  %, i.e., it explains the deconfinement of long-range interacting vortex pairs driving the BKT transition in the short-range interacting 2DHXY model.  % i.e., it explains how the BKT transition drives the low-temperature confinement of (emergent) long-range interacting vortex pairs in the short-range interacting 2DHXY model.  
Indeed, both the composite charge-hop mechanism and the emergent Coulomb physics of equation~\eqref{eq:ConstrainedElectrolytePDF2} also exist in the 2DXY model [to first order in the case of equation~\eqref{eq:ConstrainedElectrolytePDF2}] but the emergent Coulomb physics is softened by the non-linear cosine couplings of the 2DXY model.  As a result, equation~\eqref{eq:ConstrainedElectrolytePDF2} cannot be written exactly in the 2DXY case.  %, and 2DXY vortex separation does not turn on precisely at the transition, as demonstrated below in Sections~\ref{sec:TopologicalNonergodicity} and \ref{sec:TSF}.  

We additionally emphasize that the mechanics of the composite 2DHXY charge hops are due to the 2DHXY charges being phase-difference vortices -- a result of the local $2\pi$-modular XY symmetry that couples the vortices to the spin waves.  The auxiliary gauge field also locally propagates the long-range Coulomb interactions throughout the generalized lattice-field electrolyte, but the composite charge-hop mechanism is special to a system of phase-difference vortices.  It is also interesting to note that multiple topological sectors describe any given charge configuration in both cases, but that this becomes a countably infinite set of topological sectors in the case of the generalized lattice-field electrolyte -- also due to an absence of topological-defect--spin-wave coupling.  %Topological sectors can therefore be independently sampled during a Markov process, as in \sect{sec:Nonergodicity}.
%Due to the decoupling of the topological defects and auxiliary gauge field, adoption of the generalized lattice-field electrolyte will therefore allow us to isolate and uniquely manipulate the global topological defects with global-defect dynamics.  
For completeness, we also present the Boltzmann distribution of the generalized lattice-field electrolyte in a spin-like representation~\cite{Faulkner2015TSFandErgodicityBreaking} in Appendix~\ref{sec:MaggsToVillain}.  This further elucidates the connection with the two-dimensional XY models -- in particular, by highlighting the equivalence with Villain's approximation to the 2DXY model~\cite{Villain1975TheoryOneAndTwoDimensionalMagnets}.  %For completeness, we use a continuum analogue the Coulomb marginal density [equation~\eqref{eq:RealElectrolyteCoulombPDF}] to present a variant on the Salzberg--Prager argument for the phase transition in Appendix~\ref{sec:PhaseTransition}. 

\begin{figure*}
\includegraphics[width=\linewidth]{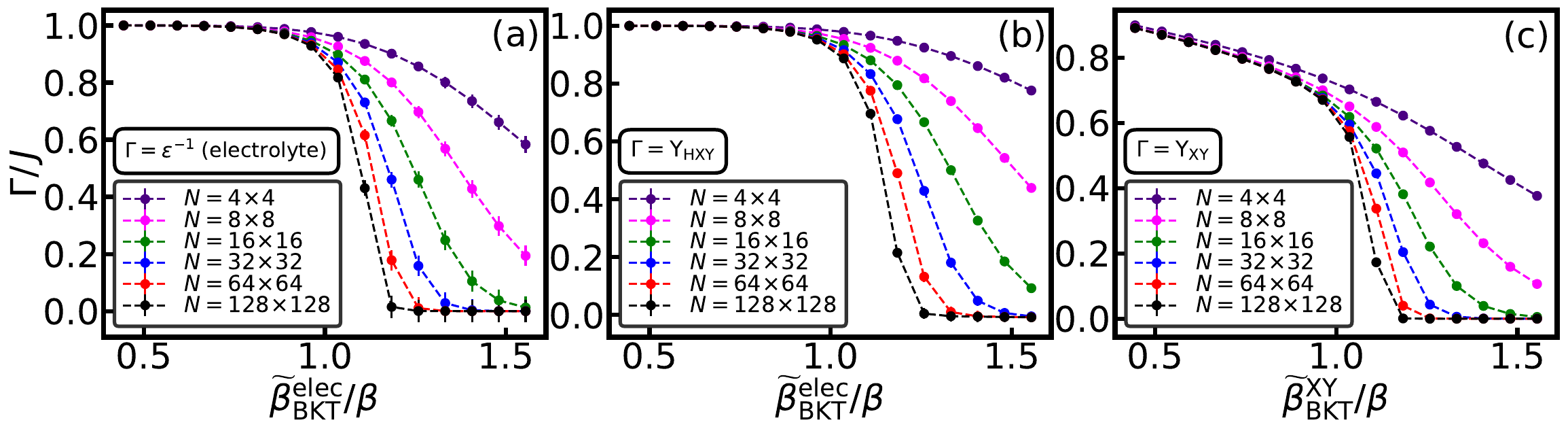}
\caption{The response functions defined in Section~\ref{sec:SpinStiffness} capture the BKT transition with a universal jump at the phase transition.  Against various system sizes $N$ and reduced temperatures $\tilde{\beta}_{\rm BKT}^{\rm model} / \beta$, estimates are presented of the inverse electric permittivity of the 2D generalized lattice-field electrolyte $\epsilon^{-1}$ [(a)] and helicity modulus $\Upsilon$ of the 2DHXY model [(b)] and the 2DXY model [(c)] -- with $\tilde{\beta}_{\rm BKT}^{\rm elec} := 1 / (1.351 J)$ and $\tilde{\beta}_{\rm BKT}^{\rm XY} := 1 / (0.887 J)$ the approximate inverse BKT transition temperatures of the corresponding model (we use $\tilde{\beta}_{\rm BKT}^{\rm elec}$ for the 2DHXY model -- see Section~\ref{sec:HXYModel}).  Results are consistent with a universal jump from $2 / \pi / \beta$ to zero at the transition from the low- to the high-temperature phase, with nonzero values defining topological order (in the thermodynamic limit) within the framework of Jos\'{e} {\it et al}.~\cite{Jose1977Renormalization} and Nelson \& Kosterlitz~\cite{Nelson1977UniversalJump}.  Data in (a) / (b-c) are averaged over $1536$ / $384$ simulations.  $10^5$ equilibration samples were discarded from each simulation.  Dashed lines are guides to the eye.  The sum in equation~\eqref{eq:InverseVacuumPermittivityHXY} is cut off at $\hat{n} = 11$.}
\label{fig:ResponseFunctions}
\end{figure*}

\subsection{Spin stiffness and inverse electric permittivity}
\label{sec:SpinStiffness}

Response functions that universally capture the BKT transition can now be derived and compared.  For the 2DXY models, BKT demonstrated a phase transition characterized by a spin-spin correlation function [equation~\eqref{eq:XYCorrelationFunction}] with power-law decay at low temperature and exponential decay at high temperature~\cite{Berezinskii1973DestructionLongRangeOrder,Kosterlitz1973OrderingMetastability}.  Jos\'{e} \emph{et al}.~\cite{Jose1977Renormalization} and Nelson \& Kosterlitz~\cite{Nelson1977UniversalJump} then framed this phase transition in terms of the \emph{spin stiffness} $\beta \Upsilon$, where the \emph{helicity modulus} 
\begin{align}
    \Upsilon(\beta, N) := \lim_{\Delta \to 0} \frac{1}{N} \frac{\partial^2 F(\beta, N, \Delta)}{\partial \Delta^2}
\end{align}
measures the system response following a small global perturbation $\Delta > 0$ to the absolute phase difference $\varphi(\rvec \oplus a \mathbf{e}_\mu) - \varphi(\rvec) \,\, \forall \rvec \in D$ for some $\mu \in \{ x, y \}$, with $F(\beta, N, \Delta)$ the free energy under the influence of the perturbation $\Delta$.  In the thermodynamic limit, this is equivalent to the system response following an externally applied global spin twist along the $\mu \in \{ x, y \}$ direction.  %Jos\'{e} \emph{et al}.~\cite{Jose1977Renormalization} presented an extensive renormalization-group analysis [built on by Nelson \& Kosterlitz~\cite{Nelson1977UniversalJump}] showing 
The extensive renormalization-group analyses of Jos\'{e} \emph{et al}.~\cite{Jose1977Renormalization} and Nelson \& Kosterlitz~\cite{Nelson1977UniversalJump} combined to show 
that the thermodynamic spin stiffness performs a \emph{universal jump} from $2 / \pi$ to zero at the transition from the low- to the high-temperature phase, with nonzero spin stiffness defining topological order within their framework.  This connected the spin-spin-correlation results with a thermodynamic response function and was supported by %a finite-size scaling analysis of subsequent numerical simulations~\cite{Minnhagen2003DirectEvidence}.  
subsequent numerical work~\cite{Minnhagen2003DirectEvidence}.

For the generalized lattice-field electrolyte, the inverse electric permittivity 
\begin{align}
    \epsilon^{-1}(\beta, N) := \lim_{\Delta \to 0} \frac{1}{N} \frac{\partial^2 F(\beta, N, \Delta)}{\partial \Delta^2}
\end{align}
similarly measures the system response following a small global perturbation $\Delta / (\epsilon_0 a) > 0$ to the electric field $E_\mu(\rvec) \,\, \forall \rvec \in D$ for some $\mu \in \{ x, y \}$ (i.e., a small externally applied homogeneous electric field along the $\mu \in \{ x, y \}$ direction).  This response function reduces to 
\begin{align}
    \epsilon^{-1} = \epsilon_0^{-1} \left( 1 - \frac{\beta \epsilon_0 L^2}{2}{\rm Var} \left[ \Ebar \right] \right)
    \label{eq:EvaluatedInversePermittivity}
\end{align}
and experiences the same universal jump (as $\Upsilon$) at the phase transition.  Indeed, the 2DHXY helicity modulus 
\begin{align}
    \Upsilon_{\rm HXY} = [\epsilon_0^{\rm HXY}]^{-1} \left( 1 - \frac{\beta \epsilon_0^{\rm HXY} L^2}{2}{\rm Var} \left[ \Ebar \right] \right)
    \label{eq:EvaluatedHelicityHXY}
\end{align}
can be analogously viewed (in thermodynamic limit) as the system response following a small externally applied homogeneous emergent (electric) field, where 
\begin{align}
    [\epsilon_0^{\rm HXY}]^{-1} := 2J \sum_{\hat{n} = 1}^\infty (-1)^{\hat{n} + 1}\mathbb{E}\cos \left[ \frac{\hat{n} a}{J} E_\mu(\rvec_0) \right]
    \label{eq:InverseVacuumPermittivityHXY}
\end{align}
is the \emph{2DHXY inverse vacuum permittivity}, i.e., the inverse vacuum permittivity of the constrained lattice-field electrolyte described by the Boltzmann distribution in equations~\eqref{eq:ConstrainedElectrolytePDF1} and \eqref{eq:ConstrainedElectrolytePDF2} (for any $\rvec_0 \in D$, $\mu \in \{ x, y \}$).  Upon comparing equations~\eqref{eq:EvaluatedInversePermittivity} and \eqref{eq:EvaluatedHelicityHXY}, the 2DHXY vacuum permittivity $\epsilon_0^{\rm HXY}$ becomes analogous to the vacuum permittivity $\epsilon_0$ (set to $1 / J$ in Section~\ref{sec:GeneralisedLatticeElectrolyte} above) of the generalized lattice-field electrolyte, but with a temperature dependence that reflects the 2DHXY coupling between the topological defects and the spin waves/auxiliary gauge field [as enforced by the Heaviside functions in equations~\eqref{eq:ConstrainedElectrolytePDF1} and \eqref{eq:ConstrainedElectrolytePDF2}].  In the future, it will be interesting to compare the behaviour of the 2DHXY model with that of the generalized lattice-field electrolyte with the temperature-dependent inverse vacuum permittivity given by equation~\eqref{eq:InverseVacuumPermittivityHXY}.  For completeness, the 2DXY helicity modulus 
\begin{align}
    \Upsilon_{\rm XY} = [\epsilon_0^{\rm XY}]^{-1} \left( 1 - \frac{\beta \epsilon_0^{\rm XY} N}{2}{\rm Var} \left[ \mathbf{j} \right] \right)
    \label{eq:HelicityModulusXY}
\end{align}
is also of the same form (to first order) as the inverse electric permittivity of the generalized lattice-field electrolyte.  Here, 
\begin{align}
    [\epsilon_0^{\rm XY}]^{-1} := -\frac{1}{2N} \mathbb{E} U_{\rm XY}(\hvec = 0)
\end{align}
is the \emph{2DXY inverse vacuum permittivity} and 
\begin{align}
    \mathbf{j} := \frac{J}{N} \begin{pmatrix} \sum_{\langle \rvec, \rvec' \rangle_x} \sin (\varphi_\rvec - \varphi_{\rvec'}) \\ \sum_{\langle \rvec, \rvec' \rangle_y} \sin (\varphi_\rvec - \varphi_{\rvec'}) \end{pmatrix}
\label{eq:MacroJosephsonCurrentXY}
\end{align}
is the (non-normalized) \emph{macroscopic Josephson current}~\footnote{This is ``non-normalized'' as it has not been divided by the magnetic flux quantum.  The normalized version has units $Q / T$.}, with the sum $\sum_{\langle \rvec, \rvec' \rangle_{x / y}}$ indicating a sum over nearest-neighbor lattice sites along the $x / y$ direction.  Upon comparing equations~\eqref{eq:EvaluatedInversePermittivity} and \eqref{eq:HelicityModulusXY}, the macroscopic Josephson current $\mathbf{j}$ maps (to first order) to the harmonic mode $\Ebar$ of the generalized lattice-field electrolyte.  

\begin{figure*}
\includegraphics[width=\linewidth]{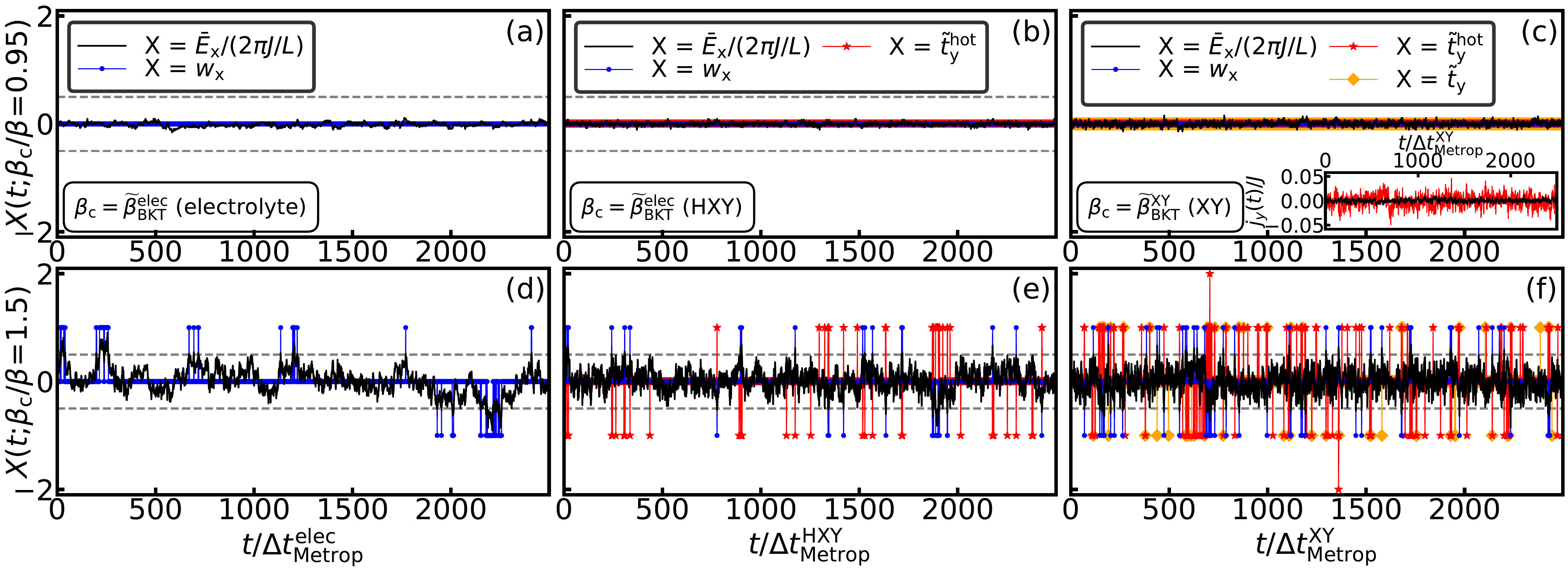}
\caption{Trace plots of the topological sector (and various analogous quantities) reflect the strikingly different topological characteristics of the low- and high-temperature phases.  Various trace plots are presented for the 2D generalized lattice-field electrolyte [(a), (d)], the 2DHXY model [(b), (e)] and the 2DXY model [(c), (f)] at low [(a)-(c)] and high [(d)-(f)] temperature -- all under local Metropolis electric-field/spin dynamics.  Each black output (excluding that in the inset) is the $x$-component of the normalized harmonic mode of the electric/emergent field $\bar{E}_x/ (2\pi J / L)$ [defined by equation~\eqref{eq:EbarDefinition}].  Each blue output is the $x$-component of the topological sector $w_x$ [defined below equation~\eqref{eq:OriginIndependentPolarisation}].  Each orange output is the $y$-component of the global twist-relaxation field $\tilde{t}_y$ [defined in Sections~\ref{sec:HXYModel} and \ref{sec:HarmonicMode}].  Each red output (excluding that in the inset) is the $y$-component of the non-annealed analogue of the global twist-relaxation field $\tilde{t}_y^{\rm hot}$ [defined in Section~\ref{sec:TSF}].  Horizontal gray lines at $\pm 1 / 2$ denote the bounds on the interval in equation~\eqref{eq:OriginIndependentPolarisationConstraint}.  Only the harmonic mode $\Ebar$ displays nonzero fluctuations at the lower temperatures (indicating the existence of local topological defects) whereas all quantities fluctuate at the higher temperatures.  This reflects the suppression/excitement of nonzero topological sectors in the low/high-temperature phase for systems constrained to local Metropolis dynamics.  The legends in (a/b/c) also apply to (d/e/f) but not to the inset.  $\tilde{\beta}_{\rm BKT}^{\rm elec} := 1 / (1.351 J)$ and $\tilde{\beta}_{\rm BKT}^{\rm XY} := 1 / (0.887 J)$ are the approximate inverse BKT transition temperatures of the corresponding model (we use $\tilde{\beta}_{\rm BKT}^{\rm elec}$ for the 2DHXY model -- see Section~\ref{sec:HXYModel}).  $\Delta t_{\rm Metrop}^{\rm elec}$ / $\Delta t_{\rm Metrop}^{\rm HXY}$ / $\Delta t_{\rm Metrop}^{\rm XY}$ is the Metropolis Monte Carlo time step of the electrolyte / 2DHXY model / 2DXY model (defined as the elapsed simulation time between $N$ attempted local-field updates).  Inset: The $y$-component of the normalized macroscopic Josephson current $j_y / J$ [defined by equation~\eqref{eq:MacroJosephsonCurrentXY}] of the 2DXY model at $\widetilde{\beta}_{\rm BKT}^{\rm XY} / \beta = 0.95$ (low temperature, black) and $\widetilde{\beta}_{\rm BKT}^{\rm XY} / \beta = 1.5$ (high temperature, red).  Its simulation variance is larger at the higher temperature and its largest values coincide with large values of $|\tilde{t}_y^{\rm hot}|$.  $N = 64 \times 64$ in all simulations [(a)-(f) and inset].  $10^4$ equilibration samples were discarded from each simulation.}
\label{fig:TopologicalSectorDynamics}
\end{figure*}

Figure~\ref{fig:ResponseFunctions} presents estimates of the inverse electric permittivity and helicity modulus of the generalized lattice-field electrolyte and two-dimensional XY models, at various system sizes and temperatures.  The results are consistent with each response function tending to the universal jump in the thermodynamic limit.

\subsection{Topological-sector fluctuations}
\label{sec:TSF}

The response functions of Section~\ref{sec:SpinStiffness} were hugely successful in characterizing both the phase transition and the notion of topological order.  By construction, however, they cannot discern whether this topological order breaks some form of topological ergodicity -- an important question as broken ergodicity typically accompanies the onset of order.  This was resolved by reframing the topological order in terms of fluctuations in the topological sector $\wvec$~\cite{Faulkner2015TSFandErgodicityBreaking,Faulkner2017AnElectricFieldRepresentation}. %outlined in Section~\ref{sec:HarmonicMode}.  
%%Global topological defects are excited in the high-temperature phase, resulting in fluctuations in the topological sector of the generalized lattice-field electrolyte~\cite{Faulkner2015TSFandErgodicityBreaking}. 
%We are now in a position to describe the topological order/nonergodicity of the low-temperature phase.  
%When restricted to local Brownian electric-field dynamics, the global topological defects of the generalized lattice-field electrolyte -- the topological sector $\wvec$ -- are frozen/excited in the low/high-temperature phase (this corresponds to simulations of the system with the local Metropolis dynamics described in Section~\ref{sec:GeneralisedLatticeElectrolyte})~\cite{Faulkner2015TSFandErgodicityBreaking}.  

As reviewed in detail below in Section~\ref{sec:TopologicalNonergodicity}, an absence of \emph{topological-sector fluctuations} $\sqrt{\langle s_{\wvec}^2 \rangle}$ is a characteristic of the low-temperature phase for systems restricted to local Brownian dynamics~\footnote{Topological-sector fluctuations were not defined as precisely in \cite{Faulkner2015TSFandErgodicityBreaking,Faulkner2017AnElectricFieldRepresentation}, despite being consistent with the present definition.}.  This is reflected in figures~\ref{fig:TopologicalSectorDynamics}(a) and (d), which present trace plots of (the $x$ components of) the harmonic mode $\Ebar$ and topological sector $\wvec$ of the $N = 64\times 64$ two-dimensional generalized lattice-field electrolyte at low [(a)] and high [(d)] temperature -- simulated using the local Metropolis algorithm described in Section~\ref{sec:GeneralisedLatticeElectrolyte}.  The topological sector is zero throughout the low-temperature simulation -- consistent with charge confinement in the low-temperature phase -- while nonzero topological sectors are visible at high temperature.  This reflects nonzero topological sectors requiring the separation of a neutral pair of charges through a distance greater than $L / 2$ along either Cartesian dimension.  %As the charge concentration falls to zero at low temperature, screening becomes negligible and 
%As charge screening is negligible at low temperature, 
%the configurational free-energy barrier against such configurations diverges logarithmically with the linear system size $L$, 
%The configurational free-energy barrier against such configurations diverges logarithmically with the linear system size $L$ at low temperature, 
%meaning that topological-sector fluctuations via local Brownian electric-field dynamics are absent in the low-temperature phase~\footnote{The continuum-space analogue of the lattice Green's function $G(\rvec, \rvec')$ in equation~\eqref{eq:LatticeGreensFunctionSolution} is proportional to $\ln \left[ \metric(\rvec, \rvec') \right]$.}.  
As the lattice Green's function $G(\rvec, \rvec')$ in equation~\eqref{eq:LatticeGreensFunctionSolution} scales like $- \ln \left[ \metric(\rvec, \rvec') / a \right]$ at large separation distances $\metric(\rvec, \rvec')$~\footnote{The continuum-space analogue of the lattice Green's function $G(\rvec, \rvec')$ in equation~\eqref{eq:LatticeGreensFunctionSolution} is proportional to $-\ln \left[ \metric(\rvec, \rvec') / a \right]$.}, the configurational free-energy barrier against such configurations diverges logarithmically with the linear system size $L$ at low temperature, 
meaning that topological-sector fluctuations via local Brownian electric-field dynamics are absent in the low-temperature phase.  As the charge concentration increases with temperature, however, entropy and charge screening %reduce the configurational free-energy barrier to finite values in the high-temperature phase, resulting in 
lead to topological-sector fluctuations via local Brownian electric-field dynamics in the high-temperature phase.  %This can also be seen in the context of the potential-difference arguments made around equation~\eqref{eq:LowEnergyHarmonicMode}, and 
This was verified numerically by a comprehensive analysis of the topological sectors under these local Metropolis/Brownian dynamics~\cite{Faulkner2015TSFandErgodicityBreaking} [see in particular figures 3 and 4 of \cite{Faulkner2015TSFandErgodicityBreaking}].

Analogous fluctuations are present in the XY models.  In the 2DHXY model, the global-twist relaxation field $\tilde{\tvec}$ defines the global topological defects $\tvec = -\tilde{\tvec}$, mapping precisely to the topological sector $\wvec = \glb t_y, -t_x \grb^T$ [see equation~\eqref{eq:TSGlobalTwistRelationship}] in the emergent-electrolyte representation.  For the case of the 2DXY model, the global twist-relaxation field does not always correspond to the global topological defects %(as defined by applying the 2DHXY potential to the 2DXY decomposition recipe outlined in  Section~\ref{sec:HXYModel})   
defined in Sections~\ref{sec:HXYModel} and \ref{sec:HarmonicMode} (due to the non-linear cosine couplings)  
but both its $\wvec$ and $\tilde{\tvec}$ dynamics nonetheless mimic the $\wvec$ dynamics of the generalized lattice-field electrolyte and 2DHXY model.  This is reflected in figure~\ref{fig:TopologicalSectorDynamics}, where we also present various trace plots for the 2DHXY model [(b), (e)] and the 2DXY model [(c), (f)] at low [(b)-(c)] and high [(e)-(f)] temperature -- all under local Metropolis spin dynamics for $N = 64\times 64$ lattice sites.  Each black output (excluding that in the inset) is the $x$-component of the normalized harmonic mode of the emergent field $\bar{E}_x/ (2\pi J / L)$.  Each blue output is the $x$-component of the topological sector $w_x$.  Each orange output is the $y$-component of the global twist-relaxation field $\tilde{t}_y$.  Each red output (excluding that in the inset) is the $y$-component of the non-annealed analogue of the global twist-relaxation field $\tilde{t}_y^{\rm hot}$ (defined analogously to the global twist-relaxation field $\tilde{t}_y$ but without the annealing step described in Section~\ref{sec:HXYModel}).  As for the generalized lattice-field electrolyte, only the harmonic mode $\Ebar$ displays nonzero fluctuations at the lower temperatures (indicating the existence of local topological defects) whereas all quantities fluctuate at the higher temperatures.  This reflects the suppression/excitement of nonzero topological sectors in the low/high-temperature phase for systems constrained to local Brownian spin dynamics.  As explored in detail below, this follows from the comprehensive analysis of the topological sectors of the generalized lattice-field electrolyte~\cite{Faulkner2015TSFandErgodicityBreaking} combined with renormalization-group arguments.  %Moreover, one can argue the case for each global twist-relaxation field ($\tilde{\tvec}$ and $\tilde{\tvec}^{\rm hot}$) in terms of externally applied global spin twists in combination with the potential-difference arguments made around equation~\eqref{eq:LowEnergyHarmonicMode} (i.e., without renormalization-group arguments).  
This is an example of the utility of the quadratic analogues of the 2DXY model, including the mapping to the generalized lattice-field electrolyte.
%This reflects the suppression/excitement of nonzero topological sectors in the low/high-temperature phase of the 2D generalized lattice-field electrolyte constrained to local Brownian electric-field dynamics [we cannot explicitly make analogous statements for the XY models as comprehensive analyses (demonstrating this precisely) have not been performed].  Note that we observe zero/nonzero fluctuations in the 2DXY topological sector at low/high temperature, despite this not being predicted theoretically due to the non-linear cosine couplings of the 2DXY potential.  This is an example of the utility of the quadratic analogues of the 2DXY model.

For completeness, we also present trace plots of the ($y$ component of the) macroscopic Josephson current $\mathbf{j}$ [defined in equation~\eqref{eq:MacroJosephsonCurrentXY}] in the inset of figure~\ref{fig:TopologicalSectorDynamics}(c) at low (black) and high (red) temperature [compare equations~\eqref{eq:EvaluatedInversePermittivity} and \eqref{eq:HelicityModulusXY} to recall that $\mathbf{j}$ maps (to first order) to the harmonic mode $\Ebar$ of the generalized lattice-field electrolyte].  Its simulation variance is larger at the higher temperature and its largest values coincide with large values of $|\tilde{t}_y^{\rm hot}|$.

%\subsection{Topological order/nonergodicity}
\subsection{Topological order}
\label{sec:TopologicalNonergodicity}

We are now in a position to reframe the topological order (of the low-temperature phase) as a topological nonergodicity.  In the absence of charges, the system-size dependence falls out of the Coulomb marginal density [equation~\eqref{eq:RealElectrolyteCoulombPDF}] of the generalized lattice-field electrolyte:
\begin{align}
\pi_c \left(\wvec \, | \, \{\rho(\rvec) = 0\} \right) \propto \exp \left( - 2 \pi^2 \beta J \| \wvec \|^2 \right) .
\label{eq:RealElectrolyteTSPDF}
\end{align}
This is a consequence of the matching dimensionalities of the electric field and lattice, and is analogous to the probability of 2DXY global-twist events scaling like $\exp(-2 \pi^2 \beta J)$ in the absence of other excitations (see Section~\ref{sec:GlobalTwists}) -- though in contrast with this 2DXY analogue, equation~\eqref{eq:RealElectrolyteTSPDF} holds at finite $N$ and in the presence of auxiliary-gauge fluctuations (showcasing again the utility of the mapping to the generalized lattice-field electrolyte).  It follows that, even in the restricted ensemble of equation~\eqref{eq:RealElectrolyteTSPDF}, the expected topological susceptibility %$\chi_{\wrm} (L, T) := \beta L^2 \mathbb{E} \glb \Ebar_{\wrm} - \mathbb{E} \Ebar_{\wrm} \grb^2 / J$ 
$\chi_\wvec (\beta, L) := \beta L^2 {\rm Var} \left[ \Ebar_{\wrm} \right] / (2J) = 2 \pi^2 \beta J {\rm Var} \left[ \wvec \right]$ is small but non-negligible at low temperature:
\begin{align}
\chi_\wvec \left(\beta, L \, | \, \{ \rho(\rvec) = 0 \} \right) &= \frac{\beta L^2}{2J} \frac{16 \pi^2 J^2 e^{- 2 \pi^2 \beta J} / L^2 + \dots}{1 + 4 e^{- 2 \pi^2 \beta J} +\dots} \nonumber\\
  & \sim 8 \pi^2 \beta J \exp \glb - 2 \pi^2 \beta J \grb
\label{eq:ZeroChargeTSF}
\end{align}
%as $N \to \infty$ and 
for $1 / (\beta J) \ll 2 \pi^2$ (this result also holds for the XY models restricted to global topological defects, i.e., no local topological defects or spin waves).  The Boltzmann models of both the restricted and non-restricted ensembles therefore predict topological-sector fluctuations -- signalled by nonzero expected topological susceptibility -- at all nonzero temperatures.  This is in spite of topological-sector fluctuations being absent (in the thermodynamic limit) at low temperature in systems simulated using local Metropolis dynamics, as reflected in figure~\ref{fig:TopologicalSectorDynamics} and demonstrated explicitly by the comprehensive analysis of the topological sectors~\cite{Faulkner2015TSFandErgodicityBreaking} referenced in Section~\ref{sec:TSF} (in combination with renormalization-group arguments in the case of the XY models).  %This suggests a breaking of topological ergodicity at low temperature.  

\begin{figure*}
\includegraphics[width=\linewidth]{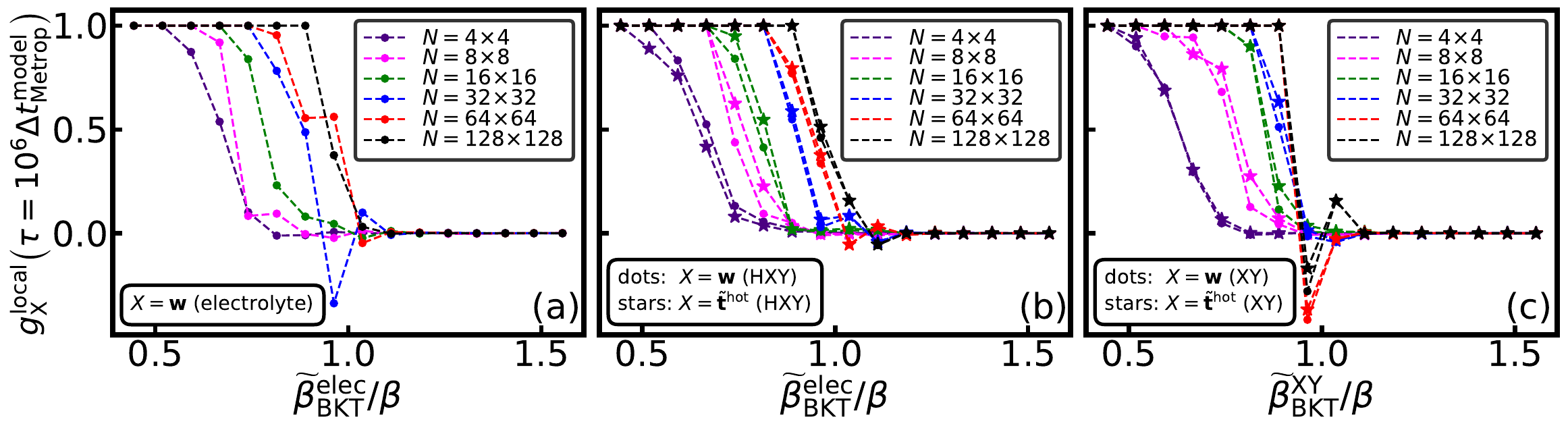}
\caption{For systems restricted to local Metropolis electric-field/spin dynamics, the finite topological stability $g_\wvec^{\rm local}(\beta, \tau, N)$ [defined in equation~\eqref{eq:TopSuscRatioFn}] characterizes the topological nonergodicity of the low-temperature phase.  For simulation timescale $\tau = 10^6 \Delta t_{\rm Metrop}^{\rm model}$ and at various system sizes $N$ and reduced temperatures $\tilde{\beta}_{\rm BKT}^{\rm model} / \beta$, estimates of the quantity are presented for the 2D generalized lattice-field electrolyte [(a)], 2DHXY model [(b)] and 2DXY model [(c)] ($\Delta t_{\rm Metrop}^{\rm model}$ is the Metropolis time step of the corresponding model; $\tilde{\beta}_{\rm BKT}^{\rm elec} := 1 / (1.351 J)$ and $\tilde{\beta}_{\rm BKT}^{\rm XY} := 1 / (0.887 J)$ are the approximate inverse BKT transition temperatures of the corresponding model (we use $\tilde{\beta}_{\rm BKT}^{\rm elec}$ for the 2DHXY model -- see Section~\ref{sec:HXYModel})).  Estimates of the analogous quantity $g_{\tilde{\tvec}^{\rm hot}}^{\rm local}(\beta, \tau = 10^6 \Delta t_{\rm Metrop}^{\rm model}, N)$ are also presented (at various system sizes and temperatures) for the 2DHXY [(b)] and 2DXY [(c)] models restricted to local Metropolis dynamics, where $g_{\tilde{\tvec}^{\rm hot}}(\beta, \tau, N)$ is defined analogously to the finite topological stability in equation~\eqref{eq:TopSuscRatioFn}.  Results are consistent with each function being one/zero in the low/high-temperature phase.  This supports the hypothesis in equation~\eqref{eq:TopologicalOrderMetrop} and is consistent with topological nonergodicity (under local Metropolis dynamics) in the low-temperature phase.  %In addition, estimates of the analogous quantities $g_{\Ebar}^{\rm local}(\beta, \tau, N)$ and $g_\mathbf{j}^{\rm local}(\beta, \tau, N)$ are presented in figures~\ref{fig:TopOrder}(a-b) and (c) for the generalized lattice-field electrolyte/2DHXY model and 2DXY model restricted to local Metropolis dynamics, where again $g_{\Ebar}(\beta, \tau, N)$ and $g_\mathbf{j}(\beta, \tau, N)$ are defined analogously to the finite topological stability in equation~\eqref{eq:TopSuscRatioFn}.  The results suggest that all quantities are zero for all system sizes and nonzero temperatures -- as the response functions of Section~\ref{sec:SpinStiffness} cannot discern topological ergodicity.  
Each function is estimated from the ratio of two simulation estimates.  Errors are therefore large and not shown.  Data in (a) / (b)-(c) are averaged over $1536$ / $384$ simulations.  $10^5$ equilibration samples were discarded from each simulation.  Dashed lines are guides to the eye.}
\label{fig:TopOrder}
\end{figure*}

The above discrepancy between Metropolis simulations and predictions of the Boltzmann model suggests a loss of ergodicity at low temperature (as such discrepancies are the essence of broken ergodicity).  We characterize the discrepancy via the notion of \emph{long-time topological stability} 
%As stated in Section~\ref{sec:IsingModelAndSSB} but applied to this more general case, discrepancies between simulation results and predictions of the Boltzmann model are the essence  of ergodicity (rather than symmetry) breaking.  We are able to similarly characterize the above Metropolis results via the \emph{long-time topological stability} 
\begin{align}
\gamma_\wvec(\beta) := \lim_{\tau \to \infty} \lim_{N \to \infty} g_\wvec (\beta, \tau, N) 
\label{eq:TopologicalOrderDef}
\end{align}
where the \emph{finite topological stability}  
\begin{align}
    g_\wvec(\beta, \tau, N) := 1 - \sqrt{\frac{\langle s_{\wvec}^2(\beta, \tau, N) \rangle}{{\rm Var} \left[ \wvec \right](\beta, N)}} 
\label{eq:TopSuscRatioFn}
\end{align}
measures discrepancies between the topological-sector fluctuations $\sqrt{\langle s_{\wvec}^2 \rangle}$ and their expected value $\sqrt{{\rm Var} \left[ \wvec \right]}$.  Indeed, with $\gamma_\wvec(\beta) = 1$ defining the equivalent concepts of \emph{topological order} and \emph{topological nonergodicity} under the chosen dynamics, \emph{topological ergodicity} then corresponds to 
\begin{align}
    \lim_{\tau \to \infty} \lim_{N \to \infty} \langle s_{\wvec}^2 (\beta, \tau, N) \rangle = \lim_{N \to \infty} {\rm Var} \left[ \wvec \right](\beta, N) ,
\end{align}
with a topologically ergodic simulation defined by a simulation variance $s_{\wvec}^2$ that has converged to its expected value ${\rm Var} \left[ \wvec \right]$ within some small $\varepsilon > 0$ (i.e., ergodic exploration of the topological sector $\wvec \in \mathbb{Z}^2$).  For all three models, the local Metropolis simulations in figure~\ref{fig:TopologicalSectorDynamics} are then consistent with 
\begin{align}
\gamma_\wvec^{\rm local}(\beta)  = 
\begin{cases}
    1 & {\rm for} \,\, \beta > \beta_{\rm BKT} , \\
    0 & {\rm for} \,\, \beta < \beta_{\rm BKT} .
\end{cases}
\label{eq:TopologicalOrderMetrop}
\end{align}
%with the low-temperature topological order/nonergodicity due to vanishing topological-sector fluctuations in the thermodynamic limit.  
This thermodynamic limit is singular as exchanging the order of the limits in equation~\eqref{eq:TopologicalOrderDef} returns zero at all nonzero temperatures.  
As noted in Sections~\ref{sec:IsingModelAndSSB} and \ref{sec:SingularLimitGSB}, the long-time directional stability functions defined in equations~\eqref{eq:DirectionalStabilityIsingDef} and \eqref{eq:DirectionalStabilityXYDef} (which characterize general symmetry breaking in the 2D Ising and 2DXY models) are of the form of the long-time topological stability defined in equation~\eqref{eq:TopologicalOrderDef}.  Such functions may be viewed as the natural characterization of (weakly) broken ergodicity.  We note also that topological nonergodicity corresponds to asymptotically slow mixing of the topological sector~\footnote{Asymptotically slow mixing defines broken ergodicity in classical statistical physics, despite typically being described with respect to some algorithm/dynamics that is ergodic on long enough timescales at any finite system size.  This asymptotically slow mixing is sometimes referred to as \emph{weakly broken ergodicity}~\cite{Ramesh2024WeakErgodicityBreaking}.}.

%The expected variance ${\rm Var} \left[ {\bf w} \right]$ () is not analytically tractable. The simulations presented in Section~\ref{sec:TSF} can, however, be supplemented with a global-defect dynamics that guarantee nonzero simulated topological susceptibility at all nonzero temperatures~\cite{Faulkner2015TSFandErgodicityBreaking}.  These dynamics are analogous to the global-twist dynamics of the XY models, defined along the $x/y$ dimension by 
%\begin{align}
%    E_{x/y}(\rvec) \mapsto E_{x/y}(\rvec) \pm q / \sqrt{N} \,\,\,\, \textrm{for all} \,\,\,\, \rvec \in D .
%    \label{eq:DefGlobalTSdynamics}
%\end{align}
%Due to the independence of the topological defects and auxiliary gauge field (as described by the factorization of the Boltzmann distribution into the product of the two marginals in equation~\ref{eq:RealElectrolyteFactorisedPDF}) these dynamics isolate and uniquely manipulate the global topological defects, reducing to $\bar{E}_{x/y} \mapsto \bar{E}_{x/y} \pm q \sqrt{N}$.  This is in contrast with the global-twist dynamics of the XY models, which may also alter the Poisson and rotational components of the emergent electric field.

\begin{figure*}
\includegraphics[width=\linewidth]{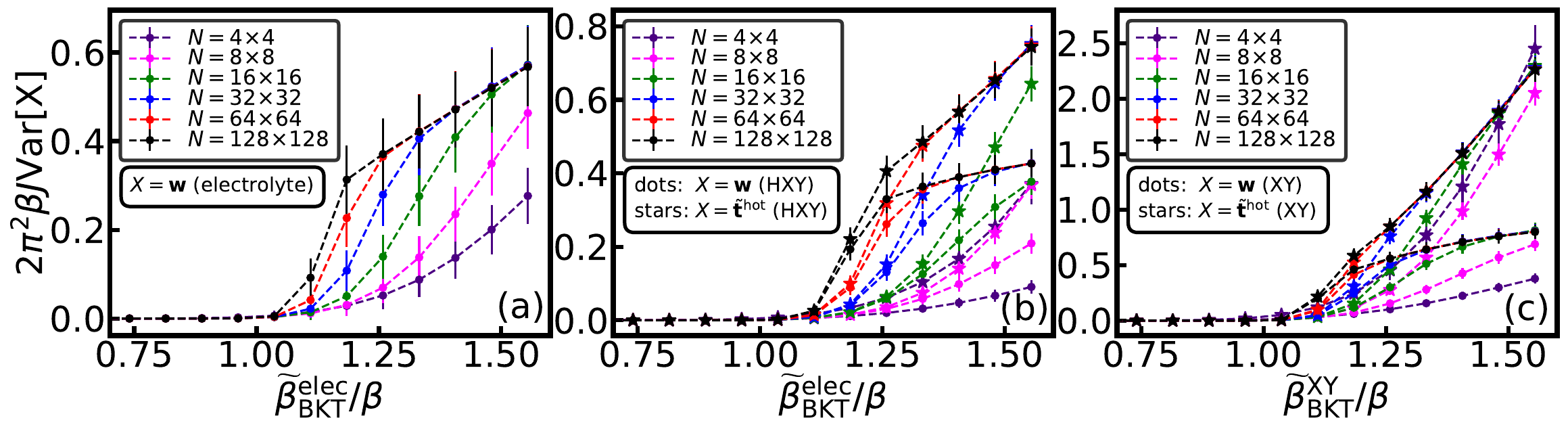}
\caption{The unbiased estimates (used in figure~\ref{fig:TopOrder} to demonstrate the topological nonergodicity of the low-temperature phase) are presented for the 2D generalized lattice-field electrolyte [(a)], 2DHXY model [(b)] and 2DXY model [(c)].  Each is presented as a function of system size $N$ and reduced temperature $\tilde{\beta}_{\rm BKT}^{\rm model} / \beta$, with $\tilde{\beta}_{\rm BKT}^{\rm elec} := 1 / (1.351 J)$ and $\tilde{\beta}_{\rm BKT}^{\rm XY} := 1 / (0.887 J)$ the approximate inverse BKT transition temperatures of the corresponding model (we use $\tilde{\beta}_{\rm BKT}^{\rm elec}$ for the 2DHXY model -- see Section~\ref{sec:HXYModel}).  Results are consistent with topological-sector fluctuations under purely local Metropolis dynamics turning on precisely at the phase transition~\cite{Faulkner2015TSFandErgodicityBreaking}.  Long-time topological stability (supported by figure~\ref{fig:TopOrder}) therefore induces topological nonergodicity throughout the low-temperature phase in systems restricted to local Metropolis dynamics.  Data in (a) / (b)-(c) are averaged over $1536$ / $384$ simulations.  $10^5$ equilibration samples were discarded from each simulation.}
\label{fig:TopologicalSusceptibilities}
\end{figure*}

The expected variance ${\rm Var} \left[ {\bf w} \right]$ is not, however, analytically tractable in the non-restricted ensemble.  %On the face of it, this presents a problem as regards supporting the hypothesis in equation~\eqref{eq:TopologicalOrder}, but simulations using local electric-field dynamics (such as those presented in Section~\eqref{sec:TSF}) can be supplemented with a global-defect dynamics defined along the $x/y$ dimension by 
To support the hypothesis in equation~\eqref{eq:TopologicalOrderMetrop}, we therefore construct supplemental global dynamics that ensure topologically ergodic simulations on non-divergent timescales -- as the squared topological-sector fluctuations $\langle s_{\wvec}^2\rangle$ then become non-biased estimators of ${\rm Var} \left[ {\bf w} \right]$ on non-divergent timescales, providing access to equation~\eqref{eq:TopSuscRatioFn} for any chosen dynamics.  For each model, this is achieved with a global dynamics that directly samples some global topological quantity.  For the XY models, this corresponds to the global-twist dynamics defined below equation~\eqref{eq:DefGlobalSpinTwists}, which directly sample $\tilde{\tvec}^{\rm hot}$.  For the generalized lattice-field electrolyte, the global dynamics are defined by the Metropolis proposal 
\begin{align}
    E_{x/y}(\rvec) \mapsto E_{x/y}(\rvec) \pm \frac{2\pi J}{L} 
    \label{eq:DefGlobalTSdynamics}
\end{align}
for all $\rvec \in D$ along each Cartesian dimension at each Monte Carlo time step.  These dynamics directly sample $\wvec$, in analogy with the global-twist dynamics of the XY models directly sampling $\tilde{\tvec}^{\rm hot}$.  Since the probability of global-defect events is non-negligible at all nonzero temperatures [see Section~\ref{sec:GlobalTwists} and equation~\eqref{eq:RealElectrolyteTSPDF}], the global-defect dynamics of each model ensure topologically ergodic simulations on non-divergent timescales (n.b., direct sampling of $\tilde{\tvec}^{\rm hot}$ leads to much improved mixing of $\wvec$ for the XY models).  For simulations with supplemental global-defect dynamics, $\langle s_{\wvec}^2\rangle$ and $\langle s_{\tilde{\tvec}^{\rm hot}}^2\rangle$ are then (respectively) non-biased estimators of ${\rm Var} \left[ {\bf w} \right]$ and ${\rm Var} \left[ \tilde{\tvec}^{\rm hot} \right]$ on non-divergent timescales.  Moreover, for the generalized lattice-field electrolyte, the independence of the topological defects and auxiliary gauge field  [as described by equation~\eqref{eq:RealElectrolyteFactorisedPDF}] means that these dynamics isolate and uniquely manipulate the global topological defects, reducing to 
\begin{align}
    \bar{E}_{x/y} \mapsto \bar{E}_{x/y} \pm \frac{2\pi J}{L} .
    \label{eq:SimplifiedGlobalTSdynamics}
\end{align}
The comprehensive analysis of the topological sectors~\cite{Faulkner2015TSFandErgodicityBreaking} was therefore independent of the auxiliary gauge field, and was also accelerated by the $\mathcal{O}(1)$ global dynamics of equation~\eqref{eq:SimplifiedGlobalTSdynamics}.  In contrast, the $\mathcal{O}(N)$ global-twist dynamics of the XY models [defined below equation~\eqref{eq:DefGlobalSpinTwists}] typically also alter the low-energy-Coulomb and rotational components of the emergent field.  

Figure~\ref{fig:TopOrder} presents estimates of the finite topological stability $g_\wvec^{\rm local}(\beta, \tau = 10^6 \Delta t_{\rm Metrop}^{\rm model}, N)$ at various system sizes and temperatures for the generalized lattice-field electrolyte [(a)], the 2DHXY model [(b)] and the 2DXY model [(c)], all restricted to local Metropolis dynamics ($\Delta t_{\rm Metrop}^{\rm model}$ is the Metropolis time step for the relevant model).  Figures~\ref{fig:TopOrder}(b) and (c) also present estimates of $g_{\tilde{\tvec}^{\rm hot}}^{\rm local}(\beta, \tau = 10^6 \Delta t_{\rm Metrop}^{\rm model}, N)$ (at various system sizes and temperatures) for the 2DHXY and 2DXY models restricted to local Metropolis dynamics, where $g_{\tilde{\tvec}^{\rm hot}}(\beta, \tau, N)$ is defined analogously to the finite topological stability in equation~\eqref{eq:TopSuscRatioFn}.  The results are consistent with each function being one/zero in the low/high-temperature phase.  This supports the hypothesis in equation~\eqref{eq:TopologicalOrderMetrop} and is consistent with topological nonergodicity (under local Metropolis dynamics) in the low-temperature phase.  %The strong fluctuations in the intermediate region are due to topological-sector fluctuations via local electric-field dynamics representing increasingly rare events \blue{(or CSD???)} -- an inevitable precursor to the loss of topological ergodicity.  The width of this region appears to decrease with increasing system size, consistent with it shrinking to zero and reflecting the sharp transition predicted by equation~\eqref{eq:TopologicalOrderMetrop}.  
Each estimate is relatively noisy in the transition region, certainly for the generalized lattice-field electrolyte and 2DXY model.  This is partially explained by each being estimated from the ratio of two simulation estimates -- but may also reflect critical slowing down at the phase transition.  

%In addition, $g_{\Ebar}^{\rm local}(\beta, \tau, N)$ and $g_\mathbf{j}^{\rm local}(\beta, \tau, N)$ are presented in figures~\ref{fig:TopOrder}(a/b) and (c) for the generalized lattice-field electrolyte/2DHXY model and 2DXY model restricted to local Metropolis dynamics, where again $g_{\Ebar}(\beta, \tau, N)$ and $g_\mathbf{j}(\beta, \tau, N)$ are defined analogously to the finite topological stability in equation~\eqref{eq:TopSuscRatioFn}.  These are the analogous stability measures of the current-based components of the response functions outlined in Section~\ref{sec:SpinStiffness}.  The data suggest that all quantities are zero for all system sizes and nonzero temperatures, demonstrating that the response functions cannot discern whether some region of thermodynamic phase space is topologically ergodic.

%We emphasize that analogous results [to equation~\eqref{eq:TopologicalOrderMetrop}] are not visible for the response functions presented in Section~\ref{sec:SpinStiffness}.  The corresponding null simulation data can be found in the data repository accompanying this paper (see Section~\ref{sec:CodeAndSimulationData}).

For completeness, figure~\ref{fig:TopologicalSusceptibilities} presents the unbiased estimates used in figure~\ref{fig:TopOrder} to demonstrate the topological nonergodicity of the low-temperature phase.  In the case of the topological susceptibility of the generalized lattice-field electrolyte, a comprehensive analysis of equivalent numerical simulation data demonstrated that topological-sector fluctuations under purely local Metropolis dynamics turn on precisely at the phase transition~\cite{Faulkner2015TSFandErgodicityBreaking}.  Long-time topological stability therefore induces topological nonergodicity throughout the low-temperature phase in systems restricted to local Metropolis dynamics, corresponding to broken ergodicity between the topological sectors.  We note that while topological nonergodicity corresponds to charge confinement, it is absent in the Salzberg--Prager and BKT theories because their charge-interaction representations do not include the degrees of freedom required to describe the topological nonergodicity -- the topological sector $\wvec$.  By renormalization-group arguments, these results hold for all global topological quantities whose expected susceptibilities are presented in figure~\ref{fig:TopologicalSusceptibilities}.  Analogous results do not hold for the current-based components ($\Ebar$ and $\mathbf{j}$) of the response functions outlined in Section~\ref{sec:SpinStiffness}.  This is because they cannot discern whether some region of thermodynamic phase space is topologically ergodic.

In addition to ensuring topologically ergodic simulations on non-divergent timescales, the global-twist dynamics also ensure $U(1)$-symmetric XY simulations on non-divergent timescales at all nonzero temperatures (see Section~\ref{sec:GlobalTwists}).  Since, to leading order, topologically ergodic simulations require ergodic exploration of $\tilde{{\bf t}}$ only over the five-element set $\{ (0, 0), \pm (1, 0), \pm (0, 1) \} \subset \mathbb{Z}^2$, we may assume that these timescales are of the same order (at low temperature, $r \in \mathbb{N}$ occurrences of non-trivial $\tilde{{\bf t}}$ values resulting from global moves lead to $r$ of the Dirac distributions described in Section~\ref{sec:GlobalTwists}).  It follows that supplemental global-twist dynamics ensure both topological ergodicity and $U(1)$ symmetry on non-divergent timescales at all nonzero temperatures.  Topological order (defined by the topological nonergodicity) therefore induces the broken $U(1)$ symmetry of Section~\ref{sec:XYModelAndGSB}.    

%\blue{In figure XX, we also present global-defect fluctuations for the 2DXY model, as defined at the end of Section~\ref{sec:HarmonicMode}.  These fluctuations turn on below the transition, consistent with 2DXY vortex separation turning on below the transition and the simulations presented in figure~\ref{fig:TopologicalSectorDynamics}(xx) in Section~\ref{sec:TSF}.  This is due to the non-linear cosine coupling essentially softening the emergent charges, as discussed in Sections~\ref{sec:HarmonicMode} and \ref{sec:GeneralisedLatticeElectrolyte}.}

\section{Discussion}
\label{sec:Discussion}

%Global-twist dynamics guarantee both topological ergodicity and $U(1)$ symmetry in the XY models, both of which are broken in the low-temperature phase.  This demonstrates that the topological nonergodicity of the low-temperature phase induces the (general) $U(1)$ symmetry breaking.

The pioneering work of Salzberg \& Prager~\cite{Salzberg1963EquationOfStateTwoDimensional}, BKT~\cite{Berezinskii1973DestructionLongRangeOrder,Kosterlitz1973OrderingMetastability}, Villain~\cite{Villain1975TheoryOneAndTwoDimensionalMagnets}, Jos\'{e} {\it et al}.~\cite{Jose1977Renormalization} and Nelson \& Kosterlitz~\cite{Nelson1977UniversalJump} described a phase transition mediated by local topological defects -- or vortices -- in the spin/condensate phase field.  This was fundamental to modelling the superfluid stiffness in the experiments of Bishop \& Reppy~\cite{Bishop1978StudySuperfluid}.  %Moreover, the Mermin--Wagner--Hohenberg theorem~\cite{Mermin1966AbsenceFerromagnetism,Hohenberg1967ExistenceOfLong-RangeOrder} had previously shown that spontaneous symmetry breaking must be absent at the transition.  The description involving local topological defects also resolved this question as to the mechanics of the transition -- characterizing it as topological in nature -- though the topological order of the low-temperature phase had not been shown to induce a topological nonergodicity.  
%This characterization of the transition as topological in nature also explained the existence of a phase transition itself -- as the Mermin--Wagner--Hohenberg theorem~\cite{Mermin1966AbsenceFerromagnetism,Hohenberg1967ExistenceOfLong-RangeOrder} had previously demonstrated an absence of spontaneous symmetry breaking -- though the topological order of the low-temperature phase had not been shown to induce a topological nonergodicity.  
The characterization in terms of a low-temperature topological ordering also demonstrated the existence of a phase transition itself -- a huge breakthrough as the Mermin--Wagner--Hohenberg theorem~\cite{Mermin1966AbsenceFerromagnetism,Hohenberg1967ExistenceOfLong-RangeOrder} had previously demonstrated an absence of spontaneous symmetry breaking [reflected in the expected norm of the $U(1)$ order parameter going to zero in the thermodynamic limit at all nonzero temperatures].  
This first body of work led, however, to three interconnected paradoxes.  %i) The \emph{long-range} interacting Salzberg--Prager and \emph{short-range} interacting BKT pictures exhibit the same phase transition.  
i) The prototypical \emph{short-range} interacting 2DXY spin model maps to an electrolyte of \emph{long-range} interacting electrostatic charges.  ii) The onset of order typically induces a nonergodicity under certain dynamics, but the topological ordering had not been shown to break any form of topological ergodicity.  iii) The absence of spontaneous symmetry breaking seemed to contradict the numerous experimental observations consistent with broken symmetry at low temperature~\cite{Baity2016EffectiveTwoDimensionalThickness,Shi2016EvidenceCorrelatedDynamics,Bishop1978StudySuperfluid,Bramwell2015PhaseOrder,Wolf1981TwoDimensionalPhaseTransition,Resnick1981KosterlitzThouless,Bramwell1993Magnetization,Huang1994MagnetismFewMonolayersLimit,Elmers1996CriticalPhenomenaTwoDimensionalMagnet,BedoyaPinto2021Intrinsic2DXYFerromagnetism,Hadzibabic2006KosterlitzThouless,Fletcher2015ConnectingBKTAndBEC,Christodoulou2021Observation}.

%Vallat, Beck~\cite{Vallat1994CoulombGas} and then Faulkner et al.~\cite{Faulkner2015TSFandErgodicityBreaking,Faulkner2017AnElectricFieldRepresentation} then re-expressed the above framework as an electrostatic-field theory on the torus.  This description included global topological defects in the spin/condensate phase field -- or the topological sector of the emergent electric field.  As presented in detail in Section~\ref{sec:EmergentElectrolyte}, these additional degrees of freedom are ergodically frozen at low temperature for systems governed by local Brownian dynamics, resolving the question of topological nonergodicity. 

%To resolve the question of whether topological order induces a topological nonergodicity, Vallat \& Beck~\cite{Vallat1994CoulombGas} and then 
To resolve the first two questions, %Faulkner, Bramwell \& Holdsworth~\cite{Faulkner2015TSFandErgodicityBreaking,Faulkner2017AnElectricFieldRepresentation} 
foundational work by Vallat \& Beck~\cite{Vallat1994CoulombGas} was taken further with an emergent electrostatic-field theory formulated on the torus~\cite{Faulkner2015TSFandErgodicityBreaking,Faulkner2017AnElectricFieldRepresentation}.  %, as reviewed in detail in Section~\ref{sec:EmergentElectrolyte}.  %This description included global topological defects in the spin/condensate phase field -- expressed as the topological sector of the emergent electric field.  As presented in detail in Section~\ref{sec:EmergentElectrolyte}, this described the topological nonergodicity, as these additional degrees of freedom are ergodically frozen at low temperature for systems governed by local Brownian dynamics~\cite{Faulkner2015TSFandErgodicityBreaking,Faulkner2017AnElectricFieldRepresentation}. 
The long-range interacting electrostatic charges of the Salzberg--Prager model map to local topological defects in the short-range interacting emergent (electric) field.  This demonstrated how the long-range charge-charge interactions emerge from the short-range spin-spin interactions -- and comparison with the lattice electric-field algorithm of Maggs \& Rossetto~\cite{Maggs2002LocalSimulationAlgorithms,Rossetto2002Thesis} elucidated the mechanism by which the emergent long-range interactions propagate throughout the short-range spin model.  Moreover, %the emergent electrostatic-field theory 
the framework included global topological defects in the spin/condensate phase field.  These additional degrees of freedom map to the topological sector of the emergent field and are nonergodically frozen at low temperature for systems governed by local Brownian dynamics~\cite{Faulkner2015TSFandErgodicityBreaking,Faulkner2017AnElectricFieldRepresentation}.  This reframed topological order as an ergodicity breaking between the topological sectors, thus resolving the question of topological nonergodicity.

%Prior to the above field theory, Bramwell \& Holdsworth described the transition in terms of the expected norm of the symmetry-breaking order parameter~\cite{Bramwell1993Magnetization,Bramwell1994Magnetization,Archambault1997MagneticFluctuations,Bramwell1998Universality}, as reviewed in Section~\ref{sec:MWHandBHtheories}.  This was fundamental to both the symmetry-breaking critical exponent~\cite{Bramwell1993Magnetization,Bramwell1994Magnetization} and descriptions of experimental measurements related to the expected norm in finite systems~\cite{Bramwell1993Magnetization,Huang1994MagnetismFewMonolayersLimit,Elmers1996CriticalPhenomenaTwoDimensionalMagnet,Chung1999EssentialFiniteSizeEffect,BedoyaPinto2021Intrinsic2DXYFerromagnetism,Venus2022RenormalizationGroup}.  
% resolving the symmetry-breaking paradox between theory and experiment -- describing experimental symmetry-breaking measurements related to the expected norm in finite systems~\cite{Bramwell1993Magnetization,Huang1994MagnetismFewMonolayersLimit,Elmers1996CriticalPhenomenaTwoDimensionalMagnet,Chung1999EssentialFiniteSizeEffect,BedoyaPinto2021Intrinsic2DXYFerromagnetism,Venus2022RenormalizationGroup}.  

Prior to the above field theory, Bramwell \& Holdsworth worked with others to develop a comprehensive framework for the expected norm of the symmetry-breaking order parameter in finite systems~\cite{Bramwell1993Magnetization,Bramwell1994Magnetization,Archambault1997MagneticFluctuations,Bramwell1998Universality}.  This included a thermodynamic critical exponent~\cite{Bramwell1993Magnetization,Bramwell1994Magnetization} and was fundamental to descriptions of experimental measurements related to the expected norm in finite systems~\cite{Bramwell1993Magnetization,Huang1994MagnetismFewMonolayersLimit,Elmers1996CriticalPhenomenaTwoDimensionalMagnet,Chung1999EssentialFiniteSizeEffect,BedoyaPinto2021Intrinsic2DXYFerromagnetism,Venus2022RenormalizationGroup}.  Recent advances~\cite{Faulkner2024SymmetryBreakingBKT} built on this to provide a complete thermodynamic theory for broken $U(1)$ symmetry -- within the concept of general symmetry breaking.  This broadened the elegant yet restrictive framework of spontaneous symmetry breaking by allowing the expected norm of the $U(1)$ order parameter to go to zero in the thermodynamic limit, provided the fluctuations in its directional phase are asymptotically smaller.  This asymptotically slow directional mixing of the $U(1)$ order parameter was demonstrated in the low-temperature BKT phase.  The low-temperature $U(1)$ order parameter therefore arbitrarily chooses some well-defined direction in the thermodynamic limit -- as in spontaneous symmetry breaking.  This resolved the third paradox and predicted negligible phase fluctuations compared to the expected norm in arbitrarily large experimental systems.  This low-temperature phase stabilization should be detectable via the Josephson current across a single Josephson junction formed from two nodes of superconducting film, the mean of the phase vectors measured at the nodes of a 2D Josephson-junction array, the magnetization vector in XY magnetic films (with a six-fold crystal field~\cite{Jose1977Renormalization}) and the orientational order parameter in the hexatic phase of colloidal films~\cite{Bernard2011TwoStepMelting,Thorneywork2017TwoDimensionalMelting}.  %The above was reviewed in detail in Section~\ref{sec:XYModelAndGSB}.

This review paper expanded on the %field theory and broken symmetry 
above concepts to elucidate fully the intimate connection between the topological nature of the BKT transition and its recently characterized symmetry-breaking properties.  Supplemental global-twist dynamics tunnel through the $U(1)$ sombrero potential via high-energy global topological defects in the emergent field.  These dynamics guarantee both topological ergodicity and $U(1)$ symmetry at all nonzero temperatures (on timescales that do not diverge with system size, in both cases) demonstrating that topological order (defined by the topological nonergodicity) induces broken $U(1)$ symmetry at low temperature~\cite{Faulkner2024SymmetryBreakingBKT}.  The field theory is the bridge connecting the two concepts because it splits the topological-defect field into its local and global components.  This complete theoretical framework connects the pioneering topological framework of the first vanguard of BKT theory~\cite{Salzberg1963EquationOfStateTwoDimensional,Mermin1966AbsenceFerromagnetism,Hohenberg1967ExistenceOfLong-RangeOrder,Berezinskii1973DestructionLongRangeOrder,Kosterlitz1973OrderingMetastability,Villain1975TheoryOneAndTwoDimensionalMagnets,Jose1977Renormalization} with the broad and diverse array of experimental measurements that are consistent with low-temperature broken symmetry~\cite{Baity2016EffectiveTwoDimensionalThickness,Shi2016EvidenceCorrelatedDynamics,Bishop1978StudySuperfluid,Bramwell2015PhaseOrder,Wolf1981TwoDimensionalPhaseTransition,Resnick1981KosterlitzThouless,Bramwell1993Magnetization,Huang1994MagnetismFewMonolayersLimit,Elmers1996CriticalPhenomenaTwoDimensionalMagnet,BedoyaPinto2021Intrinsic2DXYFerromagnetism,Hadzibabic2006KosterlitzThouless,Fletcher2015ConnectingBKTAndBEC,Christodoulou2021Observation}. 

This also has consequences for experimental measurements at the transition itself.  A corollary of the low-temperature broken symmetry is a flattening of the probability distribution of the $U(1)$ order parameter 
%The probability distribution of the magnetization necessarily flattens 
as the system transitions between its symmetric and symmetry-broken phases (cf., the order-parameter distributions in figure~\ref{fig:MagnetisationEvolution}).  For systems with local Brownian dynamics, %the resultant small configurational free-energy gradients should therefore provoke a critical slowing down at the transition.  
a critical slowing down should accompany the corresponding small probability gradients at the transition.  
This is a likely explanation of the strongly autocorrelated electrical resistance recently measured at the transition in various superconducting films~\cite{Shi2016EvidenceCorrelatedDynamics} because i) large condensate-phase differences (over long distances) induce resistance, and ii) increasingly large regions of symmetry-broken condensate-phase coherence (persisting on significant timescales) are a necessary precursor to the system-spanning symmetry-broken condensate-phase coherence of the low-temperature phase.  We suggest that this critical slowing down should also be detectable across the broad and diverse array of BKT experimental systems~\cite{Baity2016EffectiveTwoDimensionalThickness,Shi2016EvidenceCorrelatedDynamics,Bishop1978StudySuperfluid,Bramwell2015PhaseOrder,Wolf1981TwoDimensionalPhaseTransition,Resnick1981KosterlitzThouless,Bramwell1993Magnetization,Huang1994MagnetismFewMonolayersLimit,Elmers1996CriticalPhenomenaTwoDimensionalMagnet,BedoyaPinto2021Intrinsic2DXYFerromagnetism,Hadzibabic2006KosterlitzThouless,Fletcher2015ConnectingBKTAndBEC,Christodoulou2021Observation}.  Indeed, this is supported by the Metropolis simulation results presented in figures 2 and 7 of Archambault, Bramwell \& Holdsworth~\cite{Archambault1997MagneticFluctuations}.  The former indicates a flattening of the order-parameter distribution near the transition, while the estimated magnetic susceptibility presented in the latter is very noisy near the transition, suggesting that local Metropolis simulations exhibit very slow dynamics in this region.  The present paper is therefore likely to provide a platform connecting the topological character of the BKT transition with strongly correlated dynamics on long experimental timescales.  It will also be interesting to explore the poor Metropolis exploration of the asymmetric heavy tail of the low-temperature $\| \mvec \|$ distribution~\cite{Archambault1997MagneticFluctuations} suggested by both figures~\ref{fig:MagnetisationEvolution}(a)-(b) and the noisy statistics also present at low temperature in figure 7 of Archambault, Bramwell \& Holdsworth~\cite{Archambault1997MagneticFluctuations}.

In analogy with both Swendsen--Wang and Wolff simulations~\cite{Swendsen1987Nonuniversal,Wolff1989Collective} circumventing critical slowing down in the 2D Ising model, the constant-speed dynamics of event-chain Monte Carlo may indeed alleviate the above challenges of slow Metropolis dynamics along low-gradient directions of the order-parameter distribution -- certainly those due to the continuous spin-wave fluctuations~\cite{Lei2018IrreversibleMarkovChains}.  %At low temperature, this hypothesis is supported by the superdiffusive dynamics of the location of the 2DXY active spin~\cite{Kimura2017AnomalousDiffusion}, and we conjecture that it 
We conjecture that this hypothesis is particularly strong for the asymmetric-velocity process presented here, as this appears to maintain ballistic-style dynamics on long timescales.  Indeed, it will also be interesting to explore whether this process leads to faster directional mixing (than its symmetric-velocity counterpart) in the symmetry-broken phase.  At the transition, alleviating critical slowing down would be consistent with both the superdiffusive dynamics of the location of the 2DXY active spin~\cite{Kimura2017AnomalousDiffusion} and the $\sqrt{N}$ speed-up of lifted (relative to standard) Metropolis--Hastings simulations of the Curie--Weiss model at its phase transition~\cite{Bierkens2017APiecewiseDeterministic}, as the limiting process of this algorithm is a similar piecewise deterministic Markov process from Bayesian computation.  On the other hand, it is possible that event-chain Monte Carlo (as presented here) cannot circumvent the contribution from the local topological defects~\cite{Lei2018IrreversibleMarkovChains} -- in which case it could become a powerful tool for separating physical effects near the transition.  Moreover, biasing event-chain dynamics with \emph{factor fields} typically accelerates mixing~\cite{Lei2019ECMCwithFactorFields,Krauth2024HMCvsECMC} and it would be interesting to explore their impact on the local topological defects.  More broadly, this motivates further questions regarding the speed of piecewise deterministic Markov processes at phase transitions across statistical physics and computational statistics.  Furthermore, the foundational formulation of broken ergodicity as some asymptotically slow mixing was strongly influenced by Bayesian computation, demonstrating the power of cross-pollination of knowledge~\cite{Faulkner2024SamplingAlgorithms}.

\acknowledgements
It is a pleasure to thank S.~T. Bramwell, P.~C.~W. Holdsworth, A.~C. Maggs, A. Taroni, Z. Shi, D. Popovi\'{c}, S.~Livingstone and S.~Grazzi for many illuminating discussions.  The author is particularly grateful to STB and PCWH for their outstanding supervision of his doctoral research, which led to Section~\ref{sec:EmergentElectrolyte}.  The author is grateful to the reviewers for useful comments and acknowledges support from EPSRC fellowship EP/P033830/1.  Simulations were performed on BlueCrystal 4 at the Advanced Computing Research Centre (University of Bristol) and Avon at the Scientific Computing Research Technology Platform (University of Warwick).  

\appendix

\section{Code and simulation data}
\label{sec:CodeAndSimulationData}

Code is freely available on GitHub at \href{https://github.com/michaelfaulkner/super-aLby}{https://github.com/michaelfaulkner/super-aLby}, commit hash \href{https://github.com/michaelfaulkner/super-aLby/commit/0ec5116ad97ab771d26741eea8ca439b4e8e674e}{0ec5116} (Ising simulations)~\cite{FaulknerSuperAlby} and \href{https://github.com/michaelfaulkner/xy-type-models}{https://github.com/michaelfaulkner/xy-type-models}, commit hash \href{https://github.com/michaelfaulkner/xy-type-models/commit/0adf8e722ac849e616458fbbc6975b26e9bb8841}{0adf8e7} (lattice-electrolyte and XY simulations)~\cite{FaulknerXyTypeModels}.  All published data can be reproduced using these applications (as outlined in each README).  The simulation data used to make figures~\ref{fig:2dIsingMagDensityVsTime} and \ref{fig:ResponseFunctions}--\ref{fig:TopologicalSusceptibilities} are available at the \href{https://datadryad.org}{Dryad data repository} at \href{https://doi.org/0.5061/dryad.v15dv427n}{https://doi.org/0.5061/dryad.v15dv427n} \cite{Faulkner2025EmergentElectrostatics}.  Those used to make 
figures~\ref{fig:MagnetisationEvolution}--\ref{fig:GlobalTwists} are available at the University of Bristol data repository, \href{https://data.bris.ac.uk/data/}{data.bris}, at \href{https://doi.org/10.5523/bris.3ov1rl6xtshwv2iuixrbs6f39q}{https://doi.org/10.5523/bris.3ov1rl6xtshwv2iuixrbs6f39q} \cite{Faulkner2023GeneralSymmetryBreaking}.  Simulations for figures~\ref{fig:MagnetisationEvolution}, \ref{fig:ECDFs} and \ref{fig:GlobalTwists}(a) started from randomized configurations.  Others started from cold or `inherited' configurations, as outlined in the relevant README and configuration files.  Standard error estimates were used (i.e., not accounting for autocorrelation in the Markov chain) and all non-visible error bars are smaller than the marker size.

\section{Polarization}
\label{sec:Polarisation}

The relationship in equation~\eqref{eq:EbarPolarisationTopSector} is seen by splitting the sum of the $x$ component of the emergent field into separate sums over all $x$ components that enter a particular strip of plaquettes of width $a$ that wrap around the torus in the $y$ direction: 
\begin{widetext}
\begin{align}
     \frac{L^2}{a} \bar{E}_x = & \, a \sum_{\rvec \in D} E_x \left( \rvec + \frac{a}{2} \UnitX \right) \nonumber\\
     = & \, \frac{a}{2} \sum_{y = a / 2}^{L - a / 2} E_x \left( L, y \right) + \sum_{x = a / 2}^{L - a / 2}  x \sum_{y = a / 2}^{L - a / 2} \left[ E_x \left( x \ominus \frac{a}{2}, y \right) - E_x \left( x \oplus \frac{a}{2}, y \right) \right] + \left( L - \frac{a}{2} \right) \sum_{y = a / 2}^{L - a / 2} E_x \left( L, y \right) \nonumber\\
     = & \, \sum_{x = a / 2}^{L - a / 2}  x \sum_{y = a / 2}^{L - a / 2} \left[ E_x \left( x \ominus \frac{a}{2}, y \right) - E_x \left( x \oplus \frac{a}{2}, y \right) \right] + L \sum_{y = a / 2}^{L - a / 2} E_x \left( L, y \right) \nonumber\\
     = & \, - J a \sum_{x = a / 2}^{L - a / 2} x \sum_{y = a / 2}^{L - a / 2} \rho (x, y) + L \sum_{y = a / 2}^{L - a / 2} E_x \left( L, y \right) ,
\end{align}
\end{widetext}
and similarly for $L^2 \bar{E}_y / a$.

\section{Some additional discrete vector calculus}
\label{sec:DiscreteVectorCalculus}

It is useful to consider the divergence and curl of the lattice (emergent) electric field 
\begin{align}
\Evec (\rvec) = - \NablaTildePhi(\rvec) + \Ebar + \hat{\boldsymbol{\nabla}} \times \Qvec (\rvec) ,
\end{align}
as defined in equation~\eqref{eq:ElectricFieldHelmholtzHodge}.  The harmonic mode $\Ebar$ is spatially constant, so that $\hat{\boldsymbol{\nabla}} \cdot \Ebar = 0$ and $\widetilde{\boldsymbol{\nabla}} \times \Ebar = 0$.  Since 
\begin{align}
\hat{\boldsymbol{\nabla}} \cdot \hat{\boldsymbol{\nabla}} \times \Qvec (\rvec) = \varepsilon_{ijk} \hat{\boldsymbol{\nabla}}_i \hat{\boldsymbol{\nabla}}_j Q_k\left(\rvec + \frac{a}{2}\mathbf{e}_k\right) = 0 ,
\end{align}
it then follows that the divergence of the lattice electric field is 
\begin{align}
\hat{\boldsymbol{\nabla}} \cdot \Evec (\rvec) = -\hat{\boldsymbol{\nabla}} \cdot \NablaTildePhi(\rvec) = -\NablaBF^2 \phi (\rvec) .
\end{align}
Similarly, since 
\begin{align}
\left[\widetilde{\boldsymbol{\nabla}} \times \NablaTildePhi(\rvec) \right]_i = \varepsilon_{ijk} \widetilde{\boldsymbol{\nabla}}_j \widetilde{\boldsymbol{\nabla}}_k \phi(\rvec) = 0 
\end{align}
for all $i \in \{ x, y, z \}$, it follows that the curl of the lattice electric field is 
\begin{align}
\widetilde{\boldsymbol{\nabla}} \times \Evec (\rvec) = \widetilde{\boldsymbol{\nabla}} \times \hat{\boldsymbol{\nabla}} \times \Qvec (\rvec) .
\end{align}
Moreover, 
\begin{align}
\left[ \widetilde{\boldsymbol{\nabla}} \times \hat{\boldsymbol{\nabla}} \times \Qvec (\rvec) \right]_i &= \varepsilon_{ijk} \widetilde{\boldsymbol{\nabla}}_j \left[ \hat{\boldsymbol{\nabla}} \times \Qvec (\rvec) \right]_k \label{eq:CurlCurlQ} \\
&= \varepsilon_{ijk} \varepsilon_{klm} \widetilde{\boldsymbol{\nabla}}_j \hat{\boldsymbol{\nabla}}_l Q_m \left( \rvec + \frac{a}{2} \mathbf{e}_m \right) \nonumber\\
%&= (\delta_{il}\delta_{jm} - \delta_{im}\delta_{jl}) \widetilde{\boldsymbol{\nabla}}_j \hat{\boldsymbol{\nabla}}_l Q_m \left( \rvec + \frac{a}{2} \mathbf{e}_m \right) \nonumber\\
&= \hat{\boldsymbol{\nabla}}_i \left( \widetilde{\boldsymbol{\nabla}} \cdot \Qvec(\rvec) \right) - \NablaBF^2 Q_i \left( \rvec + \frac{a}{2} \mathbf{e}_i \right) \nonumber 
\end{align}
for all $i \in \{ x, y, z \}$ [using $\hat{\boldsymbol{\nabla}}_i \widetilde{\boldsymbol{\nabla}}_j f(\rvec) = \widetilde{\boldsymbol{\nabla}}_j \hat{\boldsymbol{\nabla}}_i f(\rvec)$ for all $i, j \in \{ x, y, z \}$ for any scalar test function $f$].  Since we have chosen $\Qvec(\rvec) = [0, 0, Q(\rvec)]^T$ and $\widetilde{\boldsymbol{\nabla}} \cdot \Qvec(\rvec) = 0 \,\, \forall  \rvec$ for all such vector fields on a 2D lattice, it then follows that  
\begin{align}
\widetilde{\boldsymbol{\nabla}} \times \Evec (\rvec) = - \NablaBF^2 Q(\rvec) \UnitZ .
\label{eq:CurlE}
\end{align}
This is enforced by the second Dirac object in both equation~\eqref{eq:ConstrainedElectrolytePDF1} and equation~\eqref{eq:MaggsRossettoPDF}. 

It is also helpful to show that starting from equation~\eqref{eq:CurlE} implies that the rotational component of $\Evec(\rvec)$ is $[\hat{\boldsymbol{\nabla}}_y Q(\rvec), -\hat{\boldsymbol{\nabla}}_x Q(\rvec)]^T$, as required.  For any 2D vector field $\Evec(\rvec)$ with rotational component $\hat{\boldsymbol{\nabla}} \times \Qvec'(\rvec)$ described by some $\Qvec'(\rvec) = [Q_x'(\rvec + a\UnitX / 2), Q_y'(\rvec + a\UnitY / 2), Q_z'(\rvec + a\UnitZ / 2)]^T$ on a 2D lattice, $\widetilde{\boldsymbol{\nabla}} \times \Evec (\rvec) = \hat{\boldsymbol{\nabla}} \left( \widetilde{\boldsymbol{\nabla}} \cdot \Qvec'(\rvec) \right) - \NablaBF^2 \Qvec'(\rvec)$ in general [see equation~\eqref{eq:CurlCurlQ}].  On a 2D lattice, equation~\eqref{eq:CurlE} then implies that 
\begin{align}
\NablaBF^2 Q_z'\left(\rvec + \frac{a}{2}\UnitZ \right) = \NablaBF^2 Q(\rvec),
\label{eq:CurlEz}
\end{align}
\begin{align}
\widetilde{\boldsymbol{\nabla}}_y \left[ \hat{\boldsymbol{\nabla}} \times \Qvec'(\rvec) \right]_z = 0
\label{eq:CurlEx}
\end{align}
and
\begin{align}
\widetilde{\boldsymbol{\nabla}}_x \left[ \hat{\boldsymbol{\nabla}} \times \Qvec'(\rvec) \right]_z = 0.
\label{eq:CurlEy}
\end{align}
In equation~\eqref{eq:CurlEz} we used $\hat{\boldsymbol{\nabla}}_z \left( \widetilde{\boldsymbol{\nabla}} \cdot \Qvec'(\rvec) \right) = 0$ on a 2D lattice, while equation~\eqref{eq:CurlEx} is a result of the $x$ component of equation~\eqref{eq:CurlE}: 
\begin{align}
0 = & \hat{\boldsymbol{\nabla}}_x \left( \widetilde{\boldsymbol{\nabla}} \cdot \Qvec'(\rvec) \right) - \NablaBF^2 Q_x'\left(\rvec + \frac{a}{2}\UnitX \right) \nonumber\\
= & \hat{\boldsymbol{\nabla}}_x \left(\widetilde{\boldsymbol{\nabla}}_x Q_x'\left(\rvec + \frac{a}{2}\UnitX \right) + \widetilde{\boldsymbol{\nabla}}_y Q_y'\left(\rvec + \frac{a}{2}\UnitY \right) \right) \nonumber\\
& - \NablaBF^2 Q_x'\left(\rvec + \frac{a}{2}\UnitX \right) \nonumber\\
= & \hat{\boldsymbol{\nabla}}_x \widetilde{\boldsymbol{\nabla}}_y Q_y'\left(\rvec + \frac{a}{2}\UnitY \right) - \hat{\boldsymbol{\nabla}}_y \widetilde{\boldsymbol{\nabla}}_y Q_x'\left(\rvec + \frac{a}{2}\UnitX \right) \nonumber\\
= & \widetilde{\boldsymbol{\nabla}}_y \left[ \hat{\boldsymbol{\nabla}} \times \Qvec'(\rvec) \right]_z ,
\end{align}
and analogously for equation~\eqref{eq:CurlEy} [in the second line, we used $\widetilde{\boldsymbol{\nabla}} \cdot \Qvec'(\rvec) = \widetilde{\boldsymbol{\nabla}}_x Q_x'\left(\rvec + a \UnitX / 2 \right) + \widetilde{\boldsymbol{\nabla}}_y Q_y'\left(\rvec + a \UnitY / 2 \right)$ on a 2D lattice].

Equation~\eqref{eq:CurlEz} $\Rightarrow \hat{\boldsymbol{\nabla}}_{x / y} Q_z'\left(\rvec + a\UnitZ / 2 \right) = \hat{\boldsymbol{\nabla}}_{x / y} Q(\rvec) + \textrm{const}_{x / y}$ where both arbitrary constants $\textrm{const}_{x / y}$ must be zero for the $x$ and $y$ cases to hold simultaneously.  Now since $\hat{\boldsymbol{\nabla}}_z Q_{x / y}'(\rvec + a\mathbf{e}_{x / y} / 2) = 0$ on a 2D lattice, it then follows that 
\begin{align}
\left[ \hat{\boldsymbol{\nabla}} \times \Qvec'(\rvec) \right]_x = \hat{\boldsymbol{\nabla}}_y Q(\rvec) 
\label{eq:CurlQx}
\end{align}
and
\begin{align}
\left[ \hat{\boldsymbol{\nabla}} \times \Qvec'(\rvec) \right]_y = - \hat{\boldsymbol{\nabla}}_x Q(\rvec) .
\label{eq:CurlQy}
\end{align}
In addition, equation~\eqref{eq:CurlEx}/\eqref{eq:CurlEy} implies that $\left[ \hat{\boldsymbol{\nabla}} \times \Qvec'(\rvec) \right]_z$ is a function only of $x/y$, implying that 
\begin{align}
\left[ \hat{\boldsymbol{\nabla}} \times \Qvec'(\rvec) \right]_z = \textrm{const.} 
\label{eq:CurlQz}
\end{align}
Equations~\eqref{eq:CurlQx}-\eqref{eq:CurlQz} imply that equation~\eqref{eq:CurlE} enforces the required properties in the rotational component of the (emergent) electric field $\Evec$ in equations~\eqref{eq:ConstrainedElectrolytePDF1} and \eqref{eq:MaggsRossettoPDF}.  Moreover, equations~\eqref{eq:CurlQx}-\eqref{eq:CurlQz} also imply that equation~\eqref{eq:CurlE} enforces  
\begin{align}
\exp \left( -\zeta \| \hat{\boldsymbol{\nabla}} \times \Qvec'(\rvec) \|^2 \right) \propto \exp \left( -\zeta \| \hat{\boldsymbol{\nabla}} Q(\rvec) \|^2 \right) 
\end{align}
for any fixed $\zeta \in \RR$ (i.e., including $\zeta = 1$) as required to transform (respectively) the Boltzmann distributions in equations~\eqref{eq:ConstrainedElectrolytePDF1} and \eqref{eq:MaggsRossettoPDF} to those in equations~\eqref{eq:ConstrainedElectrolytePDF2} and \eqref{eq:RealElectrolyteFactorisedPDF}.  Indeed, our arbitrary choice of $\Qvec(\rvec) = [0, 0, Q(\rvec)]^T$ gives $\left[ \hat{\boldsymbol{\nabla}} \times \Qvec (\rvec) \right]_z = 0$ on a 2D lattice, and therefore 
\begin{align}
\| \hat{\boldsymbol{\nabla}} \times \Qvec (\rvec) \|^2 = \| \hat{\boldsymbol{\nabla}} Q(\rvec) \|^2 .
\end{align}

\section{Lattice Green's function}
\label{sec:GreensFunction}

To solve the lattice Green's function defined by $a^2 \boldsymbol{\nabla}_\rvec^2 G(\rvec, \rvec') = - \mathbb{I} \left[ \rvec = \rvec' \right] \,\, \forall \, \rvec, \rvec' \in D$ in equation~\eqref{eq:LatticeGreensFunctionDefinition}, we define the $\kvec$-space lattice Green's function 
\begin{align}
\widetilde{G}_{\rvec'}(\kvec) := \sum_{\rvec} e^{-i\kvec \cdot \rvec} G(\rvec, \rvec') .
\end{align}
Since $\sum_{\kvec} e^{i\kvec \cdot (\rvec - \rvec')} / N = \mathbb{I} \left[ \rvec = \rvec' \right]$, we may now write 
\begin{align}
\sum_{\kvec} e^{i\kvec \cdot (\rvec - \rvec')} = & - a^2 \boldsymbol{\nabla}_\rvec^2 \sum_{\kvec} e^{i\kvec \cdot \rvec} \widetilde{G}_{\rvec'}(\kvec) \nonumber\\
= \, & 2 \sum_{\kvec} \left[ 2 - \cos(k_x a) - \cos(k_y a) \right] \nonumber\\
& \,\qquad \times e^{i\kvec \cdot \rvec} \widetilde{G}_{\rvec'}(\kvec) ,
\end{align}
which is solved by 
\begin{align}
    \widetilde{G}_{\rvec'}(\kvec) = 
    \begin{cases}
            \frac{\exp \left( - i\kvec \cdot \rvec' \right)}{2\left[ 2 - \cos (k_x a) - \cos (k_y a) \right]} \, & \forall \, \kvec \ne 0 , \\
            0 & {\rm for} \, \kvec = 0 .
    \end{cases}
\label{eq:KspaceLatticeGreensFunctionSoln}
\end{align}
We note again that we chose to set the $\kvec = 0$ mode in equation~\eqref{eq:KspaceLatticeGreensFunctionSoln} to zero but we could choose any real number as this $\kvec = 0$ mode does not appear in the potential for a charge-neutral system.

\section{Villain model}
\label{sec:MaggsToVillain}

Expressing the generalized lattice-field electrolyte in terms of spin-like fields demonstrates~\cite{Faulkner2015TSFandErgodicityBreaking,Faulkner2017AnElectricFieldRepresentation} the equivalence with Villain's approximation to the 2DXY model~\cite{Villain1975TheoryOneAndTwoDimensionalMagnets}.  We redefine $\DeltaPhi$ such that its $\mu \in \{ x, y \}$ component is defined as 
\begin{align}
\label{ExtendedSpinDifference}
[ \DeltaPhi ]_\mu \left( \rvec + \frac{a}{2}\mathbf{e}_\mu \right) := \varphi(\rvec\! \oplus \!a\mathbf{e}_\mu)\!-\! \varphi(\rvec)\! +\! q \hat{s}(\rvec\! \oplus \!a\mathbf{e}_\mu,\rvec) 
\end{align}
with $\{ \varphi (\rvec) \in [-q / 2, q / 2) : \rvec \in D' \}$ and $\{ \hat{s}(\rvec \oplus a\mathbf{e}_\mu, \rvec) \in \ZZ : \rvec \in D' \}$.  We then make the identification 
\begin{align}
\label{MaggsFieldIdentify}
\mathbf{E}(\rvec)\equiv \frac{1}{\epsilon_0 a}
\left( \begin{array}{c}
[ \DeltaPhi ]_y(\rvec + a \mathbf{e}_x / 2)  \\
\\
-[ \DeltaPhi ]_x(\rvec + a \mathbf{e}_y / 2)  \end{array} \right).
\end{align}
The movement of a positive charge $q$ from site $\rvec$ to site $\rvec \oplus a \mathbf{e}_{x / y}$ corresponds to a unit decrease/increase in $\hat{s}(\rvec \oplus a \mathbf{e}_{x / y} / 2 \oplus a \mathbf{e}_{y / x} / 2, \rvec \oplus a \mathbf{e}_{x / y} / 2 \ominus a \mathbf{e}_{y / x} / 2)$ [see figure~\ref{fig:MaggsRossettoDynamics}(a)]~\footnote{In \cite{Faulkner2015TSFandErgodicityBreaking}, it was erroneously stated that $\hat{s}$ components change in increments of $\pm q$.  In \cite{Faulkner2017AnElectricFieldRepresentation}, it was correctly stated that they change in increments of $\pm 1$.}.  A clockwise electric-flux rotation of $\Delta / \epsilon_0$ around some spin site $\rvec_0 \in D'$ corresponds to a decrease $\Delta$ in $\varphi(\rvec_0)$.  To ensure that $\varphi(\rvec_0)$ remains within $[-q / 2, q / 2)$, the operation $\varphi(\rvec_0) \mapsto \left[ \varphi(\rvec_0) + q / 2 \right] \!\!\! \mod (q) - q / 2$ is then applied; if this operation has an effect [i.e., if $\varphi(\rvec_0) \notin [-q / 2, q / 2)$ between the $\varphi$ rotation and the operation] each of the four $\hat{s}$ field components associated with $\rvec_0$ are updated by $\pm 1$ such that the new charge configuration equals the charge configuration before the $\varphi$ rotation.  This modulo operation is required because $\varphi(\rvec) \in [-q / 2, q / 2)$ for all $\rvec \in D'$.  The field rotation is depicted in figure~\ref{fig:MaggsRossettoDynamics}(b) where the $\varphi$ component is represented by a spin-like arrow to emphasize the analogies with the XY models.  With $q = 2\pi$ and $\epsilon_0 = 1 / J$, the Boltzmann distribution in equation~\eqref{eq:MaggsRossettoPDF} then becomes that of the Villain model:  % with an additional core-energy term:
%\begin{widetext}
%\begin{align}
%\pi_{\rm Maggs} \glb \gld s \glb \rvec \grb , \varphi \glb \rvec \grb \grd , \epsilon_{\rm core} \grb \propto & \,\, \exp \left[ -\frac{\beta J}{2} \sum_{\langle \rvec,\rvec' \rangle}  |\varphi(\rvec) - \varphi(\rvec') + 2\pi s(\rvec, \rvec')|^2 + \frac{\beta a^4}{2}\sum_{\mathbf{r}\in D}\epsilon_{\rm core}(m(\rvec)) \rho(\mathbf{r})^2 \right] .
%\end{align}
%\end{widetext}
\begin{widetext}
\begin{align}
\pi_{\rm Maggs} \glb \gld \varphi(\rvec), \hat{s}(\rvec, \rvec') \grd \grb \propto \exp \left[ -\frac{\beta J}{2} \sum_{\langle \rvec,\rvec' \rangle}  |\varphi(\rvec) - \varphi(\rvec') + 2\pi \hat{s}(\rvec, \rvec')|^2 \right] .
\end{align}
\end{widetext}

\bibliographystyle{ieeetr}
\bibliography{Faulkner}

\end{document}